\documentclass{report}

\usepackage{setspace}
\usepackage{brandeis_thesis}
\usepackage{amsmath}
\usepackage{latexsym}
\usepackage{theorem}

\newcounter{df}[section]

\newcommand{\Def}{\par \addtocounter{df}{1}%
{\bf Definition \arabic{section}.\arabic{df}. }}
\newcommand{\Rem}{\par {\bf Remark: }}
\newtheorem{Lemma}{Lemma}[section]

\newcommand{\Fin}{{\cal P}_{\mbox{fin}}}
\newcommand{\Proof}{{\bf Proof. }}
\newcommand{\eproof}{ $\Box$}

\newtheorem{lemma}{Lemma} [section]
\newtheorem{theorem}{Theorem} [section]
\newtheorem{axiom}{Axiom} [section]
\newtheorem{corollary}{Corollary} [section]

\newcommand{\Tcal}{\cal T}
\newcommand{\sql}{\sqsubseteq}

\newcommand{\N}{{\bf N}}
\newcommand{\R}{{\bf R}}
\newcommand{\Int}{{\rm Int}}
\newcommand{\Fix}{{\rm Fix}}
\newcommand{\Z}{{\bf Z}}
\newcommand{\MUB}{{\rm MUB}}
\newcommand{\Info}{{\it Info}}
\newcommand{\Neginfo}{{\it Neginfo}}
\newcommand{\muInfo}{{\it \mu Info}}
\newcommand{\Total}{{\it Total}}
\newcommand{\Sub}{{\it Sub}}

{\theorembodyfont{\rmfamily}

\newtheorem{definition}{Definition} [section]
\newtheorem{example}{Example} [section]
}

\begin{document}

%  \thesistitle{Title}{Dept}{Advisor}{Student}{Date}
%  \sigpagethree{Student}{Advisor}{Two}{Three}
%  \copyrightpage{Student}{Year}

\thesistitle{Mathematics of Domains}{Computer Science}{Harry G. Mairson}{Michael A. Bukatin}{February 2002}
\sigpagethree{Michael A. Bukatin}{Harry G. Mairson}{Ralph D. Kopperman, City College of New York}{Timothy J. Hickey, Brandeis University}

\pagenumbering{roman}   %% front matter numbered in roman
\setcounter{page}{2}    %% NOTE: \thesisabs will add one to page

\thesisack{{\em Dedicated to the memory of Anya Pogosyants and Igor Slobodkin}}
{

In chronological order, I would like to thank Alexander Shen',
who introduced me to domain theory, which he used in his remarkable
PhD Thesis on algoritmic variants of the notion 
of entropy, and Alexander Saevsky, for fruitful collaboration
in 1986-1989, which enabled my results on subdomains and
finitary retractions, presented here.

I would like to thank my advisor 
Harry Mairson for arranging  financial support
from Fall 1992 to Spring 1996 under NSF Grant CCR-9216185 and
Office of Naval Research Grant ONR N00014-93-1-1015 and for
numerous fruitful discussions. 

I am especially thankful to Joshua Scott and Svetlana Shorina,
for the joint work in my research project in
the field of analysis on domains. The results of our joint papers
are the most significant results reported in this Thesis.

The influence of Abbas Edalat, Bob Flagg, and Klaus Keimel was quite
crucial for some parts of this Thesis.

I would also like to acknowledge helpful communications with
Michael Alekhno- vich,
Alexander Artemyev, Will Clinger, 
Yuri Ershov, Nadezhda Fradkova, Carl Gunter,
Reinhold Heckmann, Michael Huth,
Ataliya Kagan, Ralph Kopperman, Elena Machkaso- va, Yuri Matiyasevich,
Steve Matthews, Albert Meyer, Michael Mislove, Robert Muller,
Simon O'Neill, Gordon Plotkin, Jeffrey Sanchez, Vladimir Sazonov,
Ross Viselman, Steve Vickers, and Mitch Wand.

}

\thesisabs{Mathematics of Domains}
           {Michael A. Bukatin}
{

Two groups of naturally arising questions in the
mathematical theory of {\em domains for
denotational semantics} are addressed. Domains are equipped with
{\em Scott topology} and
represent {\em data types}. {\em Scott continuous functions}
represent {\em computable functions} and form the most
popular continuous model of computations.

{\bf Covariant Logic of
Domains:} Domains are represented as {\em sets
of theories}, and Scott continuous functions are represented
as {\em input-output inference engines}. The questions
addressed are:
{\bf A.} What constitutes a {\em subdomain}? 
Do subdomains of a given domain $A$
form a domain? 
{\bf B.} Which {\em retractions} are {\em finitary}? 
{\bf C.} What is the essence of
generalizations
of {\em information systems} based on {\em non-reflexive logics}?
Are these generalizations
restricted to continuous domains?

{\bf Analysis on Domains:}

{\bf D.} How to describe Scott topologies via
{\bf generalized distance functions} satisfying the requirement
of Scott continuity (``abstract computability'')? The answer is 
that the 
axiom $\rho (x, x) = 0$ is incompatible with Scott continuity
of distance functions.
The resulting {\bf relaxed metrics} are studied.

{\bf E.} Is it possible to obtain Scott continuous relaxed metrics via
{\em measures} of domain subsets representing 
{\em positive} and {\em negative
information} about domain elements?
The positive answer is obtained via
the discovery of the novel class of {\bf co-continuous
valuations} on the systems of Scott open sets.

Some of these natural questions were studied earlier.
However, in each case a novel approach is presented,
and the answers are supplied with much more compelling
and clear justifications, than were known before.

}

\doublespacing      %% almost everything needs to be double spaced

\setcounter{secnumdepth}{10}

\setcounter{chapter}{-1}

\chapter{Preface}

This text represents the results of my research in mathematics
of approximation domains equipped with Scott topology.
I would be the first to classify this research as belonging
to the field of pure mathematics, yet it is being presented
as a dissertation in computer science. I think, this demands some
explanation.

\section{Accepted Practice}

There is a widely recognized body of computer science applications
of the mathematical theory of domains, reflected in dozens of monographs,
and hundreds, if not thousands, of research papers (some of the related
references are in the text). Some of the most
prominent contributors to this theory work at departments of
computer science, rather than at departments of mathematics.
Due to these reasons, it is quite often that purely mathematical
research in this field is presented as doctoral dissertations in computer
science, even in the leading schools.

For example, a famous work by Kim Wagner~\cite{Wagner} defended
as a PhD Thesis in the School of Computer Science of
Carnegie Mellon University, clearly belongs to this class
of research. As my main achievements to this moment are
in mathematical theory, and not in its applications, I believe it
appropriate to follow this practice.

\section{Recongnized Applications}

The main recognized application of domains is to serve as denotational
models of programming languages.
This application is due to Christofer Strachey, who introduced the idea
of denotational semantics, and Dana Scott, who invented approximation
domains equipped with what is now called Scott topology and
constructed extensional models of lambda-calculus by solving
reflexive domain equations like $D \cong [ D \rightarrow D ]$.
The classical textbook
describing this class of applications is~\cite{Stoy},
and there are many successors. It is widely argued that denotational
descriptions of programming languages using approximation
domains should be considered canonical definitions of those languages,
while other types of descriptions should be considered as derivative.
This viewpoint obviously has its share of opponents too.

Among more hands-on applications of this approach are verifications
of compiler correctness, uses of abstract interpretation to
perform some static analysis of programs, attempts to produce
logical frameworks based on denotational semantics, etc.

Another important application, which emerged in the nineties,
is to use domains to produce computational models for
classical mathematical structures and thus to provide the framework
for computations in such structures. This direction of research
was originated by Abbas Edalat, who used domains to generalize
the notion of Riemann integral to fractal spaces and obtained more
efficient methods to compute integrals on such spaces
(see~\cite{Edalat} and references therein).
The research group at Imperial College headed by Edalat is now
working towards constructing systematic models of this kind
for a wide class of mathematical structures
(see~\cite{EdalatSurvey} for the review of these efforts).

\section{Potential Applications}

Many researchers in this field probably have their own lists
of prospective applications, which often do not reduce to the
two classes of applications described before.
My own list includes the following.

When denotational semantics was first introduced, it was
suggested that the canonical approach would be not to produce
denotational description of existing languages, but to design
languages fitting the denotational models of interest.
While this approach has not received sufficient development,
I believe that the potential for such development exists.

Another class of applications is related to the
potentially possible development
of our abilities to perform computations in domains with
reasonable efficiency, i.e. to be able to find reasonable
approximations for a sufficiently wide class of definitions
of domain elements
and Scott continuous functions while spending realistic
amounts of resources. We call this open problem {\bf Problem A}
from now on in this text.
I cannot predict at this point, whether
such a development is actually possible. However, if it is
possible, one immediate consequence would be the possibility
of a generator of prototype implementations of programming
languages based on their denotational semantics.

The central part of this monograph develops theories of relaxed
metrics and co-continuous valuations on domains. This
development is just one manifestation of the rapid evolution
of analysis on domains taking place recently. There can be little
doubt that a lot of analogs of basic structures of functional
analysis will soon emerge for domains. If Problem A is actually
solved, these analogs should enable us to transfer
various methods of optimization known in classical mathematics
into the computational setting and should lead to the development of
the whole new class of advanced methods for machine learning.

Thus Problem A is quite crucial for the further applicability
of this field.

\section{Simplicity of Presentation and Prerequsites}

I make an emphasis on simplicity and intuitive
clarity of presentation. The techniques here are 
deliberately elementary and non-categorical,
except for those rare cases when categorical methods
are unavoidable.

The Introduction can be used as a brief tutorial for the
readers, who are not familiar with the field.
The Introduction sacrifices the completeness of presentation
of the standard information on domains for the accessibility.

Thus the knowledge of domain
theory and category theory is not strictly required to
read this text. The reader is, however, expected to be somewhat
familiar with basic notions of the following
fields: formal logic, theory
of metric and topological spaces, measure theory.

\section{Structure of the Text}

The Introduction can be thought of as a short tutorial in domain
theory and a repository of its basic definitions. It also contains
a detailed overview of other parts of the text.

Part~\ref{part:Logic} and Part~\ref{part:Analysis} constitute
the core of the Thesis. Part~\ref{part:Logic} presents our
results related to the logic of spaces of fixed points.
Part~\ref{part:Analysis} presents our results in the theory
of generalized distances and measures for domains.

\section{Electronic Coordinates}

Here are some relevant electronic coordinates. I hope they will be alive
for a while.

\noindent
My e-mail address: {\tt bukatin@@cs.brandeis.edu}

\noindent 
This thesis: {\tt http://www.cs.brandeis.edu/$\sim$bukatin/thesis.ps.gz}

\noindent
My papers in computer science: 

{\tt http://www.cs.brandeis.edu/$\sim$bukatin/papers.html}

\tableofcontents
%  \listoftables
%  \listoffigures
\newpage                %% strange KLUDGE

%  %%    ===== body of thesis =====
\pagenumbering{arabic}  %% body of thesis numbered in arabic numbers
\doublespacing          %% just in case

%  \chapter{Introduction}
%   ....(body of thesis)....
%

\part{Introduction}\label{part:Intro}
\chapter{Continuity in Computations and Scott Domains}

Here we present evidence of the
continuous nature of computations~\cite{Bukatin4}.
 
\section{Continuity in Software Engineering}
 
It seems appropriate to start with a real life observation
before dealing with mathematical models. Virtually all software
products used today contain a number of bugs, which lead to refusal
to work, crashes, and incorrect processing of data. Yet for a usable
software product these unpleasant effects are observed relatively
infrequently, as most of the time the product works satisfactory.
What we should say here is that the real product is sufficiently close
to the ideal product, close enough to be used in lieu of the unavailable
ideal product. This is a phenomenon of continuity. This continuity
results from the process of testing and amending the software
and is crucial for our ability to use any software at all.
Moreover, it is quite possible that the ideal product cannot
exist at all (e.g. if the specifications are for an undecidable
problem), yet we still can consider models which contain representatives
for such ideal products and for their real-life approximations.
 
The author so far has not encountered studies dealing with this
phenomenon mathematically. Software reliability models do not
seem to qualify,
because they always abstract from the nature of the particular software
involved and just study the dynamics of error rates.
We would like to see the difference between the
ideal software and its real-life approximation expressed semantically,
error classification and rates being built on top of such semantical
structure.
We think that such approach to software design is potentially
more promising than the concept of fully verified software.
 
\section{Continuity in Constructive Mathematics}
 
Another evidence of continuity of computations comes from effective
versions of mathematical analysis (constructive, recursive, or intuitionistic
analysis --- see, for example~\cite{BishopE:cona,GoodsteinR:61}).
All constructive functions from constructive real numbers
to constructive real numbers must be continuous. It is impossible to
properly treat computations at the points of discontinuity. If one
has to deal with non-continuous functions, they have to be partially
defined --- points of discontinuity cannot belong to their domain
(although see~\cite{ChunKuen:95}).
 
\section{Continuity in Denotational Semantics of
Programming Languages}
 
The most important and mathematically rich example of a continuous
approach to computation is denotational
semantics (see, for example, \cite{scott-strachey71,scott72s,SCOTT82,
Stoy,WINSKEL93,AbramskyS:domt}).
We briefly
introduce its main ideas here.
 
Denotational semantics of a formal language is a map $S$ from the
space of formal texts to the certain space of meanings (semantic
space). $S[ \! [P] \! ]$ is a meaning of the formal text $P$; equivalently we
say that $P$ denotes $S[ \! [P] \! ]$.
 
To provide denotational semantics of the $\lambda$-calculus
it was necessary to solve equations $D \cong [ D \rightarrow D ]$ for
semantic spaces --- the task which looked quite non-trivial
and which does not have non-trivial solutions
if we interpret $[ D \rightarrow D ]$ as the
set of all functions from $D$ to $D$.
Topological ideas gave the solution~\cite{Scott72b,Stoy}.

Continuity in computations always relates closely to the idea of formal
approximation. Interval numbers formally approximate one another, partially
defined functions do the same, etc. The relation of formal approximation
usually is a partial order, traditionally denoted as $\sqsubseteq$.
Properties vary in different approaches,
but the existence of the least defined element and of lowest
upper bounds of increasing chains (limits of sequences of better and
better formal approximations) is usually required. Topology then is
introduced in such a way, that the continuous functions are
monotonic functions preserving limits of increasing chains.
The resulting topological spaces are called {\em domains}.
 
It is natural to require of correctly defined computational
processes that the better formal approximation of an input is known,
the better formal approximation of the corresponding output should be computed
(monotonicity). Also well defined computational schemes (e.g.
in numerical methods) tend to preserve limits of formal approximations ---
we can get the result with an arbitrarily small error if we take the
input data with sufficient precision and iterate long enough.

Thus continuous functions above are good candidates to serve as
models of computable functions. The main paradigm of the domain theory is:
"continuous" is the right mathematical model for "computable". The resulting
topological spaces possess a number of remarkable properties.
Especially important for computer science applications are theorems
of fixed points for functions and functors.
The existence of canonical fixed points of continuous functions justifies all
recursive definitions of functions.
The existence of canonical fixed points of continuous functors justifies all
recursive definitions of types; in particular, they give the solution of
the equation $D \cong [ D \rightarrow D ]$, which
is interpreted as homeomorphism, and $[ D \rightarrow D ]$ --- as the space of
all continuous functions from $D$ to $D$ in the pointwise topology.

\chapter{Overview of Results}

\section{Overview of Part~\ref{part:Logic}}

The elementary covariant logical approach to domains is known
under the name of {\em information systems}. Under this approach
domain elements are thought of as {\em theories} in a
logical calculus, and continuous functions are thought of as
{\em inference engines}, deducing information about $f(x)$
from information about $x$.

This approach provides very simple and powerful intuition
about domains, thus making it easier to learn domain theory
and to find and motivate new notions and constructions.
Thus, it can be recommended as both didactic and research
framework.

A more complicated and less elementary contravariant
logical approach to domains uses the ideas of {\em Stone duality}.
We use Stone duality only as an auxiliary vehicle
in this text.

In Chapter~\ref{chap:Alg} of Part~\ref{part:Intro}
we introduce the framework of information
systems for algebraic Scott domains. We reformulate
the notion of consistency in a more technically
convenient way compared to the standard framework~\cite{SCOTT82}.

Part~\ref{part:Logic} is structured as follows.
In Chapter~\ref{chap:Hoofman} we explain the intuition
behind the generalization of this approach to the case
of continuous Scott domains by R.Hoofman~\cite{Hoofman}.
Based on this intuition we further generalize this
approach to domains of fixed points of Scott continuous
transformations of powersets.
We also discuss the significance
of the algebraic case from the logical point of view ---
algebraic Scott domains correspond to the standard logic,
while some of the more general classes of 
domains result from either Hoofman's
non-reflexive logic or Sazonov's non-finitary logic.

The subsequent chapters, which represent our earlier
results, may be also viewed as applications of
Chapter~\ref{chap:Hoofman}. Chapter~\ref{chap:Subdomains}
studies the notion of {\em subdomain} and the 
{\em domain of subdomains}
for the algebraic Scott domains. We show that the closure
operations play an exclusive role in the formation of
subdomains, as opposed to finitary retractions or projections,
and that this exclusive role can be explained by the intuition
presented in Chapter~\ref{chap:Hoofman}.

Chapter~\ref{chap:Fin} studies finitary retractions and
provides a novel simple criterion for finitarity.
Again, the results can be viewed as an illustration for
the intuition of Chapter~\ref{chap:Hoofman}.
Some of the results of Chapters~\ref{chap:Subdomains}
and~\ref{chap:Fin} might
be known as folklore, as Carl Gunter suggested to me,
since it is quite likely that somebody might have
discovered them independently. However, I was not able
to trace any written evidence of that.

\section{Overview of Part~\ref{part:Analysis}}

This part contains the core results of this Thesis.

Our research program is to develop analogs of classical
mathematical structures (metrics, measures, series,
vector spaces, etc.) for domains with Scott topology,
with hope to be able eventually to use methods of classical
continuous mathematics over the spaces of programs.

Our research program is essentially dual to the
research program initiated by Abbas Edalat, who suggested
to use domains in order to produce computational models for
classical mathematical structures and thus provide the framework
for computations in such structures~\cite{EdalatSurvey}.
We should point out that these dual research programs share
a lot of technical material.

Chapter~\ref{chap:Valuations} studies measure theory
on domains. This chapter introduces novel classes of
{\em co-continuous valuations} and CC-valuations and builds
such valuations for a large class of domains.

Chapter~\ref{chap:Metrics} studies theory of generalized
distances on domains. We discover that if we would like
our generalized distances to be Scott continuous functions
and simultaneously to describe Scott topology via these
distances, then the axiom $\rho (x, x) = 0$ cannot hold.
Accordingly, we introduce a new class of generalized
distances with values in interval numbers. We call these
distances {\em relaxed metrics} and build them here
for some domains.

Chapter~\ref{chap:muInfo} introduces a general mechanism
of building relaxed metrics from measure-like structures,
and shows how to build relaxed metrics from CC-valuations
for continuous Scott domains.

Chapter~\ref{chap:Neginfo} generalizes the results of
the previous chapter to continuous dcpo's by solving
the patologies of the behavior of negative information
for non-weakly Hausdorff spaces. The connection between
negative information and tolerances is explored.

\chapter{Main Definitions of Domain Theory}

This chapter is rather terse. Its main purpose is to
serve as a repository of the main definitions in this text.
The two subsequent chapters give important examples of
the notions given here and provide necessary intuition.

We assume that the reader is familiar with the notions
of partially ordered set (partial order), topological
space, and continuous function. In some rare cases we will
use simple categorical notions, but the reader is able to
skip them without excessive harm to the overall
comprehension.

\section{Notation}

We use $U\subset V$ as an
equivalent of $U\subseteq V\ \& \ U\ne V$.

\section{Domains}

\subsection{Directed Sets}

A partially ordered set (poset),
$(S, \sqsubseteq)$, is {\em directed} if $\forall x, y \in S.
\ \exists z \in S.\ x \sqsubseteq z, y \sqsubseteq z$.
In particular, empty sets are directed.

The notion of directed set should be considered as a generalization
of the notion of increasing sequence, $s_1 \sqsubseteq
s_2 \sqsubseteq \ldots$.

\subsection{Directed Complete Partial Orders (DCPO's)}

A poset, $(A, \sqsubseteq)$, is a {\em directed complete
partial order} or {\em dcpo} if
for any directed $S \subseteq A$,
the least upper bound $\sqcup S$ of $S$ exists in $A$.

Note that, since the empty set is directed, any dcpo $A$ must
have the least element, which we denote as $\bot$ or $\bot_A$.

Because domains without least elements are also considered lately,
differences in terminology occur at this point. Sometimes what we
call {\em directed complete partial orders} is called {\em complete
partial orders}, and a
larger class of partial orders, where only non-empty directed
sets must have least upper bounds and, thus, the least element does not
have to exist, is called {\em directed complete partial orders}.

\subsection{The ``Way Below" Relation and Compact (Finite) Elements}

Consider a dcpo $(A, \sqsubseteq)$ and $x, y \in A$.
We say that $x \ll y$ ($x$ is {\em way
below} $y$) if for any directed set $S \subseteq A$,
$y \sqsubseteq \sqcup S \Rightarrow \exists s\in S.
\ x \sqsubseteq s$.

An element $x$, such that $x \ll x$, is called {\em compact}
or {\em finite}.

It is easy to see, that $x \ll y$ implies $x \sqsubseteq y$, however
the inverse relationship is, in general, much more complicated.
For example, it is possible that $x \ll x$,
but it is also possible that $x \sqsubset y$ and $x \not\ll y$ at the
same time. A kind of transitivity, $x' \sqsubseteq x, x \ll y,
y \sqsubseteq y' \Rightarrow x' \ll y'$, does hold.

\subsection{Continuous and Algebraic DCPO's}

Consider a dcpo $(A, \sqsubseteq)$ and $K \subseteq A$.
We say that a dcpo $A$ is a {\em continuous
dcpo} with {\em basis} $K$, if for any $a\in A$, the set $K_{a} =
\{k \in K \ |\ k \ll a\}$ is directed and $a = \sqcup K_{a}$.
We call elements of $K$ {\em basic} elements.

If a continuous dcpo has a basis consisting only of compact elements,
such a dcpo is called {\em algebraic}.

\subsection{Bounded Completeness and Domains}

We say that $A$ is {\em bounded complete} if
$\forall B \subseteq A.\ (\exists a \in A.\ \forall b \in B. b \sqsubseteq
a)
\Rightarrow \sqcup_{A} B$ exists.

If an algebraic dcpo is bounded complete, it is
called an {\em algebraic Scott domain} or, simply, a {\em Scott domain}.

If a continuous dcpo is bounded complete, it is
called a {\em continuous Scott domain}~\cite{Hoofman}.

To be consistent with these definitions, if a dcpo
is bounded complete, we call it a {\em complete Scott domain}.

\subsection{Lattices}

If each subset of a partially ordered set $A$ has the least upper bound
in $A$, $A$ is called a {\em complete lattice}.

A complete lattice which is a continuous dcpo is called a
{\em continuous lattice}.

A complete lattice which is an algebraic dcpo is called an
{\em algebraic lattice}.

Note that complete lattices are precisely those complete
Scott domains, which have the top element, $\top$.
Also note that any subset of a complete Scott domain $A$ also has
the greatest lower bound in $A$.

\section{Scott Topology and Scott Continuous Functions}

\subsection{Aleksandrov and Scott Topologies}

Consider dcpo $(A, \sqsubseteq_{A})$ and $U \subseteq A$.
$U$ is {\em Aleksandrov open} if $\forall x, y \in A.
\ x \in U, x \sqsubseteq y \Rightarrow y \in U$.
An Aleksandrov open set $U$ is {\em Scott open} if
for any directed poset $S \subseteq A$,
$\sqcup S \in U \Rightarrow \exists s \in S.\ s \in U$.

It is easy to see that Aleksandrov open sets and Scott
open sets form topologies which are called,
respectively, the {\em Aleksandrov topology} and
the {\em Scott topology}.

\subsection{Continuous functions}

Consider dcpo's $(A, \sqsubseteq_{A})$ and
$(B, \sqsubseteq_{B})$ with the respective
Aleksandrov topologies. It it easy to see
that a function $f : A \rightarrow B$ is
(Aleksandrov) continuous iff it is monotonic,
i.e. $x \sqsubseteq_{A} y \Rightarrow f(x) \sqsubseteq_{B} f(y)$.

Consider dcpo's $(A, \sqsubseteq_{A})$ and
$(B, \sqsubseteq_{B})$ with the respective
Scott topologies.  It it easy to see
that a function $f : A \rightarrow B$ is
(Scott) continuous iff it is Aleksandrov continuous
and for any directed poset $S \subseteq A$,
$f(\sqcup_{A} S) = \sqcup_{B} \{f(s)\ |\ s \in S\}$.

\section{Functional Spaces}

Consider dcpo's $A$ and $B$. We define the functional
space $[A \rightarrow B]$ as the set of all Scott
continuous functions $f : A \rightarrow B$ with the
partial order $f \sqsubseteq g \Leftrightarrow
\forall x \in A.\ f(x) \sqsubseteq_B g(x)$. It is
easy to check that $[A \rightarrow B]$ is a dcpo.

\section{Retractions, Projections, And Their Pairs}

Consider dcpo $A$ and its Scott continuous transformation,
$f : A \rightarrow A$. If $f = f \circ f$, then $f$ is
called a {\em retraction} of $A$. If $f$ is such a retraction
and $f \sqsubseteq id_A$, then $f$ is called a {\em projection} of $A$
($id_A$ is the identity transformation $x \mapsto x$ of $A$).

Consider dcpo's $A$ and $B$ and a pair of Scott continuous
functions, $i : B \rightarrow A$ and $j : A \rightarrow B$,
such that $j \circ i = id_B$. Then it is easy to check
that $r = i \circ j$ is a retraction of $A$. In such
a situation we call $i$ an {\em embedding} of $B$ {\bf into} $A$,
$j$ a {\em retraction} of $A$ {\bf onto} $B$, and $\langle i, j \rangle$
an {\em embedding-retraction pair} (or simply, a {\em retraction pair}).
Sometimes, the whole pair is called a retraction of $A$ onto $B$.

If $r$ is also a projection, then the terms {\em projection onto},
and {\em embedding-projection pair} (or simply,
{\em projection pair}) are used in the corresponding situations.

\chapter{Interval Numbers} \label{sec:Interval}

In this chapter we will look at several important domains,
including interval numbers, which form the basis for
correct approximate computations with real numbers.

\section{Vertical Segments and Rays of Real Line}

All domains considered in this section are linearly ordered. We
call all linearly ordered domains {\em vertical domains}.

A vertical segment of a real line, $[A,\,B], A < B$, with
its natural order, $\sqsubseteq = \le$, is a continuous
lattice. So is a vertical ray $[A,\,+\infty]$ and the
whole real line $[-\infty,\,+\infty]$ with the same order.

Note that the last two spaces must contain $+\infty$ for
directed completeness, and that the last space
must also contain $-\infty$, because all domains have
the least element.

We will call these domains the {\em domains of lower estimates}
for reasons, which will become apparent in the next
section. We denote these domains 
as $R^+$ or $R^+_{[A,\,B]}$.

We can also consider a vertical segment of a real line
$[A,\,B]$ with the inverse order, $\sqsubseteq = \ge$.
We can also consider vertical rays $[A,\,+\infty]$,
$[-\infty,\,A]$ and the whole line $[-\infty,\,+\infty]$
with the same order.

We will call these continuous lattices the {\em domains
of upper estimates}. We denote these domains with the inverse
order as $R^-$ or $R^-_{[A,\,B]}$.

Note that when we talk about $R^+_{[A,\,B]}$ or $R^-_{[A,\,B]}$, the cases
of $A=-\infty$ and/or $B=+\infty$ are included.

\subsection{The ``Way Below" Relation}

One can easily see that $\bot_{R^+}=A$ and $\bot_{R^-}=B$.

It is also easy to see that given $a, b \in [A,\,B]$,
$a \ll_{R^+} b$ iff $a=\bot_{R^+}$ or $a<b$. Similarly,
$a \ll_{R^-} b$ iff $a=\bot_{R^-}$ or $a>b$.

This allows one to understand, why these domains are
continuous but not algebraic. 

Any subset $K$, dense
in $[A,\,B]$ in the traditional sense, can be taken as
a basis.

\subsection{Scott Topology}

Consider the domain $R^+$. Its Scott open sets are the empty set,
the whole space $R^+$, and semi-intervals 
or open rays $(a,\,B], a > A$.

For the domain $R^-$, the Scott open sets are the empty set,
the whole space $R^-$, and semi-intervals 
or open rays $[A,\,b), b < B$.

\section{Interval Numbers}

In this section we define the domain of interval
numbers belonging to $[A,\,B]$, where the cases
of $A=-\infty$ and/or $B=+\infty$ are included. 
This domain will consist of
segments $[a,\,b]$, where $A \le a \le b \le B$.
We denote this domain as $R^I$ or $R^I_{[A,\,B]}$.

Informally speaking, the segment $[a, b]$
is interpreted as a partially defined number $x$, about which
it is known that $$a \leq x \leq b.$$
If $x = [a,b]$,
we call $a$ a {\em lower
bound} or {\em lower estimate} of $x$, and $b$ an 
{\em upper bound} or {\em upper estimate} of $x$.

We can consider a product domain $R^{+} \times R^{-}$
with coordinate-wise order. Then $R^I$ can be defined
as a subset of $R^{+} \times R^{-}$, obtained by
by elimination of pairs $\langle a, b \rangle$, such that $a>b$,
from $R^{+} \times R^{-}$. $R^I$ inherits its partial
order from $R^{+} \times R^{-}$.

More specifically,
if interval $[c, d]$ is within interval $[a, b]$, that is
$a \leq c \leq d \leq b$, than $[c, d]$ gives more information
about a number, than $[a, b]$. In such a case, we say that
$[a, b]$ {\em approximates} $[c, d]$, and write
$[a, b] \sqsubseteq_{R^I} [c, d]$.

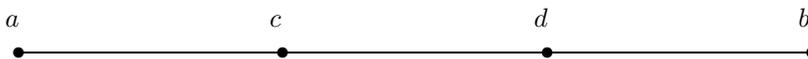
\begin{picture}(340,40)
\put(15,20){$a$}
\put(20,10){\circle*{4}}
\put(20,10){\line(1,0){100}}
\put(115,20){$c$}
\put(120,10){\circle*{4}}
\put(120,10){\line(1,0){100}}
\put(215,20){$d$}
\put(220,10){\circle*{4}}
\put(220,10){\line(1,0){100}}
\put(315,20){$b$}
\put(320,10){\circle*{4}}
\end{picture}

Consider the case, where $A=0$ and $B$ is finite.
Interval numbers $[x, y]$ can be thought of as a triangle
on the plane.

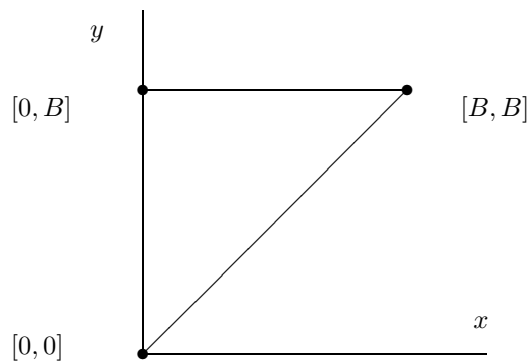
\begin{picture}(340,150)
\put(100,10){\line(1,0){130}}
\put(100,10){\line(0,1){130}}
\put(100,110){\line(1,0){100}}
\put(100,10){\line(1,1){100}}
\put(100,10){\circle*{4}}
\put(100,110){\circle*{4}}
\put(200,110){\circle*{4}}
\put(50,10){$[0,0]$}
\put(50,100){$[0,B]$}
\put(80,130){$y$}
\put(225,20){$x$}
\put(220,100){$[B,B]$}
\end{picture}

We flip this triangle so that the best defined interval numbers,
like $[a, a]$, are on the top, and the least defined one, $[0, B]$,
is on the bottom.

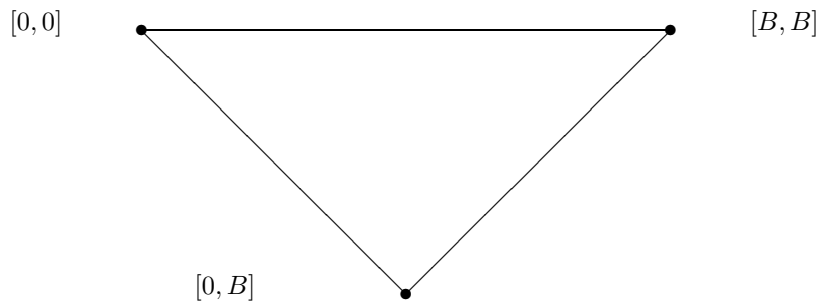
\begin{picture}(340,150)
\put(170,10){\line(1,1){100}}
\put(170,10){\line(-1,1){100}}
\put(70,110){\line(1,0){200}}
\put(70,110){\circle*{4}}
\put(270,110){\circle*{4}}
\put(170,10){\circle*{4}}
\put(90,10){$[0,B]$}
\put(20,110){$[0,0]$}
\put(300,110){$[B,B]$}
\end{picture}

\subsection{The ``Way Below" Relation}

One can easily see that $\bot^{R^I}=[A,\,B]$.

It also easy to see that $[a,\,b] \ll [c,\,d]$ iff
either $a=A$ and $b=B$, or $a=c=A$ and $b>d$,
or $b=d=B$ and $a<c$, or $a<c$ and $b>d$.

It is easy to see that $R^I$ is a continuous Scott
domain. Consider any set $Q$, dense in $[A,\,B]$ in the
traditional sense, and take elements $[A,\,b]$, $[a,\,B]$,
and $[a,\,b]$, such that $a < b$ and $a,b \in Q$, as a basis
$K \subseteq R^I$.

\subsection{Scott topology}

We consider the basis $K$ built above and build
the base of Scott topology. 

For segments
$[A,\,b] \in K$ take sets
$V^A_b = \{[x,y]\ |\ A \le  x \leq y < b\}$.
For segments $[a,\,B] \in K$ take sets
$V^B_a = \{[x,y]\ |\ a < x \leq y \leq B\}$.
For segments $[a,\,b] \in K$ take sets
$V_{a,b} = \{[x,y]\ |\ a < x \leq y < b\}$.

Sets $V^A_b$, $V^B_a$, and $V_{a,b}$ form the
base of Scott topology.

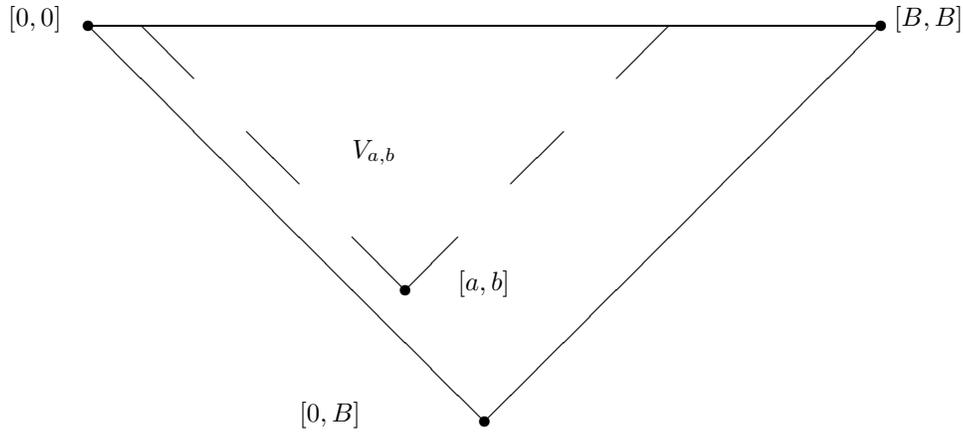
\begin{picture}(346,200)
\put(155,10){\line(1,1){150}}
\put(155,10){\line(-1,1){150}}
\put(5,160){\line(1,0){300}}
\put(5,160){\circle*{4}}
\put(305,160){\circle*{4}}
\put(155,10){\circle*{4}}
\put(85,10){$[0,B]$}
\put(-25,160){$[0,0]$}
\put(310,160){$[B,B]$}
\put(125,60){\line(1,1){20}}
\put(165,100){\line(1,1){20}}
\put(205,140){\line(1,1){20}}
\put(125,60){\line(-1,1){20}}
\put(85,100){\line(-1,1){20}}
\put(45,140){\line(-1,1){20}}
\put(125,60){\circle*{4}}
\put(145,60){$[a,b]$}
\put(105,110){$V_{a,b}$}
\end{picture}

\chapter{Algebraic Information Systems and Domains}\label{chap:Alg}

This chapter presents the approach of information systems
in the case of algebraic Scott domains
as can be found, e.g. in~\cite{SCOTT82,LarsenKG:usiiss,LARSON91}.

The algebraic information systems and domains described in this chapter
reflect the standard notions of inference and theories
in traditional formal theories.

\section{Information Systems and Domains}

The approach of information systems describes an approximation domain
as a set of {\em theories} in a logical calculus. There are two
ways to follow this approach. One could consider a given approximation
domain, which obeys certain specific axioms, and then try to
build a logical calculus such that its theories form an isomorphic
domain. This approach is very useful, when one needs to investigate
which class of domains is covered by a specific variant
of information systems, or when one starts with a given domain to begin
with.

However, for didactic purposes, a different approach has proven much
more valuable. This approach was used in the first key paper on
information systems by Dana Scott~\cite{SCOTT82} and works as follows.
We presume that there is some approximation domain on the background,
but we do not specify it precisely. Then, having this hypothetical
domain in mind, we reason about the desired properties of our logical
systems and impose appropriate axioms describing the behavior of
these systems. Then we define domains as sets of theories and
establish their properties as theorems rather than postulating
them as axioms.

\subsection{UNKNOWN as a Truth Value}

An information system is a {\em logical calculus} of 
{\em elementary statements} and their {\em finite conjuctions}.
These elementary statements and finite
conjunctions should be viewed as  
{\em continuous predicates} of a special kind
on the elements of the domain in question. 
These predicates map domains elements to a two-element set
of truth values, however these truth values are not ordinary
{\bf true} and {\bf false}, but {\bf true} and {\bf unknown}.
This crucial feature is usually not emphasized enough, but
one needs to keep it in mind.

The reason for this distinction is that we view a domain element, $x$,
as a dynamic object, about which only some of the information
is known at any given time
of the computational process, but which later can be supplemented with
more information, and thus replaced with $y$, $x \sqsubseteq y$.
The information, which was {\bf unknown} about $x$, can become
{\bf true} about $y$. This dynamic process is, however, viewed as
{\em monotonic} with respect to time --- once a piece of information
becomes {\bf true} about an element, it remains this way further on.

It is possible to talk about {\bf false} pieces of information about
$x$ --- these are such pieces of information which are not {\bf true}
about any $y$, such that $x \sqsubseteq y$, that is, the pieces
which cannot become {\bf true} about $x$ during its arbitrary monotonic
evolution. The {\bf false} truth values will be treated via
{\em consistency} mechanism, however they will always remain auxiliary,
and usually can be easily eliminated from the scene if necessary,
while {\bf true} and {\bf unknown} truth values are essential.

\subsection{Example: Formal Theories}

The most natural example one should keep in mind is any traditional
formal theory, where elementary statements are all formulas,
and the domain in question is the domain of all theories
ordered by ordinary set-theoretic inclusion.

If a statement belongs to a theory, we will say that it is
{\bf true} in (or, if you wish, ``about'') this theory,
otherwise it is {\bf unknown} in this theory. If the theory
cannot be refined to a larger non-contradictory
theory to include a specific
statement, we might wish to say that this statement
is {\bf false} in (``about'') this theory.

In the traditional formal theories any deduction is thought of
as a formal text of finite length. Hence any entailment of
a statement or a contradiction from a set of statements is
made on the basis of a finite subset of this set. This {\em finitarity}
property will be reflected in the definitions below.

\subsection{Consistency and Entailment}

Let us denote the set of {\em elementary statements} as $D$,
and the set of all its finite subsets as $\Fin (D)$.
We interpret a finite subset, 
$u \subseteq D$, $u = \{d_1, \cdots, d_n\}$,
as a {\em finite conjuction} of statements $d_1, \cdots, d_n$.

Keeping the hypothetical approximation domain at the background,
we would like to call a finite conjuction, $u$, of elementary statements
{\em consistent}, if there is a domain element, $x$, about which all
statements of $u$ are true.
 
It is traditional to assume that $\emptyset$ is consistent, that is,
that the domain in question is non-empty~\cite{SCOTT82}. 
We do follow this tradition here.
It is also traditional to assume that any single elementary
statement is true about some domain element, that is, $\{d\}$ is
consistent for any
elementary statement $d$. This means that one considers
contradictory elementary statements to be ``junk" and wants to exclude
them from an information system. We would like to depart from this
convention, as it seems to gain nothing, and makes it more difficult to
talk
about some natural examples, like the one considered in the
previous subsection, and also makes it impossible to talk about
effective structures on domains in full generality (the point of view,
advocated, in particular, by V.Yu.Sazonov; our style of
information system is, in fact, intermediate between the traditional one
and the style of Sazonov~\cite{Sazonov} and seems to be the
most convenient).

We would like to say, that a finite conjuction, $u$, of elementary
statements
{\em entails} an elementary statement, $d$, if whenever $u$ is
true about a domain element, $x$, statement $d$ is also true about $x$.

Again, it is traditional to consider only consistent conjuctions,
$u$, in the context of entailment. However, it is much more convenient
to assume that a conjuction, which is not consistent, entails everything.
It is technically convenient to introduce a special {\bf false} statement,
$\nabla$, which entails everything and follows from all
inconsistent conjunctions.

Taking all this into account, the following definition becomes
quite natural.

\begin{definition}
The tuple $A=(D_A, \nabla_A, \vdash_A)$, where $D_A$ is a set of
distinctive {\em tokens} ({\em elementary statements}),
$\nabla_A \in D_A$ (the {\em false statement}),
$\vdash_A \subseteq \Fin (D_A) \times \Fin (D_A)$ (the
{\em entailment} relation) is called an {\em algebraic
information system} if
\begin{enumerate}
\item $\emptyset \not\vdash_A \{\nabla_A\}$ (non-degeneracy; 
              a calculus must admit at least one non-contradictory theory);
\item $\forall u, v \in \Fin (D_A).\ v \subseteq u 
              \Rightarrow u \vdash_A v$ (reflexivity of entailment
              and conjunction elimination);
\item $\forall u, v_1, \ldots, v_n, w \in \Fin (D_A).\ u \vdash_A v_1,
              \ldots, u \vdash_A v_n,
              v_1 \cup \ldots \cup v_n \vdash_A w 
              \Rightarrow u \vdash_A w$
              (transitivity of entailment and conjunction introduction);
\item $\forall u \in \Fin (D_A).\ \{\nabla_A\} \vdash_A u$
              (contradiction entails everything).
\end{enumerate}
\end{definition}

The equivalence between this and the classical definition of
information system will be shown in Subsection~\ref{sec:our_and_classic}.

\subsection{Theories as Domain Elements}

\begin{definition}
{\em Domain} $|A|$ associated with an algebraic information
system $A$ is the set of theories,
\end{definition}

$\{x \subseteq D_A \ |$
  \begin{enumerate}
     \item $u \subseteq x, u \vdash_A v \Rightarrow v \subseteq x$
           ($x$ is deductively closed); 
     \item $\nabla_A \not\in x$ (deductively
           closed $x$ is consistent) $\}$.
  \end{enumerate}

Informally, we will say that if $d \in x$, then $d$ is {\bf true}
about $x$, otherwise $d$ is {\bf unknown} about $x$.

\subsection{Properties of Domains}

Assume that algebraic information system $A$ is given.
Domain $|A|$ is partially ordered by set-theoretical inclusion,
$\sqsubseteq_A = \subseteq_{D_A}$. We study the properties of this order.

\subsubsection{Deductive Closure and the Least Element}

\begin{definition}
  Subset $x \subseteq D_A$ is called (finitely) {\em consistent},
  if for any finite $u \subseteq x$, $u \not\vdash_A \{\nabla_A\}$.
\end{definition}

\begin{definition}
  Given subset $x \subseteq D_A$, we call
  subset $\overline{x} \subseteq D_A$,
  $\overline{x} = \{ d \in D_A \ |$ there is finite $u \subseteq x$,
  such that $u \vdash_A \{d\} \}$, the {\em deductive closure} of $x$.
\end{definition}

\begin{lemma}
  If subset $x \subseteq D_A$ is consistent, then $\overline{x}$
  is a domain element.
\end{lemma}

\Proof
  This simple proof is still quite instructive, so we go through
  it here. First of all, we want to show that $\overline{x}$ is deductively
  closed, that is if $u \subseteq \overline{x}$ and $u \vdash_A v$
  then $v \subseteq \overline{x}$. 

  Formula $u \vdash_A v$ implies finiteness
  of $u$ and $v$, hence consider $u = \{d_1, \ldots, d_n\}$.
  By definition of $\overline{x}$, there are $u_1, \ldots, u_n
  \subseteq x$, such that $u_1 \vdash_A \{d_1\}, 
  \ldots, u_n \vdash_A \{d_n\}$.
  Consider $u' = u_1 \cup \ldots \cup u_n$. Then, combining the
  rule of conjuction elimination and the rule 
  of transitivity, we obtain
  $u' \vdash_A \{d_1\}, \ldots, u' \vdash_A \{d_n\}$,
  applying the same combination of rules
  in their conjuction introduction and reflexivity incarnations
  once again, we obtain $u' \vdash_A u$,
  applying transitivity 
  once again we obtain $u' \vdash_A v$, and from this and
  $u' \subseteq x$ one
  concludes that $v \subseteq \overline{x}$.

  Consistency follows straightforwardly from the two previous
  definitions.  
\eproof

\begin{lemma}
  $x \subseteq y \Rightarrow \overline{x} \subseteq \overline{y}$.
\end{lemma}

\begin{lemma} \label{sec:bottom_exists}
  Since $\emptyset$ is consistent, $\bot_A = \overline{\emptyset}$ is
  the least domain element.
\end{lemma}

\subsubsection{Directed Completeness}

\begin{lemma} \label{sec:domain_dir_complete}
  Domain $|A|$ associated with an algebraic information system
  $A$ is directed complete.
\end{lemma}

\Proof
  Consider a directed $S \subseteq |A|$. For the case
  of $S = \emptyset$, Lemma~\ref{sec:bottom_exists} establishes the
  the existence of $\bot_A = \sqcup \emptyset$.

  Assume that $S= \{s_i\ |\ i \in I\}$ is
  a non-empty directed set
  of elements of domain $|A|$. Let us establish that
  $\bigcup_{i \in I} s_i$ is an element of $|A|$.
  From this it would be easy to see that $\sqcup_{i \in I} s_i =
  \bigcup_{i \in I} s_i$.

  First, establish the deductive closeness of $\bigcup_{i \in I} s_i$.
  Assume that $u = \{d_1, \ldots, d_n\} \subseteq \bigcup_{i \in I} s_i$.
  Then $d_1 \in s_{i_1}, \ldots, d_n \in s_{i_n}$. By directness
  of $(s_i, i \in I)$, there is such $i \in I$, that
  $s_{i_1} \subseteq s_i, \ldots, s_{i_n} \subseteq s_i$.
  Then $u \subseteq s_i$, and by deductive closeness of $s_i$,
  if $u \vdash_A v$, then $v \subseteq s_i$. Hence if $u \vdash_A v$,
  then $v \subseteq \bigcup_{i \in I} s_i$.

  Consistency is trivial.
\eproof

\subsubsection{``Way Below'' Relation and Algebraicity}

\begin{lemma}
  Given an algebraic information system, $A$, an element, $x$,
of associated domain $|A|$ is {\em compact} if and only if
there is a finite set, $u \subseteq D_A$, such that
$x = \overline{u}$.
\end{lemma}

\Proof
  Assume, that $x = \overline{\{d_1, \ldots, d_n\}}$ and
  that $x \sqsubseteq \sqcup S$, where $S$ is a directed
  set. By the proof of Lemma~\ref{sec:domain_dir_complete},
  $\sqcup S = \bigcup_{s \in S} s$, hence, since $d_1, \ldots,
  d_n \in x$, there are $s_1, ..., s_n \in S$, such that
  $d_1 \in s_1, \ldots, d_n \in s_n$. Hence, by directness
  of $S$, there is $s \in S$, such that $d_1 \in s, \ldots,
  d_n \in s$. Hence, by the deductive closeness of $s$,
  $x \subseteq s$, and $x \sqsubseteq s$.

  Conversely, assume that $x$ cannot be represented as a deductive
  closure of its finite subset. Then consider set 
  $U = \{u\ |\ u \subseteq x, u$ finite$\}$. It is
  easy to see, that $S = \{\overline{u}\ |\ u \in U\}$ is
  a directed set, and that $x = \sqcup S$, but for no $u \in U$,
  $x \sqsubseteq \overline{u}$, hence $x$ is not compact.
\eproof

The second part of
the proof of the previous Lemma implies that
the set of compact elements characterized there
possesses the properties of a basis of dcpo and,
hence, the following Lemma
holds.

\begin{lemma}
  Domain $|A|$ associated with algebraic information system
  $A$ is an algebraic dcpo.  
\end{lemma}

\subsubsection{Bounded Completeness}

\begin{lemma}
  Domain $|A|$ associated with algebraic information system
  $A$ is bounded complete.  
\end{lemma}

\Proof
  Consider $Y \subseteq |A|$, such that there is $x \in |A|$,
  such that for all $y \in Y$, $y \sqsubseteq x$.
  Then $\bigcup_{y \in Y} y$ is a consistent set, and
  $\overline{\bigcup_{y \in Y} y}$ 
  is the desired least upper bound of $Y$.
\eproof

\subsubsection{Algebraic Scott Domains}

The discource of this section can be summarized by the following
Theorem.

\begin{theorem}
  The domain $|A|$ associated with algebraic information system
  $A$ is an algebraic Scott domain.  
\end{theorem}

\subsection{Representing Algebraic Scott Domains}\label{sec:represent}

Here we show that algebraic information systems describe
exactly algebraic Scott domains.

\begin{theorem}
  For any algebraic Scott domain $(X, \sqsubseteq_X)$, there is
  an algebraic information system $A$, such that the
  partial order $(|A|, \sqsubseteq_A)$ is isomorphic to
  partial order $(X, \sqsubseteq_X)$.
\end{theorem}

\Proof
  Denote the set of compact elements of $X$ as $K$. Take
  $D_A = K \cup \{\nabla_A\}$, where $\nabla_A$ is a new
  token. 
  For finite subsets $u, v \subseteq D_A$,
  we say that \linebreak
  $u \vdash_A v$ if
  \begin{enumerate}
  \item either $u$ is unbounded subset of $X$, or $\nabla_A \in u$;
  \item or $u$ has an upper bound in $X$, and for any $k \in v$,
        $k \sqsubseteq_X \sqcup_X u$.
  \end{enumerate}

  First, we must prove that $A$ is an algebraic information system.
  It is a simple proof, we only show transitivity axiom here,
  which is only non-trivial if $u$ has an upper bound in $X$.
  Then assumption, that $u \vdash_A v_1, \ldots, u \vdash_A v_n$,
  means that any element $k$ of any of these $v_i$'s is less or
  equal to $\sqcup_X u$. In turn, $v_1 \cup \ldots \cup v_n \vdash_A w$,
  implies that all elements of $w$ are under the least upper bound
  of all those $k$'s. But $\sqcup_X u$ is (some) upper bound of those
  $k$'s, hence all elements of $w$ are under $\sqcup_X u$.

  Now we have to establish an order-preserving isomorphism between
  $X$ and $|A|$. With every $x \in X$ we associate $K_x = \{k \in K \ |\ k
  \sqsubseteq x\}$. 
  Now we are going to prove that all $K_x$
  are elements of $|A|$, and every element of $|A|$ equals to
  $K_x$ for some $x \in X$. Then, due to the fact
  that $x \sqsubseteq y \Leftrightarrow K_x \subseteq K_y$,
  the relation $x \leftrightarrow K_x$ would yield the necessary
  isomorphism.

  It is easy to establish that $K_x$ is an element, by using the
  directness of $K_x$.

  In the opposite direction, consider $S \subseteq K$, which is
  an element of $|A|$. First of all, let us show that $S$ is directed.
  By consistency of $S$ any finite $u \subseteq S$
  is bounded. By bounded completeness of $X$, there is 
  $\sqcup_X u$. Then, from the directness of
  $K_{\sqcup_X u}$, we immediately obtain the existence of compact
  $k \in K_{\sqcup_X u}$, which is an upper bound of $u$ (of course,
  this immediately implies $k = \sqcup_X u$, i.e. the least upper
  bound of a finite bounded set of compact elements is compact).
  By definition of $\vdash_A$, $u \vdash_A \{k\}$. Hence,
  the deductive closeness of $S$ implies that $S$ is directed.

  Now, using the fact that $X$ is directed complete, consider
  $x \in X$, such that $x = \sqcup_X S$. Let us prove that
  $S = K_x$. The only non-trivial direction is to
  prove $K_x \subseteq S$. Consider compact element
  $k \in K_x$. The definition of ``way-below'' relation and of element
  $x$, together with the facts $k \ll k$ and $k \sqsubseteq x$,
  imply that there is $k' \in S$, such that $k \sqsubseteq k'$.
  Then deductive closeness of $S$ implies $k \in S$.
\eproof

\subsection{Equivalence between Classical Information
            Systems and Our Definition} \label{sec:our_and_classic}

\begin{definition}
\label{def1}
$A=(D'_{A}, Con'_{A}, \vdash'_{A})$, where $D'_{A}$ is
a set of distinctive {\em tokens} ({\em assertions}),
$Con'_{A} \subseteq \Fin (D'_{A})$ (the {\em consistent} finite
conjunctions of assertions), $\vdash'_{A} \subseteq Con'_{A} \times D'_{A}$
(the {\em entailment} relation), is called a {\em classical information system}
if
\newcounter{n1}
\begin{list}
  {(\roman{n1})}
  {\usecounter{n1} \setlength{\topsep}{0pt} \setlength{\itemsep}{0pt}
  \setlength{\leftmargin}{1in}}
  \item $\emptyset \in Con'_{A}$ (non-degeneracy; empty domains are barred);
  \item $\forall d \in D'_{A}. \{d\} \in Con'_{A}$ (no ``junk'');
  \item $\forall u \subseteq v \in Con'_{A}. u \in Con'_{A}$;
  \item $\forall d \in u \in Con'_{A}. u \vdash' _{A} d$
        (reflexivity of entailment);
  \item $\forall u \in Con'_{A}, d \in D'_{A}. u \vdash' _{A} d \Rightarrow
         u \cup \{d\} \in Con'_{A}$ (entailment preserves consistency);
  \item $\forall u, v \in Con'_{A}, d \in D'_{A}.
         (\forall d' \in v. u \vdash' _{A} d'), v \vdash' _{A} d
         \Rightarrow u \vdash' _{A} d$ (transitivity of entailment).
\end{list}
\end{definition}

The domain $|A|$ associated with a classical information system $A$
is defined as a set of $x \in D'_A$ satisfying
the consistency condition, $\forall u \subseteq x.\ u$ is
finite $\Rightarrow u \in Con'_A$, and the condition of deductive
closeness, $\forall u \subseteq x, d \in D'_A.\ u \vdash'_A d
\Rightarrow d \in x$. Of course, $\sqsubseteq_{|A|}$ is defined
to equal $\subseteq$.

All such domains are algebraic Scott domains, and for any algebraic
Scott domain $X$ it is possible to build a classical information
system $A$, such that $|A|$ is isomorphic to $X$. So everything
is very similar to our version of algebraic information systems.

Now we are going to establish an even stronger equivalence
between these two versions. We are going to build two
translations as follows. Given an algebraic information system
$A=(D_A, \nabla_A, \vdash_A)$, we will build a classical
information system, $B = C(A)$, and given a classical
information system, $B=(D'_B, Con'_B, \vdash'_B)$,
we will build an algebraic information system $A=S(B)$,
such that the following properties hold.

For domains, $|C(A)|=|A|$ and $|S(B)|=|B|$, where the set
equalities take place, as opposed to mere isomorphisms. For information
systems, $C(S(B))=B$ and if we consider $A_c = S(C(A))$,
then $D_{A_c} = D_A \setminus \{d \in D_A\ |\ d \vdash_A \nabla_A \&
d \neq \nabla_A\}$
and $\vdash_{A_c}$ is obtained by restricting $\vdash_A$ on
$\Fin (D_{A_c}) \times \Fin (D_{A_c})$. Basically, the only
disturbance that $S \circ C$ cannot avoid is that other ``junk" tokens
equivalent to $\nabla_A$ die.

The translations are defined by the following formulas.

$D'_{C(A)} = \{d \in D_A \ |\ \{d\}\not\vdash_A \nabla_A\}$;
$Con'_{C(A)} = \{u \in \Fin (D_A)\ |\ u\not\vdash_A \nabla_A\}$;
if $u \in Con'_{C(A)}, d \in D'_{C(A)}$, then
$u \vdash'_{C(A)} d \Leftrightarrow u \vdash_A \{d\}$.

$D_{S(B)} = D'_B \cup \{\nabla_{S(B)}\}$ (under assumption
$\nabla_{S(B)} \not\in D'_B$). Consider $u,v \in \Fin (D_{S(B)}),
d_1, \ldots, d_n \in D_{S(B)}$.
If $u \not\in Con'_A$ then $u \vdash_{S(B)} v$.
If $u \in Con'_A$ then $u \vdash_{S(B)} \{d_1, \ldots, d_n\}
\Leftrightarrow u \vdash'_B d_1, \ldots, u \vdash'_B d_n$.

It is easy to check that if $A$ is an algebraic information
system and $B$ is a classical information system, then
$C(A)$ is a classical information system, $S(B)$ is
an algebraic information system, and our claims above hold.

\section{Approximable Mappings and Continuous Functions}

Information systems allow to think about domain elements
as theories. Likewise, this approach allows to think about
Scott continuous functions as special inference engines,
which infer information about output, $f(x)$, from information
about input, $x$.

Unfortunately, these {\em input-output inference engines} are called
{\em approximable mappings} for their property to approximate
one another. While this is an important property, Scott continuous
functions thought of as graphs, or, in fact, any functions to
domains approximate one another. 
Moreover, the ``mappings'' in question are not
even functions. However,
at this point of the development of the field one
does not have much of a choice, but to follow the
accepted terminology.

The following definition takes into account the intuition,
that consistent information about input should produce
consistent information about output, and that ``native''
inference relations of input and output domains can be used
by an input-output inference engine.

\begin{definition}
  Given two information systems, $A$ and $B$, relation
  $f \subseteq \Fin(D_A) \times \Fin(D_B)$ is called
  an {\em input-output inference relation} or
  an {\em approximable mapping}
  between $A$ and $B$, if the following axioms hold:
  \begin{enumerate}
    \item $\emptyset f \emptyset$ (non-triviliality; minimalistic
                                   version of $u f \emptyset$);
    \item $\forall u \in \Fin(D_A).\ u \not\vdash_A
                \{\nabla_A\} \Rightarrow \neg (u f
                \{\nabla_B\})$ (preservation of consistency);
    \item $\{\nabla_A\} f \{\nabla_B\}$ (inconsistency
                about an input allows to infer anything about
                the corresponding output);
    \item $\forall u \in \Fin(D_A), v_1, \ldots, v_n \in \Fin(D_B).\ u
                f v_1, \ldots u f v_n \Rightarrow
                u f (v_1 \cup \ldots \cup v_n)$ (accumulation of
                    output information, i.e. conjunction introduction);
    \item $\forall u', u \in \Fin(D_A), v, v' \in \Fin(D_B).\ u' \vdash_A
                u, u f v, v \vdash_B v' \Rightarrow u' f v'$
                (transitivity with ``native'' inference relations
                 in $A$ and $B$).
    \end{enumerate}
\end{definition}

\subsection{Scott Continuous Functions}

An approximable mapping, $f$, between information systems $A$ and $B$
naturally gives rise to a function, $|f| : |A| \rightarrow |B|$,
where $|f|(x)$ is computed by inferring all information from $x$
using $f$.

We will see that $|f|$ is always Scott continuous, and that for any
Scott continuous function $g: |A| \rightarrow |B|$ one can find
an approximable mapping $\widehat g$ between $A$ and $B$, such that
$|\widehat g| = g$, $\widehat{|f|} = f$. Together with the results about
identity maps and composition this yields an equivalence between
the category of algebraic information systems and approximable mappings
and the category of algebraic Scott domains and Scott continuous functions.

\begin{definition}
  Given an approximable mapping $f$ between $A$ and $B$,
  define the associated function $|f|: |A|\rightarrow |B|$
  by formula $|f|(x) = 
  \{d \in D_B\ |\ \exists u \subseteq x.\ u f \{d\}\}$. 
\end{definition} 

\begin{theorem}
  The function $|f|$ is correctly defined and Scott continuous.
\end{theorem}

\Proof
  First, of all, to prove correctness we need to show,
  that $|f|(x) \in |B|$. We need to show deductive closeness
  and consistency of $|f|(x)$. Assume, that $v = \{d_1, \ldots, d_n\}
  \subseteq |f|(x)$. Then there are $u_1, \ldots, u_n \subseteq x$,
  such that $u_1 f \{d_1\}, \ldots, u_n f \{d_n\}$. Then, by
  axioms of information systems and approximable mappings,
  $u f v$, where $u = u_1 \cup \ldots \cup u_n$. Hence, if
  $v \vdash_B v'$, by axioms of approximable mapping $u f v'$,
  from which it is easy to infer that $v' \subseteq |f|(x)$.
  Likewise, if $\nabla_B$ would belong to $|f|(x)$, there would
  be some $u \subseteq x$, $u f \{\nabla_B\}$, contradicting the
  combination of consistency of $x$ and, hence, $u$, and the preservation
  of consistency axiom for $f$.

  Monotonicity of $|f|$ is obvious. Given a directed set $S \subseteq |A|$,
  we show that $|f|(\sqcup_A S) = \bigcup_{s \in S} |f|(s)$,
  thus establishing preservation of least upper bounds of directed
  sets and Scott continuity. The less trivial part is to
  show $|f|(\sqcup_A S) \subseteq \bigcup_{s \in S} |f|(s)$.
  Recall that $\sqcup_A S = \bigcup_{s \in S} s$.
  Assume, that $d \in |f|(\bigcup_{s \in S} s)$, that is there is
  $v = \{d_1, \ldots, d_n\} \in \bigcup_{s \in S} s$, such that
  $v f \{d\}$. Then there are $s_1, \ldots, s_n \in S$, such
  that $d_1 \in s_1, \ldots, d_n \in s_n$. Then, by directness
  of $S$, we can find such $s \in S$, that $v \subseteq s$.
  Hence $d \in |f|(s)$ and belongs to the union in question.
\eproof

\begin{definition}
  Given a Scott continuous function $g: |A|\rightarrow |B|$,
  define the associated relation $\widehat g \subseteq
  \Fin (D_A) \times \Fin (D_B)$
  by formula $u \widehat g v \Leftrightarrow
  \forall x \in |A|.\ u \subseteq x \Rightarrow v \subseteq g(x)$.
\end{definition}

It is easy to check that $\widehat g$ is an approximable mapping.

If $u \in \Fin (D_A)$ is consistent, 
then $u \widehat g v \Leftrightarrow v \subseteq
g(\overline{u})$, otherwise $u \widehat g v$ for all $v \in \Fin (D_B)$.

This and the Scott continuity of $g$ allow to establish
easily, that $|\widehat g| = g$, $\widehat{|f|} = f$.

Moreover, $f \subseteq g$ iff $|f| \sqsubseteq_{[A \rightarrow B]} |g|$.

\subsection{Identity and Composition}

It is easy to see that $|\vdash_A|=id_A$,
where $\forall x \in |A|.\ id_A(x)=x$.

Define the {\em composition} of approximable mappings
$f \subseteq \Fin (D_A) \times \Fin (D_B)$ and
$g \subseteq \Fin (D_B) \times \Fin (D_C)$ as
$g \circ f = h \subseteq \Fin (D_A) \times \Fin(D_C)$,
such that $u h w \Leftrightarrow \exists v \in \Fin (D_B).\ u f v, v g w$.

One can think of $g \circ f$ as an input-output inference
engine obtained by hooking the input of engine $g$ to the
output of engine $f$: $\quad \Bigg( \quad \leftarrow g \leftarrow f \leftarrow
\quad \Bigg)$.

It is easy to check that $|g \circ f| = |g| \circ |f|$.

\section{Functional Spaces}

Consider algebraic information systems $A$ and $B$.
We are going to define an information system, $[A \rightarrow B]$,
such that $|[A \rightarrow B]|$ would consist precisely of
all approximable mappings between $A$ and $B$.

Since there is an isomorphism between approximable mappings
and Scott continuous functions, and since the set-theoretical
inclusion of approximable mappings corresponds to the partial
order on the domain of Scott continuous functions,
$[|A| \rightarrow |B|]$, we will use the information system
$[A \rightarrow B]$ to represent this functional space.

Take $D_{[A \rightarrow B]} = \Fin (D_A) \times \Fin (D_B)$.
Take $\nabla_{[A \rightarrow B]} = \langle \emptyset, 
\{\nabla_B\} \rangle$.

Say, that
$\{\langle u_1, v_1 \rangle, \ldots, \langle u_n, v_n \rangle \}
\vdash_{[A \rightarrow B]}
\{\langle u'_1, v'_1 \rangle, \ldots, \langle u'_m, v'_m \rangle \}$,
where $n, m \geq 0$,
if
\begin{enumerate}
 \item for any $i \in \{1, \ldots, m\}$,
    $\bigcup_{\{j \in \{1, \ldots, n\}\ |\ u'_i \vdash_A u_j\}} 
                                                 v_j \vdash_B v'_i$
    or $u'_i \vdash_A \nabla_A$; or
 \item $\exists I \subseteq \{1, \ldots, n\}.\ \bigcup_{i \in I}
           u_i \not\vdash_A \nabla_A, \bigcup_{i \in I} v_i \vdash_B \nabla_B$.
\end{enumerate}

We leave the necessary correctness checks to the reader.

\section{Effective Domains and Computations} \label{sec:Effective}

We say that an algebraic information system $A$ is {\em effective},
if $D_A$ and $\vdash_A$ are recursively enumerable. The
corresponding domain $|A|$ is called {\em effective} too.

An important example is produced by the {\em effective domain
of arithmetic theories} in any of the usual systems of arithmetic.
This example shows why our degree of generality is
the right one.

Usually people give a more restrictive definition,
which is equivalent to $D_A$ and $\vdash_A$ being recursive.
One of the reasons for this is the inconvenience of the definition
of a classical information system. Indeed, 
if $A$ is an effective algebraic information system, then
$Con'_{C(A)}$ is co-recursively enumerable, and
$\vdash'_{C(A)}$ has to be defined effectively on a
co-recursively enumerable domain of definition, which is
not a trivial undertaking, and, in any case, the result
would be awkward.

An element $x \in |A|$ is called {\em computable} if it is
a recursively enumerable set. Note that this condition
is equivalent to the recursive enumerability of $K_x$
under our construction from Section~\ref{sec:represent}.

If $f$ is a computable element of a functional domain,
$|[A \rightarrow B]|$, we call it a {\em computable function}.
Observe that computable functions map computable elements to
computable elements and, moreover, transform a recursive
enumeration of $x$ into the recursive enumeration of $|f|(x)$.

We should note here that the problems of the correctness of
definition for effective domains and computable elements
are not decidable in general. For example, it is often
impossible to develop a procedure deciding whether a given
effective domain is not empty, or whether a given recursive
enumeration of a computable element does not contain $\nabla$.

An implementation of a computable element is some computational
device, which recursively enumerates the tokens of this element. The issues
of more effective implementation of some classes of
computable elements (defined by some restricted classes
of formulas) are very important from the practical viewpoint
and constitute Problem A.

\part{Logic of Fixed Points for Domains}\label{part:Logic}

\chapter{Non-reflexive Logics for Non-algebraic
Domains}\label{chap:Hoofman}

In the previous chapter we saw that information systems
based on the ordinary logic correspond to algebraic
Scott domains. Hence, in order to generalize the logical approach
to larger classes of dcpo's, one has to modify the logic
of inference and/or the notion of theory.

In this chapter we concern ourselves with bounded complete
dcpo's. Some progress for some classes of non-bounded
complete spaces was achieved by Abramsky~\cite{Abramsky}.

This chapter develops our ideas from~\cite{Bukatin2,Bukatin3}
and Appendix 1 of~\cite{Bukatin4}

\section{Non-reflexive Logic}

The seminal paper~\cite{Hoofman} by R.Hoofman generalized
the logical approach to non-algebraic continuous domains.
Hoofman replaced
the ordinary relexivity rule, $A\vdash A$, with a
weaker property of {\em inverse modus ponens}: $A\vdash C \Rightarrow
\exists B.\ A\vdash B, B \vdash C$. His paper is a well-written one,
but at the same time 
the actual and simple reasons for his construction to
work are hidden in its later sections. Thus, the
average reader gets
an impression that it is a miracle construction,
and the real intuition behind this paper is lost.

In particular, another important feature, which allows to
exploit non-relexivity, is underemphasized. This feature is
the change in the notion of a theory (i.e. domain element).
A theory, $x$, is traditionally a consistent, deductively closed
set of statements. Hoofman adds an additional requirement
of {\em inverse deductive closeness}, which is trivial for a
reflexive situation:
$\forall d \in D_{A}. d \in x \Rightarrow
\exists u \subseteq_{fin} x. u \vdash_{A} d$.

We will soon see, that this feature is essential for his
approach, while the {\em inverse modus ponens rule} is
important only in order to maintain continuity
of the resulting domain. In fact,
we generalize the results of~\cite{Hoofman} to arbitrary
{\em spaces of fixed points of Scott continuous
transformations of algebraic Scott domains}, by omitting both the inverse
and the standard rule of {\em modus ponens}.

\subsection{Correspondence between Properties of Scott Continuous
Functions and Inference Rules}

Scott continuous functions are equivalent to input-output
inference engines (approximable mappings). In particular,
given an algebraic information system $A$, Scott continuous
transformations $|f|$ of the corresponding domain $|A|$ are
equivalent to the generalized inferences
$f \subseteq \Fin(D_A) \times \Fin(D_A)$, which are
transitive with the standard $\vdash_A$ and respect
consistency and inconsistency.

\subsubsection{Reflexivity}

Consider the reflexive $f$, $v \subseteq u \Rightarrow u f v$.
Since $u f v, v \vdash_A w \Rightarrow u f w$, if
$f$ is reflexive, then from $\forall u \in \Fin(D_A).\ u f u$
we can infer $\vdash_A \subseteq f$.
Hence if $f$ is reflexive, then $\forall x \in |A|.\ x \sqsubseteq_A |f|(x)$.

Conversely, $\forall x \in |A|.\ x \sqsubseteq_A |f|(x)$ implies
that $f$ is reflexive.

\subsubsection{Transitivity}

It is easy to see, that $\forall u,v,w \in \Fin(D_A).\ u f v,
v f w \Rightarrow u f w$ is equivalent to the condition $|f| \circ |f|
\sqsubseteq_{[A \rightarrow A]} |f|$.

\subsubsection{Inverse Transitivity}

It is easy to see, that the Hoofman rule of inverse transitivity,
$\forall u, w \in \Fin(D_A).\ u f w \Rightarrow 
\exists v \in \Fin(D_A).\ u f v, v f w$ is equivalent to
$|f| \sqsubseteq_{[A \rightarrow A]} |f| \circ |f|$.

\subsubsection{Retractions}

The previous two paragraphs imply that $f$ is transitive and
inversely transtive if and only if $|f|$ is a retraction:
$|f| = |f| \circ |f|$.

\subsubsection{Inverse Reflexivity}

The rule $\forall x \in |A|.\ |f|(x) \sqsubseteq_A x$
is equivalent to the following rule of ``inverse reflexivity":
$u f v \Rightarrow u \vdash v$, which will be transformed into
$u f v \Rightarrow v \subseteq u$ in some simple cases.

\subsubsection{Closures and Projections}

It is easy to see, that reflexivity implies inverse transitivity,
and that inverse reflexivity implies transitivity.

Hence, the combination of reflexivity and transitivity yields
precisely {\em closures} 
(such retractions $|f|$, that $x \sqsubseteq_A |f|(x)$),
and the combination of inverse reflexivity and
inverse transitivity yields precisely {\em projections}
(such retractions $|f|$, that $|f|(x) \sqsubseteq_A x$).

\subsection{Fixed Points as Theories}

Consider the definition of $|f|$: $|f|(x) = \{d\ |\ \exists u \in x.\ u
f \{d\}\}$. So $|f|(x)$ consists of tokens which are inferrable
from $x$ via inference engine $f$.

Hence, the deductive closeness of $x$ with respect to inference $f$,
$\forall u, v \in \Fin(D_A).$ $u \in x, u f v \Rightarrow v \in x$
is equivalent to $|f|(x) \sqsubseteq_A x$.

Similarly, the {\em inverse deductive closeness} of
$x$ with respect to inference $f$,
namely $\forall v \in \Fin(D_A).\ v \in x \Rightarrow
\exists u \in \Fin(D_A).\ u \in x, u f v$, is equivalent
to $x \sqsubseteq |f|(x)$.

Hence, together the deductive closeness and the inverse
deductive closeness of $x$ with respect to inference $f$
is equivalent to $x$ being a fixed point of $|f|$:
$|f|(x) = x$. 

\subsection{Domains of Fixed Points}

The previous paragraph suggests the
following procedure. Consider an
algebraic information system $A$ and replace its entailment
relation with an approximable mapping $f$ between $A$ and $A$.

Then replace the notion of theory with the consistent
subset $x \subseteq D_A$, such that $x$ is deductively closed
and inversely deductively closed with respect to $f$.
The result is the domain $|A_f|$
of fixed points of contionuous transformation $|f|$,
$|A_f| = \Fix(f) \subseteq |A|$.

We can call $|A_f|$ a {\em fixed-point subdomain} of $|A|$.

However, we want to introduce a more general notion of
a fixed-point subdomain.

\subsection{Adding the Top Element}

Consider an algebraic information system
$A=(D_A, \nabla_A, \vdash_A)$ and the corresponding domain $|A|$.
Consider an element $\nabla_{A_\top} \not\in D_A$ and define
the algebraic information system $A_\top$ as follows.

Take $D_{A_\top} = D_A \bigcup \{\nabla_{A_\top}\}$.
Of course, we take $\nabla_{A_\top}$ as the $\nabla$ of the
new system. We set $u \vdash_{A_\top} v$ iff either
$u \vdash_A v$ or $\nabla_{A_\top} \in u$.

Then $|A_\top| = |A| \bigcup \{\top\}$, 
where $\top = D_{A_\top} = \overline{\{\nabla_A\}}$
is the new top element.

\subsection{General Notion of Fixed-Point Subdomain}\label{sec:FixSub}

In this chapter we consider approximable mappings as
generalized entailment relations. Sometimes we would
like an approximable mapping $f$ to entail the
contradiction from some of finite conjuctions of statements
from $D_A$, noncontradictory under $\vdash_A$. 
In order to formalize such a situation we
have to consider approximable mappings $f$ from $A$ to
$A_\top$.

The general notion of a fixed-point subdomain of $A$ is
the set of fixed points of Scott continuous function
$|f|:|A| \rightarrow |A_\top|$, $\Fix(f) = \{ x \in A\ |\ x=|f|(x)\}$. 
Equivalently, one can
consider fixed points of Scott continuous transformations $|f|$ of
domain $A_\top$, such that $|f|(\top)=\top$.

If, for the sake of tradition, one wants to impose
the requirement that the subdomains are non-empty, one
has to add the requirement $|f|(\bot) \neq \top$.

\subsection{Application: Removing the Compact Top Element}

Assume that an algebraic Scott domain $|B|$ has the compact
top element $\top_B \neq \bot_B$ (in particular, $|A_\top|$ has the compact
top element, $\overline{\{\nabla_{A_\top}\}}$). Then consider
$|f|:|B| \rightarrow |B_\top|$, such that
$|f|(\top_B)=\top_{B_\top}$ and $|f|(x)=x$ for all other $x \in |B|$.

Then $\Fix (f) = B \setminus \{\top_B\}$. 

The compactness of $\top_B$ is important, since a Scott continuous
function on an algebraic Scott domain
is completly defined by its values on compact elements.

In terms of entailment, $u f v$ iff either $u \vdash_B v$ or
$\top_B \subseteq \overline{u}$. It is easy to see, that
since $\top_B \neq \bot_B$, the resulting system
$B_{\not\top} = (D_B, \nabla_B, f \bigcap (\Fin(D_B) \times \Fin(D_B)))$
is an algebraic information system,
and $|B_{\not\top}|=B \setminus \{\top_B\}$.

\subsection{Powersets and Qualitative Domains}

Consider set $D$, such that $\nabla_A \not\in D$,
and define $D_A = D \bigcup \{\nabla_A\}$ and
$u \vdash^m_A v$ iff $v \subseteq u$ or $\nabla_A \in u$.
The resulting {\em minimal} algebraic information
system $A^m=(D_A,\nabla_A,\vdash^m_A)$ defines the powerset
of $D$ as its domain $|A^m|$.

If we consider $W \subset \Fin (D_A)$, such that $\emptyset \not\in W$,
$\{\nabla_A\} \in W$,
and modify the previous construction so that
$u \vdash_A v$ iff $v \subseteq u$ or $\exists w \in W.\ w \subseteq u$,
then we obtain a {\em qualitative domain} (see~\cite{Hoofman}).
We can modify the construction of cutting the compact top
element above to obtain any qualitative domain from the
powerset domain.

\subsection{General Notion of Finitary Information System
            And Finitary Scott Domain}\label{sec:FinInfo}

Consider an algebraic information system $A=(D_A,\nabla_A,\vdash_A)$ and
an approximable mapping $f$ from $A$ to $A_\top$.
We agreed that $f$ describes
a fixed-point subdomain of $|A|$. Observe that $f$ also
describes a fixed-point subdomain of $|A^m|$.
Moreover, the sets of fixed points of corresponding
Scott continuous functions $|f|:|A|\rightarrow |A_\top|$
and $|f|:|A^m|\rightarrow |A^m_\top|$ coincide.

This leads us to a general definition of what we call
a {\em finitary information system} and the definition
of the corresponding
{\em finitary Scott domain}.
We will respect the rule $|f|(\bot) \neq \top$
meaning the non-emptyness of the resulting domains.
These definitions will be respectively a streamlined equivalent of $A^m$
together with an approximable mapping $f$ between
$A^m$ and $A^m_\top$ and a description of the set of fixed points of
the corresponding function $|f|$.

\begin{definition}
The tuple $A=(D_A, \nabla_A, \vdash'_A)$, where $D_A$ is a set of
distinctive {\em tokens} ({\em elementary statements}),
$\nabla_A \in D_A$ (the {\em false statement}),
$\vdash'_A \subseteq \Fin (D_A) \times \Fin (D_A)$ (the
{\em entailment} relation) is called a {\em finitary
information system} if
\begin{enumerate}
\item $\emptyset \not\vdash'_A \{\nabla_A\}$ (non-degeneracy;
              a calculus must admit at least one non-contradictory theory);
\item $\emptyset \vdash'_A \emptyset$ (non-triviliality);
\item $\forall u, v_1, \ldots, v_n \in \Fin(D_A).\ u
                \vdash'_A v_1, \ldots u \vdash'_A v_n \Rightarrow
                u \vdash'_A (v_1 \cup \ldots \cup v_n)$ (accumulation of
                    output information);
\item $\forall u', u, v, v' \in \Fin(D_A).\ u' \subseteq
                u, u' \vdash'_A v', v \subseteq v' \Rightarrow u \vdash'_A v$;
\item $\forall u, v \in \Fin (D_A).\ u \vdash'_A \{\nabla_A\}
              \Rightarrow u \vdash'_A v$;
\item $\forall u, v \in \Fin (D_A).\ \nabla_A \in u 
              \Rightarrow u \vdash'_A v$.
\end{enumerate}
\end{definition}

\begin{definition}
The {\em finitary Scott domain} $|A|$ associated with a finitary information
system $A$ is the set of theories,
\end{definition}

$\{x \subseteq D_A \ |$
  \begin{enumerate}
     \item $u \subseteq x, u \vdash'_A v \Rightarrow v \subseteq x$
           ($x$ is deductively closed);
     \item $v \subseteq x \Rightarrow \exists u \in
           \Fin(D_A).\  u \vdash'_A v, u \subseteq x$
           ($x$ is inversely deductively closed);
     \item $\nabla_A \not\in x$ (deductively
           closed $x$ is consistent) $\}$.
  \end{enumerate}

In order to check the correctness of our discourse,
one has to consider information systems $A^m=(D_A,\nabla_A,\vdash^m_A)$
and the corresponding system $A^m_\top$. Given a finitary
information system $A=(D_A,\nabla_A,\vdash'_A)$, one should
define $f\subseteq \Fin(D_A) \times \Fin(D_A \bigcup 
\{\nabla_{A^m_\top}\})$ via $u f v \Leftrightarrow u \vdash'_A v$ or
$\nabla_A \in u$ and establish that $f$ is an approximable mapping,
such that $\neg (\emptyset f \{\nabla_{A^m_\top}\})$.
Then given an approximable mapping $f$ between $A^m$ and
$A^m_\top$,
such that $\neg (\emptyset f \{\nabla_{A^m_\top}\})$,
one should define $\vdash'_A$ via $u \vdash'_A v \Leftrightarrow$
$u f v$ and $\nabla_{A^m_\top} \not\in v$ and establish that
$A=(D_A,\nabla_A,\vdash'_A)$ is a finitary information system.

Finally one should observe that we have just defined one-to-one
correspondence between all possible entailment relations in
finitary information systems
$(D_A,\nabla_A,\vdash'_A)$ and all approximable mappings
between $A^m$ and $A^m_\top$,
such that $\neg (\emptyset f \{\nabla_{A^m_\top}\})$,
and that under that correspondence
finitary Scott domains are exactly the sets of fixed points of
the respective functions $|f|$.

\subsection{Domains of Fixed-Point Subdomains}

In the subsections~\ref{sec:FixSub} and~\ref{sec:FinInfo} 
we defined a finitary information
system and the corresponding finitary Scott domain of
fixed points of $|f|$
for any approximable mapping $f$,
such that $\neg (\emptyset f \{\nabla_{A_\top}\})$,
between any algebraic information system $A=(D_A,\nabla_A,\vdash_A)$
and the corresponding system $A_\top$.

The approximable mapping $f$, such that $\emptyset f \{\nabla_{A_\top}\}$,
corresponding to Scott continuous function $|f|$,
such that $|f|(\bot) = \top_{A_\top}$, is the top
compact element of the domain $[A\rightarrow A_\top]$,
and, hence, it can be removed by the technique described
above, yielding the algebraic Scott domain
$|[A\rightarrow A_\top]_{\not\top}|$.

Because these mappings $f$ are in one-to-one correspondence
with fixed-point subdomains of $|A|$, we can take the domain
$|[A\rightarrow A_\top]_{\not\top}|$ as representing fixed-point subdomains
of $|A|$. Of course, different $f$ might have the same
set of fixed points, and hence the corresponding fixed-point
domains will coincide as sets, but their underlying
entailment relations will differ, so they should be considered
different subdomains.

Hence we call $|[A\rightarrow A_\top]_{\not\top}|$ the {\em domain of
fixed-point subdomains} of $|A|$.

\subsection{Discussion}

One should observe that $f$ is not an arbitrary new
entailment relation, but is closely related to the
original $\vdash_A$, namely $f$ is transitive with
respect to $\vdash_A$ due to the axioms of approximable
mappings. This property is responsible for
the fact that all elements of the resulting
fixed-point subdomain belong to the original domain.

We can think about a fixed-point subdomain as
the result of some elements of the original
domain being destroyed. There are three mechanisms
of such a destruction. An element can lose its
deductive closeness or consistency. When these
two mechanisms are involved, we still remain
within the realm of reflexive logic and algebraic
subdomains result. This important case is covered in details
in Chapter~\ref{chap:Subdomains}.

The third mechanism of destruction of elements of the original
domain is the loss of inverse deductive closeness.
When this happens we, in general, go beyond
ordinary reflexive logic. However, there are cases
of non-reflexive $f$'s, where we still can remain
within the realm of reflexive logic. The price for
this is the distortion of the resulting subdomain:
instead of the proper fixed-point subdomain we
only obtain a domain isomorphic to this
fixed-point subdomain. One such case, namely the
case of finitary retractions, is considered in details
in Chapter~\ref{chap:Fin}.

We could have considered a different definition of
a fixed-point subdomain of $|A|$, namely any finitary
information system determined by arbitrary approximable
mapping $f$ from $A^m$ to $A^m_\top$, such that
$\Fix(|f|)$ is a non-empty subset of $|A|$. Then, however, it is unlikely
that we could form a domain of such subdomains, although
I did not try to prove such a result. Even if we could
do so, such domains of subdomains would not have to
be isomorphic for isomorphic domains $|A|$ and $|B|$ if
the cardinalities of $D_A$ and $D_B$ differ.
Hence this alternative definition is unsatisfactory.

\subsection{Results for Continuous Scott Domains}

Now the results on continuous information systems by Hoofman
can be easily explained. He retained transitivity of $f$ and
replaced its reflexivity by inverse transitivity. This
resulted in $|f|$ being a retraction.

It is well known that retractions of continuous lattices
are continuous lattices, and that all continuous lattices
can be obtained as retractions of powersets, and these
results can be generalized for continuous Scott domains,
since the only thing which distinguish them from continuous
lattices is that continuous Scott domains do not have to
possess the top element.

These facts explain the success of Hoofman's program for
continuous Scott domains.

\section{Infinitary Logic}

In~\cite{Sazonov} Sazonov and Sviridenko introduced another
generalization of logic. They kept reflexivity and transitivity,
but removed the finitarity requirement that an infinite
set of statements $x$ only infers those statements, which
the finite subsets $u \subseteq x$ infer.

Their approach describes exactly all bounded complete
partial orders. Directed completeness is equivalent to
the following requirement of {\em partial finitarity}:
a set of statements $x$ cannot infer the contradiction,
unless one of its finite subsets $u \subseteq x$ infers
the contradiction.

Hence partially finitary systems of Sazonov and Sviridenko
exactly correspond to complete Scott domains.

Sazonov and Sviridenko also gave characterizations of
some subclasses of complete Scott domains in their system,
in particular, they described a class of their logic
corresponding to continuous Scott domains.

\section{Open Issues}

\subsection{The Class of Finitary Scott Domains}

What is the class of domains described as fixed-point domains
is an important open question. It is well known that
the set of fixed points of a Scott continuous transformation
of a complete lattice is a complete lattice. It was
traditionally thought that the converse is also true,
namely that any complete lattice can be obtained as
a set of fixed points of a Scott continuous transformation of
a sufficiently large powerset.
For example,
Exercise 18.4.3(ii)(1) on page 491 of the famous
textbook on lambda calculus by Barendregt~\cite{Barendregt}
asks to establish this for the case of countable bases of
the respective Scott topologies.

Hence we expected that our approach would allow us to
obtain all complete Scott domains. We thus expected to follow
the terminology of continuous information systems by
Hoofman and call the fixed-point information systems developed
in this chapter {\em complete information systems}.

However, our analysis of literature and numerous conversations
with experts in the field convinced us that the problem
stated in the textbook by Barendregt is, in fact, open.
Our attempts to solve it were, so far, unsuccessful.
Hence we opted for the less committing terminology of
{\em finitary information systems} and {\em finitary Scott domains}.

\subsection{Important Subclasses}

What subclasses of finitary domains will be obtained,
if we require $f$ to be transitive or, more strongly,
inversely reflexive? The conjecture is that we still
obtain all finitary Scott domains as sets of fixed
points of the corresponding functions $|f|$.

Transitivity is, of course, a very desirable property
of a logical system, so it is of interest to check whether
it restricts generality. These questions come in at least
two flavors: for $f$ defining a continuous transformation
of $|A^m|$ and for $f$ defining a continuous transformation
of an arbitrary algebraic Scott domain $|A|$.

\subsection{Approximable Mappings And Other Issues
for Finitary Information Systems}

Hoofman successfully introduced the notion of
continuous approximable mapping for continuous
information systems based on the fact, that given
spaces $A$ and $B$ and their retractions $r_A:A \rightarrow A$
and $r_B:B \rightarrow B$, one can build a retraction
$(A \rightarrow B) \rightarrow (A \rightarrow B)$, namely
$f \mapsto r_B \circ f \circ r_A$. Scott continuous
functions $f:A \rightarrow B$,
such that $f = r_B \circ f \circ r_A$, are in one-to-one
correspondence with Scott continuous functions
$\Fix(r_A) \rightarrow \Fix(r_B)$.

Unfortunately, no such simple solution seems possible
for finitary information systems. Yet, a satisfactory
notion of finitary approximable mapping seems to be
necessary for further successful studies of
finitary information systems.

There is a possibly simpler variant of this problem if $f$ is
required to be transitive.

\subsection{Possible Translation between Nonreflexive and
            Nonfinitary Logics}

In~\cite{Sazonov} Sazonov and Sviridenko gave a translation
between their logic and the logic of Hoofman for
continuous Scott domains.

Naturally there is a question whether such a translation
is possible for a larger class of domains.

It also might be of interest to combine the features
of nonreflexive and nonfinitary logic in one system.

\chapter{Subdomains for the Algebraic Case}\label{chap:Subdomains}

This chapter represents our results obtained in 1986-1988 and
presented in~\cite{Bukatin1}. With the introduction of 
our new version of the definition of algebraic
information system and our ideas in nonreflexive logic
the presentation is greatly simplified.

For the duration of this chapter ``domain" means algebraic
Scott domain, ``subdomain" means algebraic Scott subdomain,
and ``information system" means algebraic information system.

\section{The Brief History of the Question}

In this chapter, we study the question ``Which subsets of a domain
should be considered subdomains?'' for the domains described by
algebraic information systems.
One criterion for a good answer
is a requirement that all subdomains of domain $|A|$
should themselves form a domain, $|\Sub(A)|$. We would also like to be
able to solve domain equations such as
$|X| \cong |\Sub(X)| + \ldots$.
It will be shown in the next few pages
that the framework of domains as abstract cpo's does not
provide sufficient hints for the solution, but information
systems do.

The systematic use of the operation of taking
a certain subset of a domain to form another domain and the systematic
consideration of domains, whose elements are domains,
first appear in the mid-'70's~\cite{Scott76,ShamirWadge}.
It is notable that nobody has tried
to consider arbitrary subsets of domain $|A|$ that would satisfy
a particular version of domain axioms, although at the first glance
this would seem to be the most intuitive and general version of a subdomain
definition in the framework of domains as abstract cpo's.
The reason why nobody has considered such a definition is that
such arbitrary subsets might satisfy the domain axioms for random causes,
and their collection would be totally unmanageable; in particular, one cannot
hope to form a domain, $|\Sub(A)|$, with these subsets as elements.

The next idea, which originated by Dana Scott in~\cite{Scott76}, 
is to define a subdomain as a
set of fixed points of a {\em retraction} belonging to a certain class.
There are two problems here: a) What are the reasons for using retractions?
b) What is the appropriate class of retractions? None of these problems
receives a clear answer in the framework of domains as abstract cpo's.
On the use of retractions  Scott writes: ``it seems almost an
accident that the idea [to use the retractions] can be
applied'' (see~\cite{Scott76}, p.540).

For the algebraic case, Scott suggests using {\em closure operations},
that is, retractions $|r|$, such that
$|r| \sqsupseteq id$.
This suggestion can be motivated by the
theorem (see~\cite{Scott76}, Theorem 5.1)
saying that the fixed points of a closure operation [on an algebraic lattice]
form an algebraic lattice. Although this is not true for arbitrary
retractions or projections (i.e. retractions $|r|$, such that
$|r| \sqsubseteq id$), if we consider {\em finitary} retractions or
projections ({\em finitary}  means  algebraicity of the set
of fixed points), then this argument no longer favors closure operations.

In the paper entitled ``Data Types as Objects"~\cite{ShamirWadge},
Shamir and Wadge
suggest that to describe the
systems of polymorphic types, one should consider data types
that incorporate their own subtypes as  elements, and  allow
an element of a data type to belong
simultaneously to different subtypes of this
data type. 

Let us briefly present the main features of subtypes in~\cite{ShamirWadge}.
The subtype
$|Y| = \{ x \in |D| \ |\ x \sqsubseteq y\}$ is associated
with each element $y$ of a data type $|D|$. We can view
such subtype as the set of fixed points of
the (finitary, in the algebraic case)
projection $x \mapsto x \sqcap y$. Subtypes are ordered simply by
inclusion. Thus, this simple approach yields the result that
$|D| \cong |\Sub(D)|$ for every domain $|D|$. Each element represents
the subtype of its approximations, and an element $x$ belongs to all
subtypes $|Y|$ of $|D|$ that are represented by $y$'s, such that
$x \sqsubseteq y$.

At the end of the Introduction to~\cite{Bukatin1}, we discussed
the rather adverse relations between retraction-based approaches to
subtyping (~\cite{Scott76,ShamirWadge}, and this chapter are
based on retractions) and the dominant form of modern
theories of typing  that deals with types of intensional objects
(the type polymorphism in ML is, probably, the most widely known example).

\section{Algebraic Subdomains Correspond to Closures}

When we considered this problem in~\cite{Bukatin1},
we were using classical information systems. We wrote:
``Let us consider informally, what should we do to an information
system $A$ to obtain a subdomain $|B| \subseteq |A|$.
We would like to destroy
some elements of $|A|$. The elements are consistent
and deductively closed sets of assertions in $A$. Therefore, some
elements of $|A|$ must lose their consistency or deductive closure
to be destroyed. Also notice, that the loss of deductive
closure means intensification of entailment."

Then we figured out, that inconsistency can be treated
as entailment of contradiction in $A_\top$, and hence the
loss of consistency can be also treated as
strengthening of entailment. The need to keep considering
separately entailment and consistency in a number of
technical situations, nevertheless, complicated our
whole discourse.

With the definition of algebraic information system
introduced in this Thesis, the matters become much
simpler.

\begin{definition}
We say that for algebraic information systems $A$ and $B$,
domain $|B|$ is an {\em (algebraic) subdomain} of
domain $|A|$ if $|B| \subseteq |A|$.
\end{definition}

\begin{lemma}[Contravariance]
For algebraic information systems $A$ and $B$, such that $D_A = D_B$,
$|B| \subseteq |A|$ iff $\vdash_A \subseteq \vdash_B$.
\end{lemma}

It is easy to see that the case when $D_A \neq D_B$
does not lead to new domains $|B|$, so we do not
consider this case further.

Since every algebraic information system is a finitary
information system, the discourse from the previous
chapter applies. Namely, we can see that if $\vdash_A \subseteq \vdash_B$,
the approximable mapping $f$ from $A$ to $A_\top$, which
corresponds to $\vdash_B$, is reflexive. 
 
Since $B$ is
an algebraic information system,  the restriction of
$f$ to $A \rightarrow A$ approximable mapping (or its extension
to $A_\top \rightarrow A_\top$) is also transitive,
so the 
corresponding restriction or extension of $|f|$ would be a closure.
Also the non-triviality condition $|f|(\bot_A) \neq \top_{A_\top}$ holds.

\begin{definition}
An approximable mapping $f$ from an algebraic information
system $A$ to the algebraic information
system $A_\top$ is called a {\em generalized non-trivial
closure}, if
\begin{enumerate}
\item $\forall u, v \in_{fin} \Fin(D_A).\ u \vdash_A v \Rightarrow
      u f v$ (reflexivity of the restriction on $A$);
\item $\forall u, v, w \in_{fin} \Fin(D_A).\ u f v, v f w \Rightarrow
      u f w$ (transitivity of the restriction on $A$);
\item $\neg (\emptyset f \{\nabla_{A_\top}\})$ (non-triviality);
\end{enumerate}
\end{definition}

\begin{lemma}
Such generalized non-trivial closures from $|[A \rightarrow A_\top]|$
are in one-to-one
correspodence with algebraic subdomains of $|A|$. Moreover,
for closures $f$ and $g$ and the corresponding subdomains $|F|$ and $|G|$,
$|F| \subseteq |G|$ iff $g \subseteq f$.
\end{lemma}

\section{A Metatheorem on Reflexive Transitive Closure}

We are about to build the domain of algebraic subdomains
of an algebraic Scott domain $|A|$ as an algebraic subdomain
of the algebraic Scott domain $|[A \rightarrow A_\top]|$.
Our new setting allows us to do it
in a systematic way compared to~\cite{Bukatin1}.

\begin{definition}
Consider the minimal algebraic
information system $A^m = (D_A, \nabla_A, \vdash^m_A~)$ and
any binary relation $R \subseteq \Fin(D_A) \times \Fin(D_A)$.
Then define the {\em reflexive transitive closure} $\vdash^R_A
\subseteq \Fin(D_A) \times \Fin(D_A)$ as follows.
For any $u, v \in \Fin(D_A)$, we say that $u \vdash^R_A v$, if
there is a finite sequence $v_1, \ldots, v_n \in \Fin(D_A)$,
such that $v = v_n$ and the following
condition holds: 

Denote $u_0 = u$ and, inductively,
$u_i = u_{i-1} \bigcup v_i$ for all $i \in \{1, \ldots, n\}$.
Then for any $i \in \{1, \ldots, n\}$, we require that 
either $\nabla_A \in u_{i-1}$,
or $v_i \subseteq u_{i-1}$,
or there is $u' \subseteq u_{i-1}$, such that
$u' R v_i$.
\end{definition}

\begin{definition}
Consider the minimal algebraic
information system $A^m=(D_A,\nabla_A, \vdash^m_A~)$ and
any binary relation $R \subseteq \Fin(D_A) \times \Fin(D_A)$.
We say that a set $x \subseteq D_A$ is a {\em theory with respect
to $R$}, if $x$ is consistent, that is $\nabla_A \not\in x$,
and closed under R, that is if $u \subseteq x$ and $u R v$,
then $v \subseteq x$.
\end{definition}

\begin{theorem}
Consider the minimal algebraic 
information system $A^m=(D_A,\nabla_A, \vdash^m_A~)$ and
any binary relation $R \subseteq \Fin(D_A) \times \Fin(D_A)$.
If there is at least one closed theory $x$ with respect to $R$,
then $A^R=(D_A,\nabla_A,\vdash^R_A)$ is an algebraic information
system and domain $|A^R|$ is equal to the set of all theories
closed with respect to $R$.
\end{theorem}

\begin{corollary}
Consider an algebraic information system $A=(D_A,\nabla_A,\vdash_A)$
and any binary relation $R \subseteq \Fin(D_A) \times \Fin(D_A)$.
Define binary relation $R' = R \bigcup \vdash_A$.
Then if domain $|A|$ contains at least one element which
is a theory closed with respect to $R$, then
domain $|A^{R'}|$ consists of those elements of the domains $|A|$,
which are theories closed with respect to $R$.
\end{corollary}

\section{The Domain of Algebraic Subdomains}

We are now ready to study the domain of algebraic subdomains
of an algebraic Scott domain $|A|$. We denote this domain
$|\Sub(A)|$. This domain will consist of all generalized
non-trivial closures from $|[A \rightarrow A_\top]|$.
These closures will represent the corresponding algebraic
subdomains of $|A|$.

\begin{theorem}
The set $|\Sub(A)|$ of generalized non-trivial closures
from $|[A \rightarrow A_\top]|$ is an algebraic subdomain
of the algebraic Scott domain $|[A \rightarrow A_\top]|$.
Hence, $|\Sub(A)|$ is an algebraic Scott domain itself.
\end{theorem}

\Proof
Consider the algebraic information system $F=[A \rightarrow A_\top]$.
We need to define the relation $R$, which would axiomatize
properties of generalized non-trivial closures.

Consider the following three sets:

$R_r = \{\langle \emptyset, \{\langle u, v \rangle\} \rangle 
\ |\ u \vdash_A v, u, v \in \Fin(D_A)\}$.

$R_t = \{\langle \{\langle u, v \rangle, \langle v, w \rangle\},
\{\langle u, w \rangle\} \rangle
\ |\ u, v, w \in \Fin(D_A)\}$.

$R_n = \{\{\emptyset, \nabla_{A_\top}\}, \{\nabla_F\}\}$.

They axiomatize, respectively, reflexivity, transitivity,
and non-triviality.

Take $R = R_r \bigcup R_t \bigcup R_n$.

The set $\vdash_A$ is an approximable mapping from
$|F|=|[A \rightarrow A_\top]|$. (Remark: $|\vdash_A|$ is
the embedding $\lambda x.x$). $\vdash_A$ is also
a theory with respect to $R$. Hence, the corollary from
the previous section implies, that
domain $|F^{R'}|$ consists of those elements of the domains $|F|$,
which are theories closed with respect to $R$. It is easy to
see that these elements are precisely all generalized non-trivial
closures.

 Hence, the information system $F^{R'}$ is exactly the information
system $\Sub(A)$ we have been looking for.
\eproof

\begin{theorem} \label{sec:sub_contra}
For any $b, c \in |\Sub(A)|$, $b \sqsubseteq_{\Sub(A)} c$ iff
for the corresponding algebraic subdomains $|B|$ and $|C|$,
$|C| \subseteq |B|$.
\end{theorem}

The idea behind this result is simple: the stronger the
entailment, i.e. the larger the generalized closure is,
the fewer elements of $|A|$ remain intact (consistent
and deductively closed).

This differs favorably from~\cite{ShamirWadge} and from the other approaches
to subdomains as sets of fixed points of projections, where subdomains
are ordered by inclusion. Indeed, the smaller a subdomain is
as a set, the more
information we have about each of its elements,
and hence, the larger this subdomain should be informationally
($integers \sqsupseteq reals$ is desirable).

\begin{theorem}
The continuous functions
$|i|:|A|\rightarrow|\Sub(A)|$ and $|j|:|\Sub(A)|\rightarrow|A|$ defined below
yield a projection of $|\Sub(A)|$ onto $|A|$. In terms of closure operations
\begin{list}{}
  {\setlength{\topsep}{0pt} \setlength{\itemsep}{0pt}
  \setlength{\leftmargin}{1in}}
  \item $|i|(y)=\lambda x:|A|. x \sqcup_{A_{\top}} y$;
  \item $|j|(r)=|r|(\bot)$.
\end{list}
In terms of subdomains themselves
\begin{list}{}
  {\setlength{\topsep}{0pt} \setlength{\itemsep}{0pt}
  \setlength{\leftmargin}{1in}}
  \item $|i|(y)=\{x \in |A|\ |\ x \sqsupseteq y\}$;
  \item $|j|(S)=\min_{|A|}S$, where $S$ is a subdomain of $|A|$.
\end{list}
\end{theorem}

Notice that some light non-triviality is involved in the construction
of the embedding $|i|:|A|\rightarrow|\Sub(A)|$. All one-element subsets of
$|A|$ form subdomains (the corresponding generalized closure is
$|r_{y}|=\lambda x:|A_{\top}|. if\ x \sqsubseteq y\ then\ y\ else
\ \top_{A_{\top}}$). But due to Theorem~\ref{sec:sub_contra}, 
they are all incomparable
(in fact, they are total elements in $|\Sub(A)|$). Therefore,
the function $x \mapsto \{x\}$ is not monotonic.
This implies that this function is not Scott continuous and cannot
be used as an embedding.

\section{Open Issues}

\subsection{Reflexive Domain of Algebraic Subdomains}

Because $|A|$ and $|\Sub(A)|$ form an embedding-projection pair,
we can use the inverse limit construction to obtain solutions of
certain domain equations involving $\Sub$~\cite{LarsenKG:usiiss,
LARSON91,SCOTT82},
e.g. $|X|\cong|\Sub(X)|$, $|X|
\cong|\Sub(X)|+|A|$,
etc. Because $\Sub(A)$ describes theories of all possible ways to intensify
our means of entailment in $A$,
it might be of interest to study
{\em explicitly} the iterations of this construction,
$\Sub(\Sub(A))$, $\Sub^{3}(A)$, etc., 
and their limit $D$, where $|D|\cong|\Sub(D)|$.

\subsection{Relationship between Subdomains and Subtypes}

In~\cite{Bukatin1} we expressed the opinion that compile-time
and inheritance-oriented approaches to types, like subclasses
in object-oriented programming and variants of strongly
normalizing typed lambda-calculi, are incompatible with
retraction-based subtyping in domain theory. This opinion
was further confirmed by our reading of Section 10.4.4
on partial equivalence relations as types
for systems with subtyping in the textbook by Mitchell~\cite{Mitchell}. 
Mitchell explains that in such
systems entailment relation for the supertype should, in general,
be stronger, than entailment for the subtype. This and next
chapters show that this is impossible for retractions yielding
algebraic domains, at least as long as we stay with the reflexive
transitive finitary logic.

Thus the issue of developing practical programming language
with types conforming to ``data types as objects" paradigm~\cite{ShamirWadge},
where types are run-time first-class objects, and any
compile-time type checking and inference is considered to
be an optimizing partial evaluation, remains an important
open issue. As long as denotational semantics of such a language
is expressible via algebraic domains, the notion of subtyping should
probably correspond to the notion of subdomain developed in this
chapter.

\chapter{Finitary Retractions and Projections}\label{chap:Fin}

This chapter represents our results obtained in 1986-1988 and
presented in~\cite{Bukatin2}. With the introduction of 
our new version of the definition of algebraic
information system and our ideas in nonreflexive logic
this material became much more transparent.

In general, sets of fixed points of retractions of
algebraic Scott domains form continuous Scott domains.
In this chapter we study the classes of finitary
retractions and projections.

\begin{definition}
A retraction or projection of an algebraic Scott domain 
$|r|:|A| \rightarrow |A|$ is called
{\em finitary} if its set of fixed points is algebraic.
\end{definition}

For the duration of this chapter ``domain" means algebraic
Scott domain, ``subdomain" means algebraic Scott subdomain,
and ``information system" means algebraic information system.

\section{Characterization of the Classes of Finitary
         Retractions and Projections}

\subsection{Finitary Retractions}

\begin{lemma}\label{sec:fin1}
A retraction $|r|$ of an algebraic Scott domain $|A|$ is finitary
if and only if there is an algebraic Scott domain $|B|$ and
Scott continuous functions $|i|:|B| \rightarrow |A|$ and
$|j|:|A| \rightarrow |B|$, such that $\langle |i|, |j| \rangle$
is an embedding-retraction pair, such that $|r|=|i| \circ |j|$.
\end{lemma}

\Proof It is enough to establish, that the embedding
$\Fix(|r|) \rightarrow |A|$ and the map $R : |A| \rightarrow \Fix(|r|)$,
where $R(x) = |r|(x)$ for all $x \in |A|$, are Scott continuous.
This, in turn, follows from the fact that given a directed set $S$
of fixed points of $|r|$, the least upper bound of $S$ in $|A|$ is also
a fixed point of $|r|$.
\eproof

\begin{definition}\label{sec:deffin1}
We say that a retraction $|r|$ of an algebraic Scott domain $|A|$
possesses the {\em intermediate reflexive property} if

$\forall u, w \in \Fin(D_A).\ u r w \Rightarrow
        \exists v \in \Fin(D_A).\ u r v, v r v, v r w$.
\end{definition}

\begin{lemma}\label{sec:fin2}
Consider a retraction $|r|$ of an algebraic Scott domain $|A|$
possessing the intermediate reflexive property. Define
$A^r = (D^r_A, \nabla^r_A, \vdash^r_A)$ as follows.
Let $D^r_A = \{u \in \Fin(D_A)\ |\ u r u\}$.
Let $\nabla^r_A = \{\nabla_A\}$.
Let $\{u_1, \ldots, u_n\} \vdash^r_A \{v_1, \ldots, v_m\}$ iff

$(u_1 \bigcup \ldots \bigcup u_n) r (v_1 \bigcup \ldots \bigcup v_m)$ for
all $u_1, \ldots, u_n, v_1, \ldots v_m \in D^r_A$.

Then $A^r$ is an algebraic information system and $|A^r| \cong \Fix(|r|)$.
\end{lemma}

\Proof
Checking the axioms of algebraic information system for $A^r$
is straightforward. Observe that reflexivity of
$\vdash^r_A$ follows from our selection of only such
elements $u \in \Fin(D_A)$ as members of $D^r_A$, that $u r u$.

Now we use the intermediate reflexive property of $r$.
For any $x \in \Fix(|r|)$,
the set $\{u \in D^r_A\ |\ u \subseteq x\}$ belongs to $|A^r|$.
For any $y \in |A^r|$, the set $\bigcup \{u\ |\ u \in y\}$
is a fixed point of $|r|$. These are monotonic injective maps. 
This establishes the desired
isomorphism.
\eproof

\begin{lemma}\label{sec:fin3}
Given an algebraic information system $A$ and
an approximable mapping $r$ from $A$ to $A$ defining a retraction $|r|$,
the following conditions are equivalent:
\begin{enumerate}
  \item $|r|$ is finitary.
  \item $\forall u, w \in \Fin(D_A).\ u r w \Rightarrow
        \exists v \in \Fin(D_A).\ u r v, v r v, v r w$.
  \item $\forall u, w \in \Fin(D_A).\ u r w \Rightarrow
        \exists v \in \Fin(D_A).\ u r v, v r v, v \vdash_A w$.
\end{enumerate}
\end{lemma}

\Proof
$3 \Rightarrow 2$ follows from
$r$ being an approximable mapping, and $2 \Rightarrow 3$ follows
from considering $v' = v \bigcup w$ and observing that
$u r v, v r v, v r w$ implies $u r v', v' r v', v' \vdash_A w$.

$2 \Rightarrow 1$ follows from the previous lemma.

Here we prove $1 \Rightarrow 2$. Consider an algebraic information
system $B$ and approximable mappings $i$ and $j$ from the
Lemma~\ref{sec:fin1}. Consider $u, w \in \Fin(D_A)$, such that
$u r w$. Then there is $v' \in \Fin(D_B)$, such that
$u j v'$ and $v' i w$, because $r = i \circ j$. Because
$j \circ i = \vdash_B$, there is $v \in \Fin(D_A)$, such
that $v' i v$ and $v j v'$. Then $u r v$, $v r v$, and $v r w$.
\eproof

\subsection{Finitary Projections}

\begin{lemma}\label{sec:fin4}
Given an algebraic information system $A$ and
an approximable mapping $p$ from $A$ to $A$,
the following conditions are equivalent:
\begin{enumerate}
  \item $|p|$ is a finitary projection.
  \item $\forall u, w \in \Fin(D_A).\ u p w \Rightarrow
        \exists v \in \Fin(D_A).\ u \vdash_A v, v p v, v \vdash_A w$.
\end{enumerate}
\end{lemma}

\Proof
$1 \Rightarrow 2$ follows from the previous lemma and the fact
that for a projection $u p v \Rightarrow u \vdash_A v$.

To prove $2 \Rightarrow 1$, observe that in the condition 2
$u p v$ and $v p w$, so $|p|$ is a finitary retraction,
and $u \vdash_A w$, so $|p|$ is a projection.
\eproof

\section{Domains of Fixed Points of Finitary Projections and Retractions
         from the Viewpoint of Logic of Fixed Points}

We could have used our technique of fixed-point subdomains
to describe the set of fixed points of an arbitrary or
finitary retraction or projection. However, since arbitrary
retractions and projections do not possess reflexivity property,
this would move us outside of the realm of
algebraic information system even in the finitary case.

Instead, we use the fact that finitary retractions are exactly
those possessing the {\em intermediate reflexive property} that if
$\forall u, w \in \Fin(D_A).\ u r w \Rightarrow
        \exists v \in \Fin(D_A).\ u r v, v r v,$ $v r w$.
Informally, this means that there are sufficiently many finite conjuctions
$v$, which are reflexive ($v r v$) with respect to the retraction $r$,
considered as prospective entailment relation, so that any
entailment done by $r$ can be done {\em via} such a reflexive
conjuction $v$. These reflexive conjunctions serve as a backbone
of the new algebraic information system $A^r$ describing the domain
isomorphic to the domain of fixed points of $|r|$.

Then, in the case of $r$ defining a projection, we build the
domain $|A^r|$ via mechanism of {\em conjuctive completion}
and of {\em forgetting} the statements, not reflexive with
respect to $r$. In the case of $r$ defining a general finitary retraction, 
we also have to replace the native entailment by the one induced by $r$
(Lemma~\ref{sec:fin2}). It is precisely the {\em forgetting}
step, which makes it impossible to consider sets of fixed points
of finitary projections and retractions as algebraic subdomains,
as we will see later in this section.

\subsection{Conjuctive Completeness}

We say that algebraic information system $A$ is {\em (finitely) conjunctively
complete}, if for any $u \in \Fin(D_A)$, there is $d \in D_A$,
such that $\{d\} \vdash_A u$ and $u \vdash \{d\}$.

The step $D^r_A = \{u \in \Fin(D_A)\ |\ u r u\}$ 
in Lemma~\ref{sec:fin2} provides
for {\em conjuctive completion}. Actually, $A^{\vdash_A}_A$ yield
precisely the conjuctive completion of algebraic information
system $A$, and $|A^{\vdash_A}_A| \cong |A|$.

It is actually enough to add conjuctions only for such
$u \in \Fin(D_A)$, that $u r u$ and there is no such
$d \in D_A$, that $\{d\} \vdash_A u$ and $u \vdash \{d\}$,
so if we start from a conjuctively complete system to begin
with, the conjuctive completion step can be omitted. 

\subsection{Domains of Fixed Points of Finitary Projections}

When $p$ defines a finitary projection, the construction
of Lemma~\ref{sec:fin2} is rewritten as follows.

An algebraic information system
$A^p = (D^p_A, \nabla^p_A, \vdash^p_A)$ is defined by
$D^p_A = \{u \in \Fin(D_A)\ |\ u p u\}$,
$\nabla^p_A = \{\nabla_A\}$,
$\{u_1, \ldots, u_n\} \vdash^p_A \{v_1, \ldots, v_m\}$ iff
$(u_1 \bigcup \ldots \bigcup u_n) \vdash_A (v_1 \bigcup \ldots \bigcup v_m)$ for
all $u_1, \ldots, u_n, v_1, \ldots v_m \in D^p_A$.

As before, $|A^p| \cong \Fix(|p|)$.

If the information system $A$ was conjuctively complete to begin
with, the construction above could be modified as follows:

$D^p_A = \{d \in D_A\ |\ \{d\} p \{d\}\}$, $\nabla^p_A = \nabla_A$,
$u \vdash^p_A v \Leftrightarrow u \vdash_A v$ for all $u, v \in \Fin(D^p_A)$.

Then for any $x \in \Fix(|p|)$, the set $x \bigcap D^p_A$
would belong to $|A^p|$.
For any $y \in |A^p|$, the deductive closure of $y$ in $A$
is a fixed point of $|p|$. This establishes the desired
isomorphism. In this case, for any $x \in |A|$, $|p|(x)$ is
the deductive closure of $x \bigcap D^p_A$ in $A$.

\subsection{Why Finitary Projections Do Not Form Algebraic Subdomains}

We give two examples. The first example illustrates the need
for conjunctive completeness or conjuctive completion.
The second example actually shows, why forgetting does not
allow us to consider sets of fixed points of finitary projections
to be subdomains. In both examples we only list domain elements
and sets of fixed points and leave it to the reader to restore
the actual information systems and projections.

Consider the domain $|A|=\{\emptyset,\{1\},\{2\},\{1,2\}\}$.
Consider the
projection $|p|$ of $|A|$ onto $\{\emptyset,\{1,2\}\}$. While there is an
informaton system describing the domain $\{\emptyset,\{1,2\}\}$,
it cannot be obtained by a projection from our original domain,
because the original domain is not conjunctively complete.
The first version of our construction would allow us to obtain
the domain isomorphic to $\Fix(|p|)$ as $|A^p|=\{\emptyset,
\{\{1,2\}\}\}$ via a conjunctive completion.

On the other hand, if we consider an isomorphic conjunctively
complete situation, $\{\emptyset,\{1\},\{2\},\{1,2,t\}\}$
and its projection onto $\{\emptyset,\{1,2,t\}\}$,
the second version of our construction of $A^p$ would
yield $\{\emptyset,\{t\}\}$.

The set $\{\emptyset,\{1,2\}\}$ in the previous example 
can still be obtained as a subdomain of $|A|$, although not via
a projection. Now we consider a different example, namely
$|A|=\{\emptyset,\{1\},\{1,2\},\{1,3\}\}$, and its projection $|p|$
onto $\{\emptyset,\{1,2\},\{1,3\}\}$. The set $\{\emptyset,\{1,2\},\{1,3\}\}$
is not a domain determined by any algebraic information system
(consider, what would be $\overline{\{1\}}$).
The set $\{\emptyset,\{1,2\},\{1,3\}\}$
is not a subdomain of $|A|$ and cannot be obtained from $|A|$ via
a closure operation (consider, where would $\{1\}$ go under
a closure). This is a conjunctively closed situation, and the
second version of our construction of $A^p$ yields
$|A^p|=\{\emptyset,\{2\},\{3\}\}$.

The last example shows that sometimes a set of fixed points
of a finitary projection is not equal to any domain described
by an algebraic information system and, hence, cannot be
considered an algebraic subdomain of the original domain.
In other cases, like our first example, the set in question
is an algebraic subdomain, but the closure operation
describing it does not have anything in common with the
finitary projection in question.

\subsection{Domains of Fixed Points of Generalized
Nontrivial Finitary Retractions}

In order to fully incorporate the discourse of the previous
two chapters into the framework of finitary retractions, one
has to consider {\em generalized
nontrivial retractions} $|r|:|A|\rightarrow |A_\top|$,
such that $|r|(\bot) \neq \top_{A_\top}$ and $|r|(|r|(x))=|r|(x)$,
when $|r|(x) \neq \top_{A_\top}$.

In order to use the setup of Lemma~\ref{sec:fin1},
one has to consider retraction $|r'|:|A_\top|\rightarrow |A_\top|$,
such that $|r'|(\top_{A_\top})=\top_{A_\top}$ and
$|r'|(x)=|r|(x)$ for $x \in |A|$. Notice that $|B|$ has
the compact top element, and that $\Fix(|r|) \cong |B_{\not\top}|$. 

Definition~\ref{sec:deffin1} can be extended to the generalized
retraction $|r|$ without change.

Lemma~\ref{sec:fin2} and its proof are applicable to the generalized
nontrivial retraction $|r|$ without change as well.

Lemma~\ref{sec:fin3} also holds for $|r|$, but in order
to establish $1 \Rightarrow 2$ in its proof, one should
consider $|r'|$.

Lemma~\ref{sec:fin3} gives the criterion of
finitarity of generalized nontrivial retractions, and Lemma~\ref{sec:fin2} 
yields the construction of the domain isomorphic to the domain
of fixed points of a generalized nontrivial finitary retraction.

\subsection{Other Criteria of Finitarity}

The following criterion of finitarity for projections was known
before.
A projection $|p|:|A|\rightarrow |A|$ is finitary iff
$\forall x \in |A|.
\ x=|p|(x) \Rightarrow x=\sqcup\{x_{0}\in |A|_{0}\ |\ x_{0}=|p|(x_{0}), x_{0}
\sqsubseteq x\}$. This criterion can be obtained from Lemma 1(ii) 
of~\cite{Huth},
which in effect says that the set of finite elements in $\Fix(|p|)$
is $\{x_{0} \in |A|_{0}\ |\ x_{0}=|p|(x_{0})\}$.

The consideration of information systems easily produces a much nicer
criterion based on the intermediate reflexive property
(Lemma~\ref{sec:fin4}).
It does not require consideration of arbitrary non-finite elements $x$ and
allows to select finitary projections from the space of all continuous
functions rather than from the space of projections. It can be
literally rewritten in term of abstract cpo's: $\forall x_{0}, z_{0}
\in A_{0}.\ z_{0} \sqsubseteq |p|(x_{0}) \Rightarrow \exists y_{0} \in A_{0}.
\ x_{0} \sqsupseteq y_{0} \sqsupseteq z_{0}, y_{0}=|p|(y_{0})$.
Yet, one is unlikely to chose this criterion if he does not consider
information systems because it involves inequality $z_{0} \sqsubseteq
|p|(x_{0})$. Also notice that in this case we have to consider
$|p|(x_{0})$, which is generally not finite.

For retractions, the criterion
based on the intermediate reflexive property is given by
Lemma~\ref{sec:fin3}. In terms of abstract cpo's our criterion can be
rewritten as follows:
a retraction $|r|:|A|\rightarrow |A|$ is finitary iff
$\forall x_{0}, z_{0}
\in A_{0}.\ z_{0} \sqsubseteq |p|(x_{0}) \Rightarrow \exists y_{0} \in A_{0}.
\ y_{0} \sqsubseteq |p|(x_{0}), y_{0} \sqsubseteq |p|(y_{0}),
z_{0} \sqsubseteq |p|(y_{0})$.

The finite elements of $\Fix(|r|)$ are
obtained as images of the finite elements of $|A|$ under $|r|$.
Hence, in the style close to~\cite{Huth}, another criterion can be
written as follows:
a retraction $|r|:|A|\rightarrow |A|$ is finitary iff
$\forall x \in |A|.\ x=|r|(x) \Rightarrow x= \sqcup\{|r|(x_{0})\ |
\ x_{0} \in |A|_{0}, x_{0} \sqsubseteq |r|(x_{0}) \sqsubseteq x \}$.

\section{Domains of Finitary Projections and Generalized Finitary Retractions}

\subsection{Domain of Finitary Projections}

The set of all finitary projections $|A|\rightarrow|A|$ can be obtained
as the set of fixed points of
the finitary projection $|Pr|$ of $|A\rightarrow A|$. Specifically,
if $g \in |A\rightarrow A|$ define $f=|Pr|(g)$ by
$ufw$ iff $\exists v \in \Fin(D_A).\ u \vdash_A v, vgv, v \vdash_A w$.

It is easy to check that $f$ is a finitary projection, and that if $g$ is
a finitary projection then $g=|Pr|(g)$. To prove that $|Pr|$ itself is
a continuous function and a finitary projection, one should notice that
$\{(u_{1},w_{1}), \ldots, (u_{n},w_{n})\} Pr \{(u,w)\}$ iff
$\exists v.\ u \vdash_A v, v \vdash_A w,
\{(u_{1},w_{1}), \ldots, (u_{n},w_{n})\}
\vdash_{A\rightarrow A} (v,v)$.

\subsection{Domain of Generalized Finitary Retractions}

In a similar fashion the set of all generalized nontrivial finitary
retractions can be obtained as a set of fixed points of
a generalized nontrivial finitary retraction $|Ret|$
of $|A \rightarrow A_\top|$. Hence this set is an algebraic
Scott domain.

\section{Omitted Issues}

The details of construction of $|Ret|$ is omitted in the present
text. This construction is the result of mating the construction
of $|Pr|$ with the construction of the domain of subdomains.

\subsection{Limits of Projective Sequences}

It is possible to generalize the construction of the limits
of some projective sequences of domains defined by algebraic
information systems, $\{\langle |A_n|,|A_{n+1}|,|i_n|,
|j_n| \rangle\ |\ 
|i_n|:|A_n|\rightarrow|A_{n+1}|; |j_n|:|A_{n+1}|\rightarrow|A_n|;
\langle |i_n|,
|j_n|\rangle$ is an embedding-projection pair; $n=1, \ldots\}$,
from~\cite{LarsenKG:usiiss,LARSON91} to arbitrary sequences of
this kind. We do not give the details of the generalized construction
here.

\section{Open Issues}

\subsection{Issues Related to the Domain of Finitary Retractions}

I do not know, whether $|Ret|$ is unique. 

The question might
be somewhat related to the properties of the space of all retractions,
say, of a powerset. This space is a complete lattice, but it is not
algebraic and not even continuous.

\subsection{Other Subclasses of Finitary Retractions}

Here we look at two special subclasses of the class of finitary retractions,
which may deserve special attention.

Let us look at the criteria for finitarity of retractions and projections
once more. For retractions it is $urw \Rightarrow \exists v.\ urv, vrv, vrw$,
which can be rewritten as $urw \Rightarrow \exists v.\ urv, vrv, v \vdash
w$. For projections it is $upw \Rightarrow \exists v.
\ u \vdash v, vrv, v \vdash w$. There is an intermediate case
--- retractions satisfying the property that
$urw \Rightarrow \exists v.\ u \vdash v, vrv, vrw$.
This class of retractions includes both finitary projections and
closure operations, and the retractions from this class
 possess a certain weak property of minimality, hence
they can be called {\em quasi-minimal}.

Another class is obtained from the observation that while for finitary
projections finite elements of sets of fixed points are finite
in the original domain too, this does not generally 
hold for finitary retractions.
We can call the retractions possessing this property
{\em strongly finitary}. The class of strongly finitary
retractions includes the class of finitary projections, but does not
include the class of closure operations.

Both quasi-minimal retractions and strongly finitary retractions may be
worth more detailed studies.

\part{Elements of Analysis on Domains}\label{part:Analysis}
\chapter{Co-continuous Valuations} \label{chap:Valuations}

In this chapter we present our joint results with
Svetlana Shorina~\cite{BukatinShorina,BukatinShorina2}.
We study the  continuous normalized valuations $\mu$
on the systems of open sets of
domains and introduce notions of {\it co-continuity},
$\{U_i,\, i \in I\}$ is a filtered system of sets $\Rightarrow \mu(\Int
(\bigcap_{i \in I} U_i))=\inf_{i \in I} \mu (U_i)$,
and {\it strong non-degeneracy}, $U\subset V$ are open sets $\Rightarrow
\mu(U) < \mu(V)$, for such valuations.
We call the resulting class of valuations CC-valuations.
The central result of this chapter is a construction of CC-valuations
for Scott topologies on
all continuous dcpo's with countable bases. 

It seems that the notion of
co-continuity of valuations
and this result for the case of continuous Scott domains
with countable bases are both new
and belong to us~\cite{BukatinShorina,BukatinShorina2}.
The generalization of this result to
continuous dcpo's with countable bases belongs to
Klaus Keimel~\cite{Keimel}.
He worked directly with {\em completely distributive
lattices} of Scott open sets of continuous dcpo's
and used the results about completely
distributive lattices obtained by Raney in the fifties
(see Exercise 2.30 on page 204 of~\cite{Compendium}).
Here we present a proof which can be considered
a simplification of both our original proof and the proof
obtained by Keimel. This proof
also works for all continuous dcpo's
with countable bases.
A part of this proof, as predicted by Keimel,
can be considered as a special case of Raney's
results mentioned above. However, our construction is
very simple and self-contained. 

Keimel also pointed out in~\cite{Keimel} that our results are quite
surprising, because both {\em co-continuity} and
{\em strong non-degeneracy}
seem contradictory, as neither of
them can hold for the system of open sets of the
ordinary Hausdorff topology on $[0,1]$. However,
if we replace the system of open sets of this Hausdorff
topology with the system of open intervals, both conditions
would hold. We believe that the reason behind our results is that
the Scott topology is coarse enough for its system of open
sets to exhibit behaviors similar to the behaviors of
typical {\em bases of open sets} of Hausdorff topologies.

\section{CC-valuations}   \label{cc_val}
Consider a topological space $(X, \cal O)$, where ${\cal O}$ consists
of all open subsets of $X$. The
following notions of the theory of valuations can be considered
standard (for the most available presentation in a regular journal
see~\cite{Edalat}; the fundamental text in the theory of
valuations on Scott opens sets is~\cite{Jones}).

\Def A function $\mu :{\cal O} \to [0,+\infty]$ is called {\it
valuation} if
\begin{enumerate}
\item $\forall U, V \in{\cal O}. \;
U\subseteq V \Rightarrow \mu(U)\le\mu(V)$;
\item $ \forall U, V \in{\cal O}.
\; \mu(U)+\mu(V)= \mu(U \cap V)+\mu(U\cup V);$
\item $\mu(\emptyset)=0$.
\end{enumerate}
\Def A valuation $\mu$ is {\it bounded} if $\mu(X)<+\infty$. A valuation
$\mu$ is {\it normalized} if $\mu(X)=1$.
\Rem If a valuation
$\mu$ is bounded and $\mu(X) \ne 0$, then it is always easy to replace it
with a normalized valuation $\mu'(U)=\mu(U)/\mu(X)$.  \Def Define a {\it
directed system of open sets}, 
${\cal U} = \{U_i,\, i\in I\}$, as satisfying the following
condition: for any finite number of open sets
$U_{i_1},\,U_{i_2},\cdots,U_{i_n} \in {\cal U}$ 
there is $U_i,\, i\in I$, such that
$U_{i_1}\subseteq U_i, \cdots, U_{i_n}\subseteq U_i$.
\Def A valuation $\mu$
is called {\it continuous} when for any directed system of open sets
$\mu(\bigcup_{i\in I} U_i) =\sup_{i\in I} \mu(U_i)$.

%Considering $n=0$ it is easy to see that $\mu(\emptyset)=0$ if $\mu$ is
%directed continuous.

We introduce two new properties of valuations.
\Def A valuation $\mu:{\cal O} \to [0,+\infty]$ is {\it strongly
non-degenerate} if $ \forall U,\,V\in {\cal O}.\; U\subset V
\Rightarrow \mu(U)<\mu(V)$.

This is, obviously, a very strong requirement,
and we will see later that it might be reasonable to look for weaker
non-degeneracy conditions.

Consider a decreasing sequence of open sets $U_1\supseteq U_2
\supseteq\ldots$, or, more generally, a {\it filtered system of open sets}
${\cal U} = \{U_i, i \in I\}$, 
meaning that for any finite system of open sets
$U_{i_1},\cdots U_{i_n} \in {\cal U}$ 
there is $U_i,\, i\in I$, such that $U_i \subseteq
U_{i_1}, \cdots, U_i\subseteq U_{i_n}$. Consider the interior of the
intersection of these sets.  It is easy to see that for a valuation $\mu$ $$
\mu(\Int(\bigcap_{i\in  I}U_i))\le \inf_{i \in  I} \mu(U_i).$$

\Def A valuation $\mu$ is called {\it co-continuous}
 if for any filtered
system of open sets $\{U_i,\, i\in I\}$
$$ \mu(\Int(\bigcap_{i\in I}U_i))=
\inf_{i \in  I} \mu(U_i).$$

\Def A continuous, normalized, strongly non-degenerate, \linebreak
co-continuous valuation $\mu$ is called a {\it CC-valuation}.

Informally speaking, the strong non-degeneracy 
provides for non-zero contributions of compact elements
and reasonable ``pieces of space''.
The co-continuity provides for single non-compact
elements and borders $B\setminus \Int(B)$ of
``reasonable'' sets $B\subseteq A$ to have zero measures.

``Reasonable'' sets here are Aleksandrov open (i.e. upwardly
closed) sets. Thus, it is possible to consider
co-continuity as a method of dealing with
non-discreteness of Scott topology. 
We follow here the remarkable definition of a discrete
topology given by Aleksandrov: a topology is discrete if
an intersection of arbitrary family of open sets is open
(e.g. see~\cite{Aleksandrov}). Of course, if one assumes
the $T_{1}$ separation axiom, then the Aleksandrov's definition
implies that all sets are open --- the trivial version of the
definition, which is unfortunately the only one found in the
most textbooks. In this sense, Aleksandrov topology of upwardly
closed sets is discrete, but Scott topology is not.
The further development of this viewpoint might prove to be
fruitful.

We should also notice that since our valuations are bounded,
they can be extended onto closed sets via formula
$\mu(C)=\mu(A)-\mu(A\setminus C)$, and all definitions of this section can be
expressed
in the dual form. This gives us considerable benefits
in applications of these valuations
and also suggests that the generalization 
of these applications to the case of
unbounded valuations is a non-trivial undertaking.

A bounded valuation $\mu$ 
can be uniquely extended to an additive measure
defined on the ring of sets
generated from the open sets by
operations $\cap$, $\cup$, $\setminus$~\cite{Pettis}. 
We feel that it is useful to draft a
possible construction here.
We will denote the resulting additive measure also as $\mu$.

First, consider all sets $U\cap C$, where $U$ is open and $C$ is
closed. They form so-called semiring, which contains $\emptyset$,
is closed under finite intersections, and possesses the following property:
If $S=U \cap C$, $S_1=U_1 \cap C_1$, $S_1 \subset S$, then
there is a system of $n$ non-intersecting sets $S_i=U_i \cap C_i$
(in fact, it is enough to take $n=3$ here), such that $S=\bigcup_{i=1,\ldots,
n}S_i$.

Then define $\mu(U \cap C) = \mu(U) -\mu(U \setminus  C)$,
using the fact that $ U\setminus C=U\cap\overline C$ is an open set. Then
observe that this definition is correct, i.e. if
$U\cap C =U_1 \cap C_1$ then $\mu(U)-\mu(U\setminus C)=
\mu(U_1)-\mu(U_1 \setminus C_1)$, and that the resulting measure
is additive for the decomposition above: $\mu(S)=\mu(S_1)+
\ldots +\mu(S_n)$. Then the unique extension to the ring of sets
in question is the standard result of measure theory.

The issues of $\sigma$-additivity are not in the scope of this text
(interested readers are referred to~\cite{Jones,Edalat2}).
We deal with the specific infinite systems of sets we need, and mainly
focus on quite orthogonal conditions given to us by co-continuity
of $\mu$.

\section{Examples}

\subsection{Valuations Based on Weights of Basic
Elements}
\label{nonalg_noncc}
Consider a continuous dcpo $A$ with a countable basis $K$. Assign
the converging system of weights to basic elements:
$w(k) >0$, $\sum_{k\in K}w(k)=1$. Define $\mu(U)=\sum_{k\in U} w(k)$.
It is easy to see that $\mu$ is a continuous, normalized,
strongly non-degenerate valuation.

However, $\mu$ is co-continuous if and only if all basic elements are compact
(which is possible only if $A$ is algebraic). To see why this is so, we
undertake a small discourse, which will be useful
for us later in this text.

First, observe that intersection of arbitrary system of Alexandrov open (i.e.
upwardly closed) sets is Alexandrov open. Also it is a well-known fact
that $\{y\,|\, y \gg x\}$ is Scott open in a continuous dcpo.

\begin{Lemma}{\bf (Border Lemma) } \label{border_lemma}
Consider an Alexandrov open set $B\subseteq A$.
Then its interior in the Scott topology, $\Int(B)=\{y \in A\,|\, \exists x \in
B.\, x \ll y\}$. Correspondingly, the border of $B$ in the Scott topology, $B
\setminus \Int(B)=\{y \in B\,|\, \neg(\exists x\in B.\, x\ll y)\}$
\end{Lemma}

\begin{Lemma} If $K$ is a basis of compact elements, then $\mu(U)=
\sum_{k\in U} w(k)$, where $w(k)>0$, $\sum_{k\in K}w(k)=1$,
defines a CC-valuation.
\end{Lemma}
\Proof (of co-continuity) Consider a filtered system of
open sets $\{U_i,\, i\in I\}$. The set $B=\bigcap_{i\in I} U_i$
is Alexandrov open. Because $\{U_i\}$ is filtered, it is easy to see
that $\sum_{k\in B} w(k)=\inf_{i\in I} \mu(U_i)$.
Because basic elements of $K$ are compact, $k\in K,\, k\in B
\Rightarrow k\in \Int(B)$. Hence $\sum_{k\in B}w(k)=
\sum_{k \in \Int(B)}w(k)$.
\eproof
\begin{Lemma} If $K$ contains a non-compact element $x$, then $\mu(U)=
\sum_{k\in U} w(k)$, 
where $w(k)>0$, $\sum_{k\in K} w(k)=1$ is not
co-continuous.
\end{Lemma}
\Proof By the definition of basis, $x=\sqcup K_x$, where
$K_x=\{k \in K\,|\, k \ll x\}$ is a directed set. For any $k$,
consider $U_k=\{y\,|\,y \gg k\}$. Because $K_x$ is directed,
it is easy to see that $\{U_k,\, k\in K_x\}$ is filtered. Consider
$B=\bigcap_{k \in K_x} U_k$. It is easy to see that $x \in B$.

Let us show that $x \not\in \Int(B)$. If $x \in \Int(B)$ then
there is $y\in B$
such that $y\ll x$ by the Border Lemma.
Then $y\sqsubseteq k$ for some $k \in K_x$,
and since $B$ is upwardly closed, $k \in B$. This means that
$\forall k'\in K_x.\; k'\ll k$  yielding
$x\sql k$ and $k \ll x$. Therefore $k=x$ and $x$ is compact, which
is a contradiction.

Thus,
$$x \not \in \Int(B) \Rightarrow \sum_{k\in B}w(k) \ge
\sum_{k \in \Int(B)}w(k)+w(x)>\sum_{k \in \Int(B)} w(k)$$
Since $\sum_{k \in B} w(k)= \inf_{k\in K_x} \mu(U_k)$ the Lemma
is proved.
\eproof

\subsection{A Vertical Segment of Real Line} \label{vertical_segment}
Consider the segment $[0,\,1]$, $\sqsubseteq =\le$. Define
$\mu((x,\,1])=1-x$. Unfortunately, to ensure strong non-degeneracy
we have to define $\mu([0,\,1])=1+\epsilon$, $\epsilon>0$.
This is the first hint that strong non-degeneracy might
constitute too strong a restriction in many cases. In order to obtain
a normalized valuation we have to consider $\mu'(U)=\mu(U)/(1+\epsilon)$.
The resulting $\mu'$ is a CC-valuation.

\subsection{Interval Numbers}

Consider the domain $R^I$ of interval numbers belonging to the
segment $[0,\,A]$, where $A$ is finite. 
Consider a Scott open $U \subseteq R^I$.
Define $\mu_0 (U)$ as the geometric area of $U$.
This is a co-continuous valuation.

To insure strong non-degeneracy, we must concentrate some
weight on segments $[ [0,A],\ [0,0] ]$ and 
$[ [0,A],\ [A,A] ]$ and
point $[0,A]$. In order to do that, we take positive
$\epsilon, \epsilon_1, \epsilon_2$ and define
$\mu (U) = \mu_0 (U) + 
\epsilon_1 \cdot Length (U \cap [ [0,A],\ [0,0] ]) +
\epsilon_2 \cdot Length (U \cap [ [0,A],\ [A,A] ])$
when $U \neq R^I$, and $\mu (R^I) = 0.5 \cdot A^2 +
(\epsilon_1 + \epsilon_2) \cdot A + \epsilon$.

Now we can obtain a CC-valuation $\mu'$ by
defining $\mu' (U) = \mu (U) / \mu (R^I)$.

\section{Constructing CC-valuations} \label{c_cc_val}
In this section we build a CC-valuation for all continuous dcpo's with
countable bases. 
The construction generalizes the one of
Subsection~\ref{nonalg_noncc}.
We are still going to assign weights, $w(k)>0$, to
compact elements. For non-compact basic elements we proceed
as follows. We focus our attention on the pairs of non-compact
basic elements, $(k',k'')$, which do not have any compact elements between
them, and call such elements {\em continuously connected}.
We observe, that for every such pair we can construct a
special
kind of vertical chain, which ``behaves like the vertical segment
$[0,1]$ of real line''. We call such chain a {\em stick}.
We assign weights, $v(k',k'')>0$, to sticks
as well, in such a way that the sum of all $w(k)$ and all
$v(k',k'')$ is 1.

As in Subsection~\ref{nonalg_noncc}, compact elements $k$ contribute
$w(k)$ to $\mu(U)$, if $k\in U$. An intersection of the stick,
associated with a continuously connected pair $(k',k'')$,
with an open set $U$ ``behaves as either $(q,1]$ or $[q,1]$'', where
$q\in [0,1]$. Such stick contributes $(1-q) \cdot v(k',k'')$ to
$\mu(U)$. The resulting $\mu$ is the desired CC-valuation.

It is possible to
associate a complete lattice homomorphism from the lattice
of Scott open sets to $[0,1]$ with every compact
element and with every stick defined by basic continuously
connected elements, $k'$ and $k''$.
Then, as suggested by Keimel~\cite{Keimel},
all these homomorphisms together can be thought of as an
injective complete lattice homomorphism to $[0,1]^J$.
From this point of view, our construction of $\mu$
is the same as in~\cite{Keimel}.

Thus the discourse in this section yields the proof of the following:
\begin{theorem} \label{sec:main_th1} For 
any continuous dcpo $A$ with a countable basis,
there is a CC-valuation $\mu$ on the system of its Scott open sets.
\end{theorem}

\subsection{Continuous Connectivity and Sticks}
\Def \label{cont_connected}
Two elements $x \ll y$ are called {\it continuously connected}
if the set $\{ k\in A\ |\ k$ is compact, $x\ll k \ll y\}$
is empty.
\Rem This implies that $x$ and $y$ are not compact.
\begin{Lemma} \label{cont_connect_card}
If $x\ll y$ are continuously connected, then
$\{z\,|\, x\ll z \ll y\}$ has cardinality of at
least continuum.
\end{Lemma}
\Proof We use the well-known theorem on intermediate values
that $x \ll y \Rightarrow \exists z \in A.\ x\ll z \ll y$
(see~\cite{Hoofman}). Applying this theorem again and again we
build a countable system
of elements between $x$ and $y$ as follows, using rational numbers
as indices for intermediate elements:
$$ x\ll a_{1/2} \ll y,\;\,\; x\ll a_{1/4} \ll a_{1/2} \ll a_{3/4} \ll y,
\ldots $$
All these elements are non-compact and hence non-equal.
Now consider a directed set $\{a_i\,|\, i\le r\}$, where
$r$ is a real number, $0 < r < 1$. Introduce $b_r=\sqcup\{a_i\,|\,
i \le r\}$. We prove that if $r<s$ then $b_r \ll b_s$, and also that
$x\ll b_r \ll b_s \ll y$, thus obtaining the required cardinality.
Indeed it is easy to find
such $n$ and numbers $q_1,\, q_2,\,q_3,\,q_4$, that
$$x \ll a_{q_1/2^n} \sql b_r \sql a_{q_2/2^n} \ll a_{q_3/2^n}
\sql b_s \ll a_{q_4/2^n} \ll y.$$

\eproof
\Def We call the set of continuum different non-compact elements
$\{a_r\,|\, r\in (0,1)\}$ between continuously
connected $x \ll y$, built in the proof above, such that
$x \ll a_r \ll a_q \ll z  \Leftrightarrow r<q $ a (vertical) {\it stick}.

\subsection {Proof of Theorem~\ref{sec:main_th1}}
Consider a continuous dcpo $A$ with a countable basis $K$.
As discussed earlier, with every compact $k\in K$ we
associate weight $w(k)>0$, and with every continuously
connected pair $(k',k'')$, $k',k''\in K$, we associate
weight $v(k',k'')>0$ and a stick $\{a_{r}^{k',k''}\ |\ r\in (0,1)\}$.
Since $K$ is countable, we can require $\sum w(k) + \sum v(k',k'') =1$.

Whenever we have an upwardly closed (i.e. Alexandrov open)
set $U$, for any stick $\{a_{r}^{k',k''}\ |\ r\in (0,1)\}$
there is a number $q_{U}^{k',k''} \in [0,1]$, such that
$r < q_{U}^{k',k''} \Rightarrow a_{r}^{k',k''} \not\in U$ and
$q_{U}^{k',k''} < r \Rightarrow a_{r}^{k',k''} \in U$. In particular,
for a Scott open set $U$ define
$$\mu(U)=\sum_{k \in U\mbox{\small is compact}} w (k)\ +
\sum_{k',k''\in K \mbox{\small are continuously connected}} 
(1-q_{U}^{k',k''}) \cdot v(k',k'')
$$

It is easy to show that $\mu$ is a normalized valuation. The rest
follows from the following Lemmas.

\begin{Lemma}\label{sec:mu_cont} $\mu$ is continuous. \end{Lemma}
\Proof Consider a directed system of open sets, $\{U_i, i\in I\}$,
and $U= \bigcup_{i\in I} U_i$. We need to show that for any
$\epsilon >0$, there is such $U_i, i\in I$, that
$\mu(U) - \mu(U_i) <\epsilon$.

Take enough (a finite number of)
compact elements, $k_1, \ldots, k_n$, and continuously
connected pairs of basic elements,
$(k'_1, k''_1), \ldots, (k'_m, k''_m)$, so
that $w(k_1) + \ldots + w(k_n) + v(k'_1,k''_1) + \ldots +
v(k'_m,k''_m) > 1 - \epsilon /2$.
For each $k_j \in U$, take $U_{i_j},i_j \in I$, such that
$k_j \in U_{i_j}$. For each $(k'_j,k''_j)$, such that
$q_U^{k'_j,k''_j} <1$, take $U_{i'_j}, i'_j \in I$, such that
$q_{U_{i'_j}}^{k'_j,k''_j} - q_U^{k'_j,k''_j} < \epsilon /(2m)$.
An upper bound of these $U_{i_j}$ and $U_{i'_j}$ is the
desired $U_i$.
\eproof

\begin{Lemma} $\mu$ is strongly non-degenerate. \end{Lemma}
\Proof 
Let $U$ and $V$ be Scott open subsets of $A$ and $U\subset V$.
Let us prove that $V\setminus U$ contains either a compact element
or a stick between basic elements.
Take $x \in V \setminus U$.
If $x$ is compact, then we are fine. Assume that $x$ is not
compact.  We know that $x=\sqcup K_x$, $K_x=\{k \in K\,|\, k \ll x\}$ is
directed set. Since $V$ is open $\exists k \in K_x.\; k \in V$. Since
$k\sql x$ and $x\not \in U$, $ k \in V\setminus U$. If
there is $k'$ -- compact, such that $k \ll k' \ll x$, we are fine,
since $k' \in V\setminus U$. Otherwise, since any basis includes
all compact elements, $k$ and $x$ are continuously connected.

Now, as in the theorem of intermediate values $x=\sqcup\widetilde K_x$,
$\widetilde K_x=\{k' \in K\,|\, \exists k'' \in K.\; k'\ll k'' \ll
x \}$ is directed set, thus $\exists k', k''.\; k\sql k'\ll k''\ll x$,
thus $(k,k'')$ yields the desired stick.

If $k \in V\setminus U$ and $k$ is compact,
then $\mu(V)-\mu(U) \ge w(k)>0$. If the stick formed by $(k,k')$ is in
$V \setminus U$, then $\mu(V)-\mu(U)\ge v(k,k')>0$.
\eproof

\begin{Lemma} $\mu$ is co-continuous. \end{Lemma}
\Proof Recall the development in Subsection~\ref{nonalg_noncc}.
Consider a filtered system of open sets $\{U_i,\; i \in I\}$.
By Lemma~\ref{border_lemma} for $B=\bigcap_{i\in I}U_i$,
$B\setminus \Int(B)=\{ y\in B\,|\, \neg(\exists x \in B.\; x\ll y)\}$.
Notice that $B\setminus \Int(B)$, in particular, does not contain
compact elements. Another important point is that for
any stick, $q_B^{k',k''} = q_{\Int(B)}^{k',k''}$.

The further development is essentially dual to the proof
of Lemma~\ref{sec:mu_cont}.
We need to show that for any
$\epsilon >0$, there is such $U_i, i\in I$, that
$\mu(U_i) - \mu(\Int(B)) <\epsilon$.

Take enough (a finite number)
of compact elements, $k_1, \ldots, k_n$, and continuously
connected pairs of basic elements,
$(k'_1, k''_1), \ldots, (k'_m, k''_m)$, so
that $w(k_1) + \ldots + w(k_n) + v(k'_1,k''_1) + \ldots +
v(k'_m,k''_m) > 1 - \epsilon /2$.
For each $k_j \not\in \Int(B)$, take $U_{i_j},i_j \in I$, such that
$k_j \not\in U_{i_j}$. For each $(k'_j,k''_j)$, such that
$q_{\Int(B)}^{k'_j,k''_j} >0$, take $U_{i'_j}, i'_j \in I$, such that
$q_{\Int(B)}^{k'_j,k''_j} - q_{U_{i'_j}}^{k'_j,k''_j} < \epsilon
/(2m)$.
A lower bound of these $U_{i_j}$ and $U_{i'_j}$ is the
desired $U_i$.
\eproof

It should be noted that
Bob Flagg suggested and  Klaus Keimel showed that Lemma 5.3 
of~\cite{FlaKop}  can be adapted to obtain a dual
proof of
existence of CC-valuations (see~\cite{Flagg} for one presentation of this).
Klaus Keimel also noted that one can consider all pairs $k,k'$ of
basic elements, such that $k \ll k'$, instead of considering just
continuously connected pairs and compact elements.

\section{Open Problems}

The key open problems are related to fast computation
of valuations and integrals and to canonical
valuations on functional spaces and reflexive domains.

Since the similar problems exist for relaxed metrics
we will return to them at the end of the next chapter.

\chapter{Relaxed and Partial Metrics} \label{chap:Metrics}

In this chapter we present our joint results with
Joshua Scott~\cite{BukatinScott,BukSco:dist_new,BukSco:dist_old}.
We presume 
that the methods of denotational semantics 
allow us to obtain
adequate descriptions of program behavior
(e.g., see~\cite{Stoy}).
The term {\em domain} in this chapter denotes a
directed complete partial order
({\em dcpo}) equipped with the Scott topology.

The traditional paradigm of denotational semantics states
that all data types should be represented by domains and
all computable functions should be represented by
Scott continuous functions between domains. For the
purposes of this chapter all {\em continuous} functions
are Scott continuous.

Consider the typical setting in denotational semantics ---
a syntactic domain of programs, $P$, a semantic domain
of meanings, $A$, and a continuous semantic function,
$[ \! [ \ ] \! ] : P \rightarrow A$. The syntactic domain $P$
(called a syntactic lattice in~\cite{Stoy})
represents a data type of program parse trees, but we
say colloquially that programs belong to $P$.

Assume that we have a
domain representing distances, $D$, and a continuous
generalized distance
function, $\rho: A \times A \rightarrow D$. 
Assume that we can construct a {\em generalized metric
topology}, $\Tcal [ \rho ]$, on $A$ via $\rho$. 
It would be reasonable
to say that $\rho$ reflects computational properties of $A$, if
$\Tcal [ \rho ]$ is the Scott topology on $A$.

Then $\rho ([ \! [  p_{1} ] \! ] , [ \! [  p_{2} ] \! ])$
would yield a computationally meaningful distance 
between programs $p_{1}$ and $p_{2}$. The continuous
function $\rho$ cannot possess all properties of
ordinary metrics because we want $\Tcal [ \rho ]$ to be
non-Hausdorff.

\section{Axiom $\rho(x, x) = 0$ Cannot Hold}

Assume that
there is an element $0 \in D$ representing the ordinary
numerical $0$.
Let us show that $\forall x.\ \rho (x, x) = 0$ cannot be true
under reasonable assumptions. We will see
later that all other properties of ordinary metrics
can be preserved at least for all continuous dcpo's
with countable bases (see the next chapter).

It seems reasonable to assume that
any reasonable construction of $\Tcal [ \rho ]$
for any generalized 
distance function $\rho: A \times A \rightarrow D$,
should satisfy the following axiom, 
regardless of whether the distance
space $D$ is a domain, or whether $\rho$ is continuous:

\begin{axiom}\label{StandardAxiom}
For all $x, y \in A$, $\rho (x, y) = \rho (y, x) = 0$ implies that
$x$ and $y$ share the same system of open sets, i.e.
for all open sets $U \in \Tcal [ \rho ]$, $x \in U$ iff $y \in U$.
\end{axiom}

We assume this axiom for the rest of the chapter.
A topology is called $T_{0}$, if different elements do not
share the systems of open set.

\begin{corollary}\label{StandardCorollary}
If there are $x, y \in A$, such that $x \neq y$ and
$\rho (x, y) = \rho (y, x) = 0$, then $\Tcal [ \rho ]$ is
not a $T_{0}$ topology.
\end{corollary}

Let us return to our main case, where $D$ is a domain and
$\rho$ is a continuous function.

\begin{lemma}\label{SimpleLemma}
Assume that there are at least two elements $x, y \in A$,
such that $x \sqsubset_{A} y$. Assume that 
$\rho: A \times A \rightarrow D$ is a continuous function.
If $\rho (x, x) = \rho (y, y) = d \in D$, then
$\rho (x, y) = \rho (y, x) = d$.
\end{lemma}

\Proof
The continuity of $\rho$ implies its monotonicity
with respect to the both of its arguments. 
Then $x \sqsubset_{A} y$ implies
$d = \rho (x, x) \sqsubseteq_{D}
\rho (x, y) \sqsubseteq_{D} \rho (y, y) = d$. 
This yields $\rho (x, y) = d$, and, similarly, $\rho (y, x) = d$.
\eproof
 
Then we can obtain the following simple, but important result.

\begin{theorem}\label{SimpleTheorem}
Assume that there are at least two elements $x, y \in A$,
such that $x \sqsubset_{A} y$. 
Assume that 
$\rho: A \times A \rightarrow D$ is a continuous 
generalized distance function
and $\Tcal [ \rho ]$ is
a $T_{0}$ topology. Then the 
double equality $\rho (x, x) = \rho (y, y) = 0$
does not hold.
\end{theorem}

\Proof
By Lemma~\ref{SimpleLemma}, $\rho (x, x) = \rho (y, y) = 0$
would imply
$\rho (x, y) = \rho (y, x) = 0$.
Then, by Corollary~\ref{StandardCorollary},
$\Tcal [ \rho ]$ would not be $T_{0}$,
contradicting our assumptions.
\eproof

The topologies used in domain theory are usually $T_{0}$;
in particular, the Scott topology is $T_{0}$.
This justifies studying continuous generalized metrics $\rho$,
such that $\rho (x, x) = 0$ is false for some $x$, more closely.

\subsection{Intuition behind $\rho(x, x) \neq 0$}\label{sec:Intuition}

There are compelling intuitive
reasons not to expect $\rho (x, x) = 0$,
when $x$ is not a maximal element of $A$.
The computational intuition behind $\rho (x, y)$
is that the elements in question are actually
$x'$ and $y'$, $x \sqsubseteq_{A} x', y \sqsubseteq_{A} y'$, but not
all information is usually known about them.
The correctness condition $\rho (x, y) \sqsubseteq_{D} \rho (x', y')$
is provided by the monotonicity of $\rho$.

In particular, even if
$x = y$, this only means that we know the same information about
$x'$ and $y'$, but this does not mean that $x' = y'$.
Consider $x' \neq y'$, such that $x \sqsubset_{A} x'$ and
$x \sqsubset_{A} y'$. Then $\rho (x, x) \sqsubseteq_{D} \rho (x', y')$
and $\rho (x, x) \sqsubseteq_{D} \rho (y', x')$, and at least one
of $\rho (x', y')$ and
$\rho (y', x')$ is non-zero, if we want $\rho$ to yield
a $T_{0}$ topology (we do not assume symmetry yet).

\begin{example}\label{IntervalExample}
Here is an important example --- a 
continuous generalized distance on the
domain of interval numbers
$R^{I}$ --- $\rho : R^{I} \times R^{I} \rightarrow R^{I}$
(see Chapter~\ref{sec:Interval} for the definition of $R^{I}$).
Consider intervals $[a,b]$ and $[c,d]$ and set
$S = \{|x'-y'|\ \ |\ \ a \leq x' \leq b,\ c \leq y' \leq d\}$.
Define $\rho ([a,b], [c,d]) = [\min S, \max S]$.
In particular, $\rho ([a,b], [a,b]) = [0, b - a] \neq [0, 0]$,
and $\rho ([a,a], [b,b]) = [|a-b|,|a-b|]$.
\end{example}

\section{Related Work}

\subsection{Quasi-Metrics}

Quasi-metrics~\cite{SmythMB:quaurd}
and Kopperman-Flagg
generalized distances~\cite{KopFla93} are asymmetric generalized
distances. They satisfy the axiom $\rho(x, x) = 0$
and yield the Scott topology for various classes of
domains via a construction satisfying Axiom~\ref{StandardAxiom}.

Theorem~\ref{SimpleTheorem} means that if one wishes to represent
the distance spaces via domains, these asymmetric distances 
cannot be made continuous unless their nature is changed substantially.

The practice of
representing all computable functions via
continuous functions between domains suggests that 
quasi-metrics
cannot,
in general, be computed
(see Section~\ref{sec:Comput} for details in the effective setting).

\subsection{Partial Metrics}

Historically, {partial metrics} are the 
first generalized distances on domains
for which the axiom $\rho(x, x) = 0$ does not hold.
They were introduced by
Matthews~\cite{Matthews:PartialMetric, TCS::Matthews1995} and
further investigated by Vickers~\cite{Vickers:manuscr}
and O'Neill~\cite{UWARWICK:CS-RR-293}.

Partial metrics satisfy a number of additional axioms
in lieu of $\rho(x, x) = 0$ (see Section~\ref{sec:PartialMetrics}).
Matthews and Vickers state
that $\rho (x, x) \neq 0$ is caused by the fact,
that $x$ expresses a partially defined object.
The most essential component of the central construction
in this and the next chapter is a partial metric (Section~\ref{sec:Central}).

\subsection{Our Contribution}

We build partial metrics 
yielding Scott topologies for a wider class of domains 
that was known before (see the next chapter).
This class --- all continuous dcpo's with countable bases ---
is sufficiently big to solve interesting domain equations
and to define denotational semantics of at least sequential 
deterministic programming languages~\cite{Stoy}.
Actually, a recent careful analysis of 
papers~\cite{WeightedMetrics,Matthews:PartialMetric}
together had shown that an even more general result followed
easily from this papers taken together, however this fact
remained unwritten folklore known only to a few people.

We introduce the notion of
{\em relaxed metric} (Section~\ref{sec:Relaxed}), which
maintains the intuitively clear requirement to
reject axiom $\rho (x, x) = 0$, 
but does not impose the specific axioms of
partial metrics. We believe that the applicability
of these specific axioms is
more limited (Section~\ref{sec:PMIssues}).

We introduce the idea that a space of distances should be
thought of as a data type in the context
of denotational semantics
and, thus, should be represented by a domain.
We also introduce the requirement that distance functions
should be computable and, thus, Scott continuous
(the use of {\em continuous valuations}
in~\cite{UWARWICK:CS-RR-293} 
should be considered as a step in this direction).

These considerations lead to an understanding that 
relaxed metrics should map pairs of partial elements 
to {\em upper estimates} of some ``ideal'' distances,
where the {\em distance domain of upper estimates}, 
$R^{-}$, is equipped
with a dual informational order: $\sqsubseteq_{R^{-}} = \geq$.
We also consider lower estimates of ``ideal'' distances,
thus, introducing the {\em distance domain of interval
numbers}, $R^{I}$. Continuous lower estimates are useful
during actual computations of distances (Section~\ref{sec:Comput})
and for defining and computing
an induced metric structure on the space
of total elements (Theorem~\ref{OnTotals}).

There is a comparison in~\cite{Matthews:PartialMetric} 
between partial metrics and
alternative generalized distance structures such as
quasi-metrics and weighted metrics~\cite{WeightedMetrics}.
We provide what we believe to be the strongest argument
in favor of partial metrics so far --- among all those
alternatives only partial metrics can be thought of
as Scott continuous, computable functions.

\section{Relaxed Metrics}\label{sec:Relaxed}

Consider distance domains
in greater detail. 
It is
conventional to think about distances as non-negative real numbers.
When it comes to considering approximate information about reals,
it is conventional to use some kind of {\em interval numbers}.

We follow both conventions in this text. The distance domain
consists of pairs $\langle a, b \rangle$ (also denoted as $[a, b]$) of 
non-negative reals ($+\infty$ included),
such that $a \leq b$. 

Recall from the Chapter~\ref{sec:Interval}
that we denote this domain as $R^{I}$ and that
$[a, b] \sqsubseteq_{R^{I}} [c, d]$ iff $a \leq c$ and $d \leq b$.

Also recall that 
we can think about $R^{I}$ as a subset of $R^{+} \times R^{-}$,
where $\sqsubseteq_{R^{+}} = \leq$, $\sqsubseteq_{R^{-}} = \geq$,
and both ${R^{+}}$ and ${R^{-}}$ consist of 
non-negative reals  and $+\infty$. We call ${R^{+}}$ a {\em domain
of lower bounds}, and ${R^{-}}$ a {\em domain of upper bounds}.
Thus a distance function $\rho: A \times A \rightarrow R^{I}$
can be thought of as a pair of distance functions $\langle l, u \rangle$,
$l: A \times A \rightarrow R^{+}$, 
$u: A \times A \rightarrow R^{-}$.

We think about $l(x, y)$ and $u(x, y)$ as, respectively,
lower and upper bounds of some ``ideal'' distance $\sigma(x, y)$.
We do not try to formalize the ``ideal'' distances,
but we refer to them to motivate our axioms.
There are good reasons to impose the triangle inequality,
$u(x, z) \leq u(x, y) + u(y, z)$.
Assume that for our ``ideal'' distance,
the triangle inequality, 
$\sigma(x, z) \leq \sigma(x, y) + \sigma(y, z)$, holds.
If $u (x, z) > u (x, y) +
u (y, z)$, then $u (x, y) + u (y, z)$ gives a better
upper estimate for $\sigma (x, z)$ than $u (x, z)$.
This means that unless $u (x, z) \leq u (x, y) + u (y, z)$,
$u$ could be easily improved and,
hence, would be very imperfect.

This kind of reasoning is not valid for $l (x, z)$.
In fact, there are reasonable situations, when
$l(x, z) \neq 0$, but $l (x, y) = l (y, z) = 0$.
E.g., consider Example~\ref{IntervalExample} and
take $x = [2,2], y = [2,3], z = [3,3]$.

Also only $u$ plays a role
in the subsequent definition of the relaxed metric topology,
and the most important results remain true even if we take $l (x, y) = 0$.
In the last case we sometimes take $D = R^{-}$ instead of
$D = R^{I}$ making the distance domain look more like
ordinary numbers (it is important to remember, that
$\sqsubseteq_{R^{-}} = \geq$ and, hence, $0$ is the largest
element of $R^{-}$).

We also impose the symmetry axiom on the function $\rho$.
The motivation here is that we presume our ``ideal'' distance
to be symmetric, hence, we should be able to make symmetric
upper and lower estimates.

We state a definition summarizing the discourse above:

\begin{definition}\label{DefRelaxed}
A symmetric function $u: A \times A \rightarrow R^{-}$ is
called a {\em relaxed metric} when it satisfies the
triangle inequality.
A symmetric function 
$\rho: A \times A \rightarrow R^{I}$  is called
a {\em relaxed metric} when its upper part 
$u$ is a relaxed metric.
\end{definition}

\section{Relaxed Metric Topology}

An {\em open ball} with a center $x \in A$ and a real radius
$\epsilon$ is defined as $B_{x, \epsilon} =
\{y \in A \ |\ u(x, y) < \epsilon \}$.
Notice that
only upper bounds are used in this definition ---
the ball only includes those points $y$, about which we are
{\em sure} that they are not too far from $x$.

We should formulate the notion of a relaxed
metric open set more
carefully than for ordinary metrics, because it is now possible
to have a ball of a non-zero positive radius,
which does not contain its own center.

\begin{definition}
A subset $U$ of $A$ is {\em relaxed metric open} if for any
point $x \in U$, there is an $\epsilon > u (x, x)$ such that
$B_{x,\epsilon} \subseteq U$.
\end{definition}

It is easy to show that for a 
continuous relaxed metric on a dcpo 
all relaxed metric open sets are Scott open and form a topology.

\section{Partial Metrics}\label{sec:PartialMetrics}

The  distances $p$ with $p (x, x) \neq 0$ were first introduced by
Matthews~\cite{Matthews:PartialMetric, TCS::Matthews1995}.
They are known as {\em partial metrics} and obey the following
axioms:

\begin{enumerate}
\item $x=y$ iff $p (x, x) = p (x, y) = p (y, y)$.
\item $p (x, x) \leq p (x, y)$.
\item $p (x, y) = p (y, x)$.
\item $p (x, z) \leq p (x, y) + p (y, z) - p (y, y)$.
\end{enumerate}

The last axiom (due to Vickers~\cite{Vickers:manuscr}) 
implies the ordinary triangle inequality,
since the distances are non-negative. 
O'Neill found it useful to introduce
negative distances in~\cite{UWARWICK:CS-RR-293}, 
but this is avoided in the present work.

Whenever partial metrics are used to describe a partially
ordered domain, a stronger form of the first two axioms
is used:
If $x \sqsubseteq y$ then $p (x, x) = p (x, y)$,
otherwise $p (x, x) < p (x, y)$. We include 
the stronger form in
the definition of partial metrics for the purposes of this
work.

\section{Central Construction}\label{sec:Central}

Here we construct continuous relaxed metrics yielding
the Scott topology for all continuous Scott domains with countable
bases. Our construction closely resembles one
by O'Neill~\cite{UWARWICK:CS-RR-293}. We also use valuations,
but we consider continuous valuations on the
powerset of the basis instead of continuous valuations on the
domain itself. This allows us to handle a wider class of domains.

Consider a continuous Scott domain $A$ with countable basis $K$.
Enumerate
elements of $K \setminus \{\bot_{A}\}$:
 $k_{1}, ..., k_{i}, ...$. Associate weights
with all basic elements: $w(\bot_{A}) = 0$, and 
let $w(k_{i})$ form a converging sequence of strictly
positive weights. For convenience we agree
that the sum of weights of all basic elements equals 1.
For example, one might wish to consider $w(k_{i}) = 2^{-i}$
or $w(k_{i}) = \epsilon (1 + \epsilon)^{-i}, \epsilon > 0$.
Then for any $K_{0} \subseteq K$, the weight of set $K_{0}$,
$W(K_{0}) = \sum_{k \in K_{0}} w(k_{0})$, 
is well defined and
belongs to $[0, 1]$.

We have several versions of function $\rho$. For most purposes
it is enough to consider $u (x, y) = 1 - W (K_{x} \cap K_{y})$
and $l (x, y) = 0$. Sometimes it is useful to consider a better lower
bound function $l (x, y) = W (I_{x} \cup I_{y})$, where
$I_{x} = \{k \in K \ |\ k \sqcup x$ does not exist$\}$ for the
computational purposes (Section~\ref{sec:Comput}).

\begin{theorem}\label{MainTheorem} The function 
$u$ is a partial metric.
The function $\rho$ is a Scott continuous
relaxed metric. The relaxed metric
topology coincides with the Scott topology.
\end{theorem}

If, in addition, we would like the next theorem to hold,
we have to consider a different version of $\rho$ with
$u(x, y) = 
1 - W (K_{x} \cap K_{y}) - W (I_{x} \cap I_{y})$ and
$l(x, y)  = W(K_{x} \cap I_{y}) + W(K_{y} \cap I_{x})$.
The previous theorem still holds.

We introduce the notion of a {\em regular basis}.
A set of maximal elements in $A$ is denoted as $\Total (A)$ .
We say that the
basis $K$ is {\em regular} if $\forall k \in K, x \in \Total(A).\ k
\sqsubseteq x \Rightarrow k \ll x$. In particular,
if $K$ consists of compact
elements, thus making $A$ an algebraic Scott domain, $K$ is regular.

\begin{theorem}\label{OnTotals} Let $K$ be a regular basis in $A$.
Then for all $x$ and $y$ from $\Total (A)$,
$l (x, y) = u (x, y)$.
Consider $\mu : \Total (A) \times \Total (A) \rightarrow \R$,
$\mu (x, y) = l(x, y) = u(x, y)$.
Then $(\Total (A), \mu)$ is a metric space,
and its metric topology is the subspace 
topology induced by the Scott topology on $A$.
\end{theorem}

\section{Proofs of Theorems~\ref{MainTheorem} and~\ref{OnTotals}}

Here we prove relatively difficult parts of these theorems
for the case when $u(x, y) = 
1 - W (K_{x} \cap K_{y}) - W (I_{x} \cap I_{y})$ and
$l(x, y)  = W(K_{x} \cap I_{y}) + W(K_{y} \cap I_{x})$.
Lemma~\ref{LemmaOnTotals} is needed for
Theorem~\ref{OnTotals}, and other lemmas are needed
for Theorem~\ref{MainTheorem}.

\begin{lemma}[(Correctness of lower bounds)] \label{sec:Correctness}
$l(x, y) \leq u (x, y)$.
\end{lemma}
\Proof
Using $K_{x} \cap I_{x} = \emptyset$ we can rewrite $u$ and $l$.
$u (x, y) = 1 - W (K_{x} \cap K_{y}) - W (I_{x} \cap I_{y}) =
W (\overline{U})$, where $U = (K_{x} \cap K_{y}) \cup (I_{x} \cap I_{y})$.
$l (x, y) = W (V)$, where $V = (K_{x} \cap I_{y}) \cup (K_{y} \cap I_{x})$.

We want to show that $V \subseteq \overline{U}$, 
for which it is enough to show that $V \cap U = \emptyset$.
We show that $(K_{x} \cap I_{y}) \cap U = \emptyset$.
Then by symmetry the same will be true for $K_{y} \cap I_{x}$,
and hence for $V$.

$(K_{x} \cap I_{y}) \cap U = (K_{x} \cap I_{y} \cap K_{x} \cap K_{y})
\cup (K_{x} \cap I_{y} \cap I_{x} \cap I_{y})$.
But $K_{x} \cap I_{y} \cap K_{x} \cap K_{y} \subseteq I_{y} \cap  K_{y} =
\emptyset$. Similarly, $K_{x} \cap I_{y} \cap I_{x} \cap I_{y} =
\emptyset$.
\eproof

\begin{lemma}\label{LemmaOnTotals}
If $K$ is a regular basis and $x, y \in \Total(A)$,
then $l (x, y) = u (x, y)$. 
\end{lemma}
\Proof
Using the notations of the previous proof
we want to show that $\overline{U} \subseteq V$.

Let us show first that $K_{x} \cup I_{x} = K_{y} \cup I_{y} = K$.
Consider $k \in K$.
Since $x \in \Total (A)$, if $k \not\in I_{x}$,
then $k \sqsubseteq x$. Now from the regularity of K we obtain
$k \ll x$ and $k \in K_{x}$. Same for $y$.

Now, if $k \not\in U$, then $k \not\in K_{x}$ or $k\not\in K_{y}$.
Because of the symmetry it is enough to consider
$k \not\in K_{x}$. Then $k \in I_{x}$. Then, using $k \not\in U$
once again, $k \not\in I_{y}$. Then $k \in K_{y}$.
So we have $k \in K_{y} \cap I_{x} \subseteq V$.
\eproof

\begin{lemma}[(Vickers-Matthews 
triangle inequality for upper bounds)]\label{VMTriang} :

$u(x,z) \leq u(x,y) + u(y,z) - u(y,y)$.
\end{lemma}
\Proof
We want to show $1 - W (K_{x} \cap K_{z}) - W (I_{x} \cap I_{z}) \leq
1 - W (K_{x} \cap K_{y}) - W (I_{x} \cap I_{y}) + 
1 - W (K_{y} \cap K_{z}) - W (I_{y} \cap I_{z}) -
1 + W (K_{y}) + W (I_{y})$.
This is equivalent to $W (K_{x} \cap K_{y}) + W (K_{y} \cap K_{z}) +
W (I_{x} \cap I_{y}) + W (I_{y} \cap I_{z}) \leq
W (K_{y}) + W (I_{y}) + W (K_{x} \cap K_{z}) + W (I_{x} \cap I_{z})$. 

Notice that $W (K_{x} \cap K_{y}) + W (K_{y} \cap K_{z}) =
W (K_{x} \cap K_{y} \cap K_{z}) + W (K_{y} \cap (K_{x} \cup K_{z}))$,
and the similar formula holds for $I$'s.

Then the result follows from the following simple facts:
$W (K_{x} \cap K_{y} \cap K_{z}) \leq W (K_{x} \cap K_{z})$,
$W (K_{y} \cap (K_{x} \cup K_{z})) \leq W (K_{y})$, 
and the similar inequalities for $I$'s. 
\eproof

\begin{lemma} \label{sec:rho_continuous}
Function
$\rho: A \times A \rightarrow R^{I}$ is continuous.
\end{lemma}
\Proof Monotonicity of $\rho$ is trivial. 

Consider a directed set $B \subseteq A$ and some $z \in A$.
We have to show that $\rho (z, \sqcup B) = 
\sqcup_{R^{I}}  \{\rho (z,x)\ |\ x \in B \}$, which is
equivalent to 
$u (z, \sqcup B) = \inf \{u (z,x)\ |\ x \in B \}$ and
$l(z, \sqcup B) = \sup \{l (z,x)\ |\ x \in B \}$.

Rewriting this, we want to show that
$W (K_{z} \cap K_{\sqcup B}) + W (I_{z} \cap I_{\sqcup B}) =
\sup \{ W (K_{z} \cap K_{x}) + W (I_{z} \cap I_{x})\ |\ x \in B\}$ and
$W (K_{z} \cap I_{\sqcup B}) + W (I_{z} \cap K_{\sqcup B}) =
\sup \{ W (K_{z} \cap I_{x}) + W (I_{z} \cap K_{x}) |\ x \in B\}$.
Monotonicity considerations trivially yield both ``$\geq$''
inequalities, so it is enough to show ``$\leq$'' inequalities.
In fact, we will show that for any sets $C \subseteq A$ and
$D \subseteq A$, $W (C \cap K_{\sqcup B}) + W (D \cap I_{\sqcup B})
\leq \sup \{ W (C \cap K_{x}) + W (D \cap I_{x})\ |\ x \in B\}$ holds.

It is easy to show that $K_{\sqcup B} = \cup \{K_{x} \ |\ x \in B\}$
by showing first that the set $\cup \{K_{x} \ |\ x \in B\}$
is directed and $\sqcup B = \sqcup (\cup \{K_{x} \ |\ x \in B\})$.
Let us prove that $I_{\sqcup B} = \cup \{ I_{x}\ |\ x \in B \}$.
``$\supseteq$'' is trivial. Let us prove ``$\subseteq$''.
Assume that $k \not\in \cup \{ I_{x}\ |\ x \in B \}$,
i.e. $\forall x \in B.\ k \sqcup x$ exists.
It is easy to see that because $B$ is a directed set,
$\{k \sqcup x\ |\ x \in B\}$ is also directed.
Then $k \sqsubseteq \sqcup \{k \sqcup x\ |\ x \in B\} \sqsupseteq
\sqcup B$, implying existence of $k \sqcup (\sqcup B)$
and, hence, $k \not\in I_{\sqcup B}$.

Now consider enumerations of the countable or finite sets $K_{\sqcup B}$
and $I_{\sqcup B}$:
$k_{1}, ...,$ $k_{n}, ...$ and $k'_{1}, ..., k'_{n}, ...$, respectively. 
Define tail sums $S_{n} = w(k_{n}) + w(k_{n+1})$
\ $ + ...$ 
and $S'_{n} = w(k'_{n}) + w(k'_{n+1}) + ...$. 
Observe that
$(S_{n})$ and $(S'_{n})$ converge to $0$. 

Pick for every $k_{n}$ some $x_{n} \in B$ such that $k_{n} \in K_{x_{n}}$.
Pick for every $k'_{n}$ some $x'_{n} \in B$ such that $k'_{n} \in I_{x'_{n}}$.
Then using the directness of $B$, we can for any $n$ pick such
$y_{n} \in B$, that $x_{1}, ..., x_{n}, x'_{1}, ..., x'_{n}
\sqsubseteq y_{n}$. Then
$k_{1}, ..., k_{n} \in K_{y_{n}}$ and $k'_{1}, ..., k'_{n} \in I_{y_{n}}$.
It is easy to see that 
$(W(C \cap K_{\sqcup B}) + W(D \cap I_{\sqcup B})) - 
(W(C \cap K_{y_{n}}) + W(D \cap I_{y_{n}})) =
W(C \cap K_{\sqcup B}) - W(C \cap K_{y_{n}}) +
W(D \cap I_{\sqcup B}) - W(D \cap I_{y_{n}})
 < S_{n} + S'_{n}$, which
can be made as small as we like.
\eproof

\begin{lemma} 
If $B \subseteq A$ is Scott open then it is relaxed metric
open.
\end{lemma}
\Proof
Consider $x \in B$.
Because $B$ is Scott open there is a basic element $k \in B$
such that $k \ll x$. We must find $\epsilon > 0$ such that
$x \in B_{x,\epsilon } \subseteq B$. Let $\epsilon = u(x,x) +
w(k)/2$. Clearly $x \in B_{x,\epsilon }$. We claim that
$B_{x,\epsilon} \subseteq \{ y | y \gg k \} \subseteq B$. Assume, by
contradiction, that $y \gg k$ is false. Then $k \not\in K_{y}$
and $k \in K_{x}$. Then $W(K_{x} \cap K_{y}) + W(I_{x} \cap
I_{y}) + w(k) \leq W(K_{x}) + W(I_{x})$. Then $u(x,y) - w(k) =
1 - W(K_{x} \cap K_{y}) - W(I_{x} \cap I_{y}) - w(k)
\geq 1- W(K_{x}) - W(I_{x}) = u(x,x)$. Therefore $u(x,y) \geq
u(x,x) + w(k)$ and thus $y \not\in B_{x,\epsilon}$.
\eproof

\section{Computability and Continuity}\label{sec:Comput}

Consider an effective algebraic Scott domain $A$ and
its computable element $x$. We recall that this means
that
$K_{x}$ is recursively enumerable (Section~\ref{sec:Effective}).
It is easy to show that in this case $I_{x}$ must also
be recursively enumerable. However, as we discussed
in Section~\ref{sec:Effective},
one almost never should expect them to be recursive.

The actual computation of $\rho (x, y)$ goes as follows.
Start with $[0, 1]$ and go along the recursive enumerations
of $K_{x}$, $K_{y}$, $I_{x}$, and $I_{y}$. Whenever we discover
that some $k$ occurs in both $K_{x}$ and $K_{y}$,
or in both $I_{x}$ and $I_{y}$, subtract
$w(k)$ from the upper boundary. 
Whenever we discover
that some $k$ occurs in both $K_{x}$ and $I_{y}$,
or in both $I_{x}$ and $K_{y}$, add
$w(k)$ to the lower boundary. 
If this process continues
long enough, $[l (x, y), u (x, y)]$ is approximated as well as
desired.

However, there is no general way
to compute a better lower estimate for $u(x, y)$
than $l(x, y)$, or to compute a better upper
estimate for $l(x, y)$ than $u(x, y)$.
Consequently, there is no general way to determine
how close is the convergence process to the actual
values of $l(x, y)$ and $u (x, y)$, except that we know
that $u (x, y)$ is not less than 
the currently computed lower bound, and $l(x, y)$
is not greater than the currently computed
upper bound.
Of course, for large $x$ and $y$ this knowledge might provide
a lot of information, and
if the basis of our domain is regular, for total
elements $x$ and $y$ this
knowledge provides us with precise estimates 
--- i.e. if the basis is regular,
then the resulting classical metric on $\Total(A)$ can
be nicely computed.

The computational situation is very different with regard
to quasi-metrics.
Consider $u (x, y) = 1 - W (K_{x} \cap K_{y})$ and
$d (x, y) = u (x, y) - u(x, x) = W(K_{x} \setminus K_{y})$.
This is a quasi-metric in the style 
of~\cite{SmythMB:quaurd, KopFla93}, and it yields a Scott
topology~\cite{BukSco:dist_new,BukSco:dist_old}. However, as discussed
in~\cite{BukSco:dist_new,BukSco:dist_old}, 
typically $K_{x} \setminus K_{y}$
is not recursive. Moreover, one should not expect 
$K_{x} \setminus K_{y}$ or
its complement to be recursively enumerable. This precludes us
from building a generally applicable method computing
$d (x, y)$ and illustrates that it is
computationally incorrect to subtract one upper bound
from another.

\section{Axioms of Partial Metrics and Functoriality}

\subsection{Should the Axioms of Partial Metrics Hold?}\label{sec:PMIssues}

Consider relaxed metric $\rho$ and its upper part $u$.
Should we expect function $u$ to satisfy the axioms
of partial metrics?
Example~\ref{IntervalExample} describes a natural relaxed
metric on interval numbers, where $u$ is not a partial
metric. Function $u$ gives a better upper estimate
of the ``ideal'' distance between interval numbers, than
the partial metric $p ([a,b], [c,d]) = \max (b, d) -
\min (a, c)$ described in~\cite{Matthews:PartialMetric}.
For example, $u ([0,2], [1,1]) = 1$, which is what one
expects --- if we know that one of the numbers is
somewhere between $0$ and $2$, and another number equals $1$,
then we know that the distance between them is no greater
than $1$. However, $p ([0,2], [1,1]) = p([0,2], [0,2]) = 2$.

Now we describe two situations when the axioms of
partial metrics are justified.
Consider again function $d (x, y) = u (x, y) - u(x, x)$
from Section~\ref{sec:Comput}. Vickers notes in~\cite{Vickers:manuscr}
that the triangle inequality $d (x, z) \leq d (x, y) +
d (y, z)$ is equivalent to $u (x, z) \leq u (x, y) + u (y, z) -
u (y, y)$, and $d (x, y) \geq 0$ is equivalent to $u (x, x) \leq
u (x, y)$. This means that function $u$ is a partial metric
if and only if function $d$ is a quasi-metric.

Another justification comes from the consideration of
the proof of Lemma~\ref{VMTriang}. Whenever the upper
part $u (x, y)$ of a relaxed metric is based on
{\em common information} shared by $x$ and $y$ yielding a
{\em negative contribution} to the distance (we
subtract the weight of common information from the
universal distance $1$ in this chapter), both
$u (x, z) \leq u (x, y) + u (y, z) - u (y, y)$
and $u (x, x) \leq u (x, y)$ should hold.
As we shall see in full generality in the next chapter,
when $u(x,y)=1-\mu($Common
information between $x$ and $y)$,
axioms of partial metrics hold.

However, to specify function $u$ in Example~\ref{IntervalExample}, 
we use information
about $x$ and $y$,
which cannot be thought of as common information shared
by $x$ and $y$. In such case we
still expect a relaxed metric $\rho$, but its upper part
$u$ does not have to satisfy the axioms of partial metrics.

\subsection{Partial Metrics and Functoriality} \label{pm_and_func}

Here I would like to emphasize the utilitarian value of these
axioms, i.e. certain useful properties which are easy to establish in
the presence of these axioms. In all cases these considerations are closely
related to the following series of {\it open problems}: What are the
necessary and/or sufficient conditions for relaxed/partial metrics
on various classes of domains in order for all Scott open sets to be
relaxed metric open?

\subsubsection{Open Balls Should be Open Sets}

The strong Vickers--Matthews triangle inequality,
$u(x,z)\le u(x,y)+u(y,z)-u(y,y)$, is helpful in the proof
that an open ball is a relaxed open set.

The proof goes as follows. Consider $y \in B_{x,\epsilon}$.
We need to find $\delta>0$, such that
$B_{y,\delta+u(y,y)}\subseteq B_{x,\epsilon}$, i.e.
$\forall z.\; u(y,z)\le u(y,y)+\delta \Rightarrow u(x,z)<\epsilon$.
Using $u(x,z)\le u(x,y)+u(y,z)-u(y,y)$ and
$\epsilon-u(x,y)>0$ we take $\delta=(\epsilon-u(x,y))/2$ and
obtain the desired result. This is closely related to
the fact noted by  Vickers, that for $d(x,y)=u(x,y)-u(x,x)$,
$d(x,z)\le d(x,y)+d(y,z)\Leftrightarrow u(x,z)\le u(x,y)
+u(y,z)-u(y,y)$. The ordinary triangle inequality $u(x,z)\le
u(x,y)+u(y,z)$ is not enough here, because we are
looking for the ball around $y$ with the radius $u(y,y)+\delta$
and not just $\delta$.

\subsubsection{Functoriality} \label{func}

One of the open issues in the field remains to develop
a {\em functorial} approach to relaxed metrics.
This approach should facilitate our abilities to both define
canonical distances on domains and to compute these distances at a reasonable
cost.

This method is to define reasonable relaxed metrics on some specific
elementary domains, and then, given reasonable structures
(relaxed metrics or, sometimes, measures) on domains
$A$, $B$, define relaxed metric on domains $A\times B$, $A\to B$,
etc. The natural candidates for relaxed metric on $A\times B$ and $A \to B$
are
$$ \rho_{A\times B}(\langle x,y \rangle,\langle x',y' \rangle)=
\alpha \cdot \rho_A(x,x')+
\beta\cdot\rho_B(y,y'),\; \alpha,\,\beta>0,\;\alpha+\beta=1$$
$$\rho_{A \to B}(f,g)=\int_{x\in A} \rho_B(f(x),g(x))d\mu_x.$$
We also hope to be able to extend this approach to reflexive domains and to
obtain some invariance properties of $\rho$ in the process.

In all cases we would like to ensure that relaxed metric topology
and Scott topology coincide. In particular, in order for Scott open sets
to be relaxed metric open it is {\it necessary} (not sufficient)
that $u(x,x)\ge u(x,y) \Rightarrow x \sql y $.

However, for general relaxed metrics it is easy to come up with examples,
when for both $A$ and $B$ $u_A(x,x)\ge u_B(x,y) \Rightarrow x\sql y$
and $u_B(x',x')\ge u_B(x',y') \Rightarrow x' \sql y'$, but
$u_{A\times B}(\langle x,x' \rangle, \langle x,x' \rangle)
\ge u_{A \times B}(\langle x,x' \rangle,\langle y,y' \rangle)
\not \Rightarrow (x\sql y \& x'\sql y')$, and the similar situation
takes place for $A \to B$. However, the strong form of the axiom
of small self-distances,
$u(x,x) \le u(x,y)$
(namely, $ x \sql y \Rightarrow u(x,x)=u(x,y)$,
$ x \not \sql y \Rightarrow u(x,x)<u(x,y) $)
seems to eliminate these problems.

\section{Applications and Open Problems}

Let us briefly state where we stand
with regard to the applications to programs. 
We are able to
introduce relaxed metrics on a class
of domains sufficiently large 
for practical applications in the spirit
of~\cite{Stoy}.

For example, consider $X = [ \! [ {\tt while\ B\ do\ S} ] \! ]$,
and the sequence of programs, 
$P_{1} = {\tt loop}; ...;
P_{N} = {\tt if\ B\ then\ S;} P_{N-1} {\tt else\ skip\ endif}; ...$.
Define $X_{N} = [ \! [ P_{N} ] \! ]$. 
Typically $X_{N-1} \sqsubseteq  X_{N}$ and $X = \sqcup X_{N}$ hold. 
We agreed that the distances between programs will be distances
between their meanings. Assume that $M \leq N$.
Then regardless of specific weights,
$u (P_{M}, P_{M}) = u (P_{M}, P_{N})
= u (P_{M}, P)$, 
also $u (P_{M}, P_{M}) \geq u (P_{N}, P_{N}) \geq u (P, P)$,
and $u (P, P) = \inf u (P_{N}, P_{N})$. 
Of course, none of these distances has to be zero.

However, we do not know yet how to build relaxed distances so
that not only nice convergence properties are true, but also
that distances between particular pairs of programs
``look right'' --- a notion, which is more difficult to formalize,
than convergence. 
Also, we compute these distances via recursive
enumeration now, and a more efficient scheme is needed.  

Hence we think that the key open problems for relaxed distances
are the same as for measures and integrals, namely
interrelated problems of
how to compute them quickly and how to define
canonical
valuations and relaxed distances on functional spaces and reflexive domains.

In Subsection~\ref{pm_and_func} given a measure on
$A$ and a relaxed metric on $B$, we defined a functorial
relaxed metric on $[A \rightarrow B]$. Hence the problem
of functorial definitions for distances and measures are
interrelated, and in order to approach higher-order
functional spaces and reflexive domains for relaxed metrics,
we should start with finding an appropriate construction
for measures on $[A \rightarrow B]$.

It should be noted here that since we are going to
distinguish functions topologically by integrating them
over $A$, we need these measures to be based on strongly
non-degenerate valuations.

\chapter{Obtaining Generalized Distances from Valuations} \label{chap:muInfo}

In this chapter we continue to present our joint results with
Svetlana Shorina~\cite{BukatinShorina,BukatinShorina2}.

In the previous chapter we introduced a construction
of partial metrics based on
the mechanism of {\it common information}
between elements $x$ and $y$ bringing negative contribution
to $u(x,y)$. This construction was based on assigning
finite weights to basic elements and gave topologically
meaningful relaxed metrics for all continuous Scott
domains with countable bases. For such domains with regular
bases (which included all algebraic Scott domains) this
construction gave relaxed metrics which ``behaved well"
on the total elements of the domain.

Here we introduce a general method of defining
partial and relaxed metrics via information about
elements for all dcpo's via mechanism of
$\muInfo$-structures.

Then we show how to obtain such a $\muInfo$-structure
from a CC-valuation for any continuous Scott domain.
Since we know, how to construct a CC-valuation for any
continuous dcpo with countable basis, this method of
constructing partial and relaxed metrics via 
CC-valuations and the resulting $\muInfo$-structures
works for all continuous Scott domains with countable bases.

\section{Partial and Relaxed Metrics via Information}
\label{pm_via_ccval}

\subsection{$\muInfo$-structures}

Assume that there is a set ${\cal I}$ representing information
about elements of a dcpo $A$.
We choose a ring,
${\cal M}({\cal I})$, of admissible subsets of ${\cal I}$ and introduce
a measure-like  structure, $\mu$,
on ${\cal M}({\cal I})$.
We associate a set, $\Info(x) \in {\cal M}({\cal I})$,
with every $x \in A$, and call $\Info(x)$ a set of (positive)
information about $x$. We also would like to consider negative
information about $x$, 
$\Neginfo(x) \in {\cal M}({\cal I})$, --- intuitively speaking,
this is the information which cannot become
true about $x$, when $x$ is arbitrarily increased. 

\Def Given a dcpo $A$,
the tuple of $(A, {\cal I}, {\cal M}({\cal I}), \mu,
\Info, \Neginfo)$ is called a {\em $\muInfo$-structure} on $A$,
if ${\cal M}({\cal I}) \subseteq {\cal P}({\cal I})$ ---
a ring of subsets closed
with respect to $\cap,\,\cup, \setminus$ 
and including $\emptyset$ and ${\cal I}$, 
$\mu: {\cal M}({\cal I}) \to
[0,\,1]$, $\Info: A \to {\cal M}({\cal I})$,
and $\Neginfo: A \to {\cal M}({\cal I})$, and the following
axioms are satisfied:
\begin{enumerate}
\item {\bf (VALUATION AXIOMS)}
 
  \begin{enumerate}
  \item $ \mu({\cal I})=1,\; \mu(\emptyset)=0$;
  \item $ U \subseteq V \Rightarrow \mu(U)\le \mu(V)$;
  \item $ \mu(U)+\mu(V)=\mu(U\cap V)+\mu(U\cup V)$;
  \end{enumerate}
\item {\bf ($\Info$ AXIOMS)}

  \begin{enumerate}
  \item $x \sql y\Leftrightarrow \Info(x)\subseteq \Info(y)$;
  \item $x \sqsubset y\Rightarrow \Info(x) \subset \Info(y)$;
  \end{enumerate}
\item {\bf ($\Neginfo$ AXIOMS)}

  \begin{enumerate}
  \item $\Info(x) \cap \Neginfo(x) = \emptyset$;
  \item $x \sql y\Rightarrow \Neginfo(x) \subseteq \Neginfo(y)$;
  \end{enumerate}
\item {\bf (STRONG RESPECT FOR TOTALITY)} 

  $x \in \Total(A)\Rightarrow \Info(x) \cup \Neginfo(x)={\cal I}$;
\item {\bf (SCOTT CONTINUITY OF THE INDUCED RELAXED METRIC)}

  if $B$ is a directed subset of $A$ and $y\in A$, then
  \begin{enumerate}
  \item $\mu(\Info(\sqcup B) \cap \Info(y)) = sup_{x\in B} 
    (\mu (\Info(x) \cap \Info(y))$,
  \item $\mu(\Info(\sqcup B) \cap \Neginfo(y)) = sup_{x\in B} 
    (\mu (\Info(x) \cap \Neginfo(y))$,
  \item $\mu(\Neginfo(\sqcup B) \cap \Info(y)) = sup_{x\in B}
    (\mu (\Neginfo(x) \cap \Info(y))$,
  \item $\mu(\Neginfo(\sqcup B) \cap \Neginfo(y)) = sup_{x\in B}
    (\mu (\Neginfo(x) \cap \Neginfo(y))$;
  \end{enumerate}
\item {\bf (SCOTT OPEN SETS ARE RELAXED METRIC OPEN)}

  for any (basic) Scott open set $U \subseteq A$ and $x\in U$,
  there is an $\epsilon > 0$, such that $\forall y\in A.\ \mu
  (\Info(x)) - \mu (\Info(x) \cap \Info(y)) < \epsilon \Rightarrow
  y \in U$.
\end{enumerate}

We will also consider {\em deficient} $\muInfo$-structures, when
the {\bf strong respect for totality} axiom is not imposed.

In terms of lattice theory, $\mu$ is a (normalized) valuation
on a lattice ${\cal M}({\cal I})$.
The consideration of unbounded measures  is beyond the scope of this
work, and $\mu({\cal I})=1$ is assumed for convenience.
  Axioms relating $\sql$ and $\Info$ are 
in the spirit of
information systems approach~\cite{SCOTT82}, although we are not
considering any inference structure over ${\cal I}$ in this chapter.

The requirements for negative information are relatively weak, because
it is quite natural to have $\forall x \in A.\; \Neginfo(x)=
\emptyset $ if $A$ has the top element. 

The axiom that for $x \in \Total(A)$, $\Info(x) \cup \Neginfo(x)=
{\cal I}$, is desirable
because indeed, if some $i \in \Info(x)$ does not belong to
$\Info(x)$ and $x$ can not be further increased, then by our intuition
behind $\Neginfo(x)$, $i$ should belong to $\Neginfo(x)$.
However, this axiom might be too strong and will be further discussed
later.

The last two axioms are not quite
satisfactory --- they almost immediately
imply the properties, after which they are named, but they are
complicated and might be difficult to establish.
We hope, that these axioms will be replaced by something
more tractable in the future. One of the obstacles seems to be the
fact
in some valuable approaches
(in particular, in this chapter) it is not correct that $x_1 \sql x_2 \sql
\cdots$ implies that $\Info(\sqcup_{i \in \N}x_i)=\bigcup_{i \in \N}
\Info(x_i)$.

The nature of these set-theoretical representations,
${\cal I}$, of domains may vary:
one can consider sets of tokens of
information systems, powersets of domains bases,
or powersets of domains themselves, custom-made sets for
specific domains,
etc. The approach via powersets of domain bases of the
previous chapter
can be thought of as a partial case of
the approach via powersets of domains themselves adopted in this
chapter.

\subsection{Partial and Relaxed Metrics via $\muInfo$-structures}

Define the (upper estimate of the) distance between
$x$ and $y$ from $A$ as $u:\,A\times A \to \R^-$:
$$ u(x,y)=1-\mu(\Info(x) \cap \Info(y))-\mu(\Neginfo(x) \cap \Neginfo(y)).$$
I.e. the more information $x$ and $y$ have in common the smaller is the
distance between them. However a partially defined element might not have too
much information at all, so its self-distance $u(x,x)=1-
\mu(\Info(x))-\mu(\Neginfo(x))$ might be large. 

It is possible to find information
which will never made it into $\Info(x) \cap \Info(y)$ or
$\Neginfo(x) \cap \Neginfo(y)$ even when $x$ and $y$ are arbitrarily increased.
In particular, $\Info(x) \cap \Neginfo(y)$ and
$\Info(y) \cap \Neginfo(x) $ represent such information. Then we can introduce
the lower estimate of the distance
$l:\,A\times A \to \R^+$:
 $$l(x,y)=\mu(\Info(x)\cap \Neginfo(y))+
\mu(\Info(y)\cap \Neginfo(x)).$$

The proof of Lemma~\ref{sec:Correctness} is
directly applicable and yields
$(\Info(x)\cap \Neginfo(y))\cup (\Info(y)\cap \Neginfo(x)) \subseteq
{\cal I} \setminus ((\Info(x) \cap \Info(y))\cup(\Neginfo(x)\cap \Neginfo(y)))$
implying $l(x,y)\le u(x,y)$. Thus we can form an
{\bf induced relaxed metric},
$\rho:\, A\times A\to R^I$, $\rho=\langle l,u \rangle$, with 
a meaningful lower bound.

Now certain properties of partial and relaxed metrics can be established.
\begin{enumerate}
\item $u(x,x) \le u(x,y)$.\\
Indeed, axiom $U\subseteq V\Rightarrow \mu(U) \le \mu(V)$ implies
$\mu(\Info(x)\cap \Info(y)) \le \mu(\Info(x))$,
$\mu(\Neginfo(x)\cap \Neginfo(y)) \le \mu(\Neginfo(x))$, yielding
$u(x,x) \le u(x,y)$.
\item $u(x,z)\le u(x,y)+u(y,z)-u(y,y)$.\\
The proof of Lemma~\ref{VMTriang} goes through.
\begin{flushleft}
The symmetry $u(x,y)=u(y,x)$ and $l(x,y)=l(y,x)$ 
is obvious. Another helpful fact is
\end{flushleft}
\item $x \sql y \Rightarrow u(x,x)=u(x,y)$. \\ We are using
the axiom $x \sql y \Rightarrow \Info(x) \subseteq \Info(y)$,
which implies $\Info(x)\cap \Info(y) = \Info(x)$, and the same for $\Neginfo$.

\item $x\not \sql y \Rightarrow u(x,x)<u(x,y).$

This follows from the following fact:

\item Any Scott open set is relaxed metric open.

This, in turn, follows at once from the last axiom
of $\muInfo$-structure and
$\mu(\Neginfo(x)\cap \Neginfo(y)) \le \mu(\Neginfo(x))$.

\item Any relaxed metric open set is Scott open.

This follows at once from the following property:

\item The induced relaxed metric is a Scott continuous function.

This also follows immediately from the corresponding
axiom of $\muInfo$-structure.

\end{enumerate}

Thus, the following theorem is proved (cf. Theorem~\ref{MainTheorem}). 
Observe, that the
{\bf strong respect for totality} axiom is NOT used in the proof.

\begin{theorem}\label{MainTheorem_a} Let $A$ be
a dcpo with a (possibly deficient) $\muInfo$-strcuture
and the induced relaxed metric $\rho = \langle l, u \rangle $.
Then the function
$u$ is a partial metric,
the function $\rho$ is a Scott continuous
relaxed metric, and the relaxed metric
topology coincides with the Scott topology.
\end{theorem}

Now let us assume that our $\muInfo$-structure is not
deficient.

Due to the axiom
$\forall x \in \Total(A).\ \Info(x) \cup \Neginfo(x)=
{\cal I}$,
the proof of Lemma~\ref{LemmaOnTotals} would go through, yielding
$$ x,y \in \Total(A) \Rightarrow l(x,y)=u(x,y)$$
and allowing to obtain the 
following theorem (cf. Theorem~\ref{OnTotals}).

\begin{theorem}\label{OnTotals_a}
Let $A$ be
a dcpo with a $\muInfo$-strcuture
and the induced relaxed metric $\rho = \langle l, u \rangle $.
Then for all $x$ and $y$ from $\Total (A)$,
$l (x, y) = u (x, y)$.
Consider $\mu : \Total (A) \times \Total (A) \rightarrow \R$,
$\mu (x, y) = l(x, y) = u(x, y)$.
Then $(\Total (A), \mu)$ is a metric space,
and its metric topology is the subspace
topology induced by the Scott topology on $A$.
\end{theorem}

However, in the previous chapter $x \in \Total(A) \Rightarrow
\Info(x) \cup \Neginfo(x)={\cal I}$ holds under an awkward condition of the
regularity of the basis. While bases of algebraic Scott domains and
of continuous lattices can be made regular, there are important continuous
Scott domains, which cannot be given regular bases. In particular, in
$\R^I$ no element, except for $\bot$, satisfies the condition of
regularity, hence a
regular basis cannot be provided for $\R^I$.

The achievement of the construction to be described
in the Section~\ref{new_part_metric}
is that by
removing the reliance on the weights of non-compact basic
elements, it eliminates the regularity requirement and implies
$x \in \Total(A) \Rightarrow \Info(x) \cup \Neginfo(x) ={\cal I}$ for
all continuous Scott domains equipped with a CC-valuation
(which is built above for all continuous Scott domains with
countable bases) where $\Info(x)$ and $\Neginfo(x)$ are as described below in
the Subsection~\ref{new_part_metric}.

However, it might still be fruitful to consider replacing
the axiom $\forall x \in \Total(A).$ 
$\Info(x) \cup \Neginfo(x)=
{\cal I}$ by something like
$\forall x \in \Total(A).\ \mu({\cal I} \setminus
(\Info(x) \cup \Neginfo(x)))=0$.

\subsection{A Previously Known Construction}
Here we recall a construction from the previous chapter based
on a generally non-co-continuous valuation of
Subsection~\ref{nonalg_noncc}.  We will reformulate it in our terms of
$\muInfo$-structures. 
In the previous chapter it was natural to think that ${\cal I}=K$. Here
we reformulate that construction in terms of ${\cal I}=A$, thus abandoning the
condition $x \in \Total(A) \Rightarrow \Info(x)\cup \Neginfo(x)={\cal I}$
 altogether.

Define $I_x=\{y \in A\,|\, \{x,y\} {\rm \; is \; unbounded}\}$,
$P_x=\{y \in A\,|\, y\ll x\}$
(cf. $I_x=\{k \in K\,|\, \{k,x\}\, \mbox{is unbounded}\}$,
 $K_x=\{k \in K\,|\, k \ll x\}$ in~\cite{BukatinScott}).

Define $\Info(x)=P_x$, $\Neginfo(x)=I_x$. Consider a valuation $\mu$
of Subsection~\ref{nonalg_noncc}: for any $S\subset {\cal I}=A$,
$\mu(S)=\sum_{k \in S\cap K}w(k)$. $\mu$ is a continuous strongly
non-degenerate valuation, but it is not co-continuous unless
$K$ consists only of compact elements.

Because of this we cannot replace an inconvenient definition of
$\Info(x)=P_x$ by $\Info(x)=C_x=\{y \in A \,|\, y \sql x\}$
( which would restore the condition $x \in \Total(A) \Rightarrow
\Info(x) \cup \Neginfo(x)=A$) as $\mu(C_k)$ would not be equal to
$\sup_{k' \ll k} \mu(C_{k'})$ if $k$ is a non-compact basic element, leading
to the non-continuity
of the partial metric $u(x,y)$.

Also the reliance on countable system of finite
weights excludes such natural partial metrics as
 metric $u:\,\R^-_{[0,1]} \times \R^-_{[0,1]} \to \R^-$, where
$\R^-_{[0,1]}$  is the set $[0,\,1]$ equipped with
the dual partial order $\sql\,=\,\ge$, and $ u(x,y)=max(x,y)$.
We rectify all these problems in the next Subsection.

\subsection{Partial and Relaxed Metrics via CC-valuations
for Continuous Scott Domains}
\label{new_part_metric}
Assume that there is a CC-valuation $\mu(U)$ on Scott open sets of
a continuous Scott
domain $A$. Then it uniquely extends to an additive measure $\mu$
 on the ring of sets generated by the system of open sets.
 Define ${\cal I}=A$, $\Info(x)=C_x$, $\Neginfo(x)=I_x$.
It is easy to see that valuation, $\Info$, and $\Neginfo$ axioms
of $\muInfo$-structure hold.
We have $x \in \Total(A) \Rightarrow C_x \cup I_x =A$.
Thus we only need to establish $\muInfo$-structure axioms
of {\bf Scott continuity of the induced relaxed metric} and
{\bf Scott open sets are relaxed metric open} in order to
prove theorems~\ref{MainTheorem_a} and~\ref{OnTotals_a} for
our induced relaxed metric,
$u(x,y)=1-\mu(
C_x \cap C_y)-\mu(I_x\cap I_y)$, $l(x,y)=\mu(C_x \cap I_y)+
\mu(C_y \cap I_x)$.
These axioms are established by the Lemmas below.

You will  also see that for such bare-bones partial metrics, as $u(x,y)=
1-\mu(C_x \cap C_y)$, which are
nevertheless quite sufficient for topological purposes and for
domains with $\top$, only {\it co-continuity} of valuations
matters, continuity is not
important.

Observe also that since the construction in Section~\ref{nonalg_noncc}
does form a CC-valuation for algebraic Scott domains with bases of
compact elements, the construction in the previous chapter
can be considered as a partial case of our current construction if
the basis does not contain non-compact elements.

\begin{Lemma} Assume that $\mu$ is a co-continuous valuation and $B$
is a directed subset of  $A$. Then $\mu(C_{\sqcup B}\cap Q)=
\sup_{x \in B}(\mu(C_x \cap Q))$, where $Q$ is a closed
or open subset of $A$.
\end{Lemma}
\Rem Note that continuity of $\mu$ is not required here.

\Proof
 Part A: Let $Q$ be a closed subset of $A$. Then $(C_x \cap Q,\; x\in B)$
form a directed set of closed sets. We need to show that
$\overline {(\bigcup_{x \in B} C_x)\cap Q}=C_{\sqcup B} \cap Q$, then
the result will follow by co-continuity. ``$\subseteq$" is trivial.
Let us proof ``$\supseteq$". Consider $x\in C_{\sqcup B} \cap Q$.
$ x= \sqcup P_x$; $P_x=\{y\in A\, |\, y \ll x\}$ is directed set
and $P_x \subseteq C_{\sqcup B} \cap Q$. However, for $\forall y \in
P_x$, since $y \ll x$ and $x\sql \sqcup B$,
 $\exists z \in B.\; y \sql z$, hence $ y\in
C_z \cap Q$. Hence $P_x \subseteq \bigcup_{z \in B} (C_z \cap Q)=
(\bigcup_{z \in B}C_z) \cap Q$, and $x\in \overline{P_x}$ implies
$ x \in \overline{(\bigcup_{z \in B} C_z)\cap Q}$.

Part B: If $Q$ is an open set, then
$\mu(C_x \cap Q)=\mu(C_x) -\mu(C_x \setminus Q)$;
$ \mu(C_{\sqcup B} \cap Q) = \mu (C_{\sqcup B})-\mu(C_{\sqcup B}
\setminus Q) $.
The non-trivial part is to show $\mu(C_{\sqcup B} \cap Q) \le
 \sup_{x \in B} \mu(C_x \cap Q)$. This follows from $\mu(C_{\sqcup B})
 =\sup_{x \in B} \mu(C_x)$ by Part A and $\mu(C_{\sqcup B}\setminus Q)
 \ge \sup_{x \in B}\mu(C_x \setminus Q)$ by monotonicity of $\mu$.
\eproof

\begin{Lemma}
Assume that $\mu$ is a continuous valuation and $B$ is a directed subset of
$A$. Then $\mu(I_{\sqcup B} \cap Q)=\sup_{x \in B}(\mu(I_x \cap Q))$,
were $Q$ is an open or closed subset of $A$.
\end{Lemma}
\Rem Co-continuity is not needed here.

\Proof During the proof of Lemma~\ref{sec:rho_continuous}
it was established that $I_{\sqcup B}=\bigcup_{x \in B} I_x$
(think about $k \in A$ and observe that that proof goes through).

Part A: If Q is an open set, $I_{\sqcup B} \cap Q =(\bigcup_{x \in B} I_x)
\cap Q =\bigcup_{x \in B} (I_x \cap Q)$ and
$\mu(I_{\sqcup B} \cap Q)=\sup_{x\in B}( \mu(I_x \cap Q)) $ is a direct
consequence of continuity of $\mu$.

Part B:
If $Q$ is closed, let us think about $\mu(I \cap Q)$ as  $\mu(I)-
\mu(I \setminus Q)$, where both $I$ and $I \setminus Q$ are open sets.
The rest is similar to Part B of the proof of the previous Lemma.
\eproof

\begin{Lemma}    \label{Lemma63}
Assume that $\mu$ is a strongly non-degenerate valuation. Then 
the $\muInfo$-structure axiom {\bf Scott open
sets are relaxed metric open} holds.
\end{Lemma}
\Rem Neither continuity, nor co-continuity are required, and even strong
non-degeneracy condition can be probably be made weaker
(see the next section).

\Proof
Consider a Scott open set $U$ and $ x \in U$. We would like
to find $\delta>0$, such that
$\mu(C_x) - \mu(C_x \cap C_y) < \delta \Rightarrow y \in U$
(and hence,
$B_{x,\delta+u(x,x)} \subseteq U$).
Using $x=\sqcup P_x$, where $P_x=\{y\,|\, y \ll x\}$ is a directed set,
obtain $\exists y \ll x. \; y\in U$ (this includes the case of
compact $x$, $y=x$). Consider a Scott open set $\{z \in A\,|\, y \ll z \}
\ni x$ and a Scott open set $A \setminus C_x$.
Then $x \not\in (A \setminus C_x)$, hence
$(A \setminus C_x) \cup \{z \in A\, |\, y \ll z\} \supset (A \setminus C_x)$
and by strong non-degeneracy $\mu((A \setminus C_x) \cup \{z \in A\,|\, y \ll
z\})-\mu(A \setminus C_x)=\epsilon>0$. We take $\delta =\epsilon$ and prove
that
$\mu(C_x) - \mu(C_x \cap C_z) < \delta$
implies $y\sql z$, and thus
$ z \in U$. Indeed, assume $y \not \sql z$.
Then we have that $y\not \in C_z$, thus $\{z\,|\, y \ll
z\}\subseteq
(A \setminus C_z)$, thus $\mu((A \setminus C_x) \cup (A \setminus C_z))
\ge \mu(A \setminus C_x)+\epsilon$, implying
$\mu(C_x) - \mu(C_x \cap C_z) \geq \delta$.
\eproof

It is helpful to visualize the situation of this section 
via the following pictures.
They show sets $C_x$ and $I_x$, show
the ``area" accounting for $l(x,y)$ as shaded with large dots,
the ``area" accounting for $u(x,y)$ as shaded with all dots (large and
small), and show why $l(x,y) = u(x,y)$, when $x, y \in \Total(A)$.

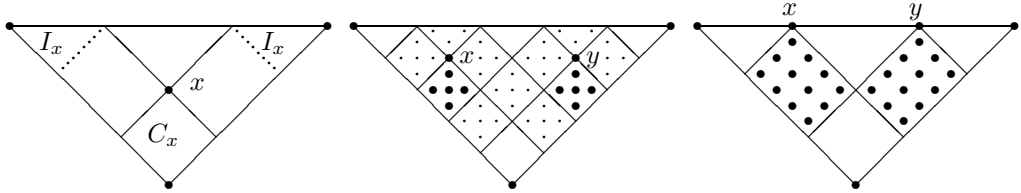
\begin{picture}(346,80)
\put(25,4){\line(1,1){60}}
\put(25,4){\line(-1,1){60}}
\put(-35,64){\line(1,0){120}}
\put(-35,64){\circle*{3}}
\put(85,64){\circle*{3}}
\put(25,4){\circle*{3}}
\put(25,40){\line(1,-1){18}}
\put(25,40){\line(-1,-1){18}}
\put(25,40){\line(1,1){24}}
\put(25,40){\line(-1,1){24}}
\put(-1,62){\circle*{1}}
\put(-3,60){\circle*{1}}
\put(-5,58){\circle*{1}}
\put(-7,56){\circle*{1}}
\put(-9,54){\circle*{1}}
\put(-11,52){\circle*{1}}
\put(-13,50){\circle*{1}}
\put(-15,48){\circle*{1}}
\put(51,62){\circle*{1}}
\put(53,60){\circle*{1}}
\put(55,58){\circle*{1}}
\put(57,56){\circle*{1}}
\put(59,54){\circle*{1}}
\put(61,52){\circle*{1}}
\put(63,50){\circle*{1}}
\put(65,48){\circle*{1}}
\put(25,40){\circle*{3}}
\put(33,40){$x$}
\put(17,20){$C_{x}$}
\put(-24,55){$I_{x}$}
\put(60,55){$I_{x}$}

\put(155,4){\line(1,1){60}}
\put(155,4){\line(-1,1){60}}
\put(95,64){\line(1,0){120}}
\put(95,64){\circle*{3}}
\put(215,64){\circle*{3}}
\put(155,4){\circle*{3}}
\put(119,64){\line(1,-1){48}}
\put(119,64){\line(-1,-1){12}}
\put(143,64){\line(1,-1){36}}
\put(143,64){\line(-1,-1){24}}
\put(167,64){\line(1,-1){24}}
\put(167,64){\line(-1,-1){36}}
\put(191,64){\line(1,-1){12}}
\put(191,64){\line(-1,-1){48}}

\put(131,52){\circle*{3}}
\put(135,50){$x$}
\put(179,52){\circle*{3}}
\put(183,50){$y$}

\put(131,46){\circle*{3}}
\put(131,40){\circle*{3}}
\put(131,34){\circle*{3}}
\put(125,40){\circle*{3}}
\put(137,40){\circle*{3}}

\put(179,46){\circle*{3}}
\put(179,40){\circle*{3}}
\put(179,34){\circle*{3}}
\put(173,40){\circle*{3}}
\put(185,40){\circle*{3}}

\put(155,46){\circle*{1}}
\put(155,40){\circle*{1}}
\put(155,34){\circle*{1}}
\put(149,40){\circle*{1}}
\put(161,40){\circle*{1}}

\put(143,34){\circle*{1}}
\put(143,28){\circle*{1}}
\put(143,22){\circle*{1}}
\put(137,28){\circle*{1}}
\put(149,28){\circle*{1}}

\put(167,34){\circle*{1}}
\put(167,28){\circle*{1}}
\put(167,22){\circle*{1}}
\put(161,28){\circle*{1}}
\put(173,28){\circle*{1}}

\put(119,58){\circle*{1}}
\put(119,52){\circle*{1}}
\put(119,46){\circle*{1}}
\put(113,52){\circle*{1}}
\put(125,52){\circle*{1}}

\put(143,58){\circle*{1}}
\put(143,52){\circle*{1}}
\put(143,46){\circle*{1}}
\put(149,52){\circle*{1}}

\put(167,58){\circle*{1}}
\put(167,52){\circle*{1}}
\put(167,46){\circle*{1}}
\put(161,52){\circle*{1}}
\put(173,52){\circle*{1}}

\put(191,58){\circle*{1}}
\put(191,52){\circle*{1}}
\put(191,46){\circle*{1}}
\put(197,52){\circle*{1}}

\put(131,58){\circle*{1}}
\put(125,62){\circle*{1}}
\put(137,62){\circle*{1}}
\put(179,58){\circle*{1}}
\put(173,62){\circle*{1}}
\put(185,62){\circle*{1}}

\put(285,4){\line(1,1){60}}
\put(285,4){\line(-1,1){60}}
\put(225,64){\line(1,0){120}}
\put(225,64){\circle*{3}}
\put(345,64){\circle*{3}}
\put(285,4){\circle*{3}}

\put(261,64){\line(1,-1){42}}
\put(261,64){\line(-1,-1){18}}
\put(309,64){\line(1,-1){18}}
\put(309,64){\line(-1,-1){42}}

\put(261,64){\circle*{3}}
\put(257,68){$x$}
\put(309,64){\circle*{3}}
\put(305,68){$y$}

\put(249,46){\circle*{3}}
\put(255,40){\circle*{3}}
\put(255,52){\circle*{3}}
\put(261,34){\circle*{3}}
\put(261,46){\circle*{3}}
\put(261,58){\circle*{3}}
\put(267,28){\circle*{3}}
\put(267,40){\circle*{3}}
\put(267,52){\circle*{3}}
\put(273,34){\circle*{3}}
\put(273,46){\circle*{3}}
\put(279,40){\circle*{3}}

\put(321,46){\circle*{3}}
\put(315,40){\circle*{3}}
\put(315,52){\circle*{3}}
\put(309,34){\circle*{3}}
\put(309,46){\circle*{3}}
\put(309,58){\circle*{3}}
\put(303,28){\circle*{3}}
\put(303,40){\circle*{3}}
\put(303,52){\circle*{3}}
\put(297,34){\circle*{3}}
\put(297,46){\circle*{3}}
\put(291,40){\circle*{3}}

\end{picture}

\section{Examples and Non-degeneracy Issues} \label{non_deg}
In this section we show some examples of ``nice" partial metrics, based
on valuations for vertical and interval domains of real
numbers. Some of these valuations are strongly non-degenerate,
while others are not, yet all examples are quite natural.

Consider the example from Subsection~\ref{vertical_segment}.
The partial metric, based on the strongly non-degenerate
valuation $\mu'$ of that example would be $u'(x,y)=
(1-\min(x,y))/(1+\epsilon)$, if $x,y>0$, and $u'(x,y)=1$,
if $x$ or $y$ equals to $0$. However, another nice valuation, $\mu''$,
can be defined on the basis of $\mu$ of Subsection~\ref{vertical_segment}:
$\mu''((x,1])=\mu((x,1])=1-x$, $\mu''([0,1])=1$. $\mu''$ is not
strongly non-degenerate, however it yields nice
partial metric $u''(x,y)=1-\min(x,y)$, yielding the Scott topology.

Now we consider several valuations and distances on the domain of interval
numbers located within the segment $[0,1]$. This domain
can be thought of as a triangle of pairs $\langle x,y \rangle$, 
$0\le x \le y \le 1$.
Various valuations can either be concentrated on $0<x\le y <1$,
or on $x=0$, $0 \le y \le 1$ and $y=1$, $0 \le x \le 1$,
or, to insure non-degeneracy, on the both of these areas with an
extra weight at $\langle 0,1 \rangle$.

The classical partial metric $u([x,y],[x',y']) =
\max(y,y')-\min(x,x')$ results from the valuation accumulated
at $x=0$, $0\le y \le 1$, and $y=0$, $0 \le x \le 1$,
namely $\mu(U)=(Length(\{x=0,\, 0\le y\le 1\} \cap U)+
Length(\{y=1,\, 0\le x \le 1\}\cap U))/2$.
Partial metrics denerated by strongly non-degenerate valuations contain
quadratic expressions.

It is our current feeling, that instead of trying to formalize
weaker non-degeneracy conditions, it is often more fruitful to change
a $\muInfo$-structure.
In particular, one can build a $\muInfo$-structure
based on ${\cal I} = [0,1] \times [0,1]$
in the situation described above.

\chapter{Negative Information and Tolerances} \label{chap:Neginfo}

In this chapter we present our later joint results with
Svetlana Shorina~\cite{BukatinShorina3,BukatinShorina4}.
In the previous chapter 
we showed how to obtain such a $\muInfo$-structure
from a CC-valuation for any continuous Scott domain.
Certain pathologies in the behavior of
$I_x=\{y\in A\,|\, \{x,y\} {\rm \; is \;
unbounded}\}$
precluded us from extending this
method beyond bounded complete domains.

In this chapter, we obtain meaningful Scott continuous relaxed metrics
for continuous dcpo's by replacing $I_x$ with
$\Neginfo(x) = \Int (J_x)$,
where $J_x = \{y\in A\,|\, x \in \Int(I_y)\}$.
This result can be understood
in terms of interplay between {\em negation duality} and
{\em Stone duality}.
Since we know how to construct a CC-valuation for any
continuous dcpo with countable basis, this method of
constructing partial and relaxed metrics via
CC-valuations and the resulting $\muInfo$-structures
works for all continuous dcpo's with countable bases.

Escardo~\cite{Escardo, Smyth2} defined a topological space $A$ to be
{\em weakly Hausdorff}, if its consistency relation is closed.
The consistency relation is given by formula
$x \uparrow y = \{\langle x, y \rangle\,|\, \exists z \in A.\, x
\sqsubseteq z \& y \sqsubseteq z\}$, where $\sqsubseteq$ is the
specialization order of $A$.

A continuous dcpo with Scott topology is
weakly Hausdorff, if and only if
negative information $I_x$ is observable, e.g. Scott open,
due to the fact that the relation
$\{\langle x, y \rangle\,|\, y \in I_x\}$ is exactly the complement of
the consistency relation.

The technique of considering $\Neginfo(x) = \Int (J_x)$ works especially well,
when a continuous dcpo $A$ satisfies the {\em Lawson condition} ---
the relative Scott and Lawson topologies on $\Total(A)$ are equal.
We obtain that the Lawson condition is equivalent to the
property $\forall x \in \Total(A).\, I_x = J_x$. This will imply
that when the Lawson condition holds,
an induced classical metric on $\Total(A)$
results.

Since the Lawson condition is
equivalent to the formula $\forall x \in \Total(A).\, I_x = J_x$,
which is a weakening of $I=J$, which is, in turn, equivalent to
$A$ being weakly Hausdorff, we call the spaces satisfying the Lawson
condition {\em very weakly Hausdorff}.

\section{Tolerances and a Smyth Conjecture}

Recently Mike Smyth~\cite{Smyth2} and Julian Webster~\cite{Webster}
advanced the approach in which tolerance is considered not as
an alternative to standard topology, but as a structure complementary 
to topology. In particular, it seems to be fruitful to equip
Scott domains with tolerances.

Sections~\ref{sec:smyth} 
and~\ref{sec:lowerb} make a small contribution to this emerging
theory.

Smyth~\cite{Smyth2} defines a tolerance as a reflexive symmetric
relation following Poincare and Zeeman. He defines a topological
tolerance space as a topological space equipped with a
tolerance relation closed in the product topology.

For a weakly Hausdorff space, $\uparrow$ is
the least closed tolerance.

The set
$I_x=\{y\in A\,|\, \{x,y\} {\rm \; is \;
unbounded}\}$ is an observable continuous representation of negative
information about $x \in A$
for a weakly Hausdorff continuous dcpo $A$ with the Scott
topology. We will see in Section~\ref{sec:bad} that, 
when $A$ is not weakly Hausdorff the largest continuous
approximation of $I_x$ is represented by
$J_x = \{y\in A\,|\, x \in \Int(I_y)\}$,
and the largest observable continuous representation of $I_x$
is represented by $J'_x = \Int (J_x)$.

Smyth conjectured, that $J$ or $J'$ is closely related to
the least symmetric closed tolerance on $A$. In this chapter we
establish that, indeed, $\{\langle x, y \rangle\,|\, y \in J'_x\}$
is the complement of this tolerance.

We also establish a relationship between this tolerance and
lower bounds of relaxed metrics on $A$.

\section{Overview of the Chapter}

\subsection{Negation Duality and Problems with $I_x$}

Certain pathologies in the behavior of $I_x$ do not allow
us to use the formula 
$\Neginfo(x) = I_x = \{y\in A\,|\,\{x,y\}$ is
unbounded$\}$
when $A$ is not bounded complete.
It is convenient to use {\em negation duality},
$x \in I_y \Leftrightarrow y \in I_x$, when analyzing the behavior
of $I_x$. Specifically, one can consider directed sets $B \subseteq A$,
and by considering $x = \sqcup B$ and $x = b \in B$, obtain the
following Lemma:

\begin{Lemma} For any dcpo $A$, 
$(\forall y \in A.\, I_y \mbox{ is Scott open})
\Leftrightarrow 
(\forall B \subseteq A.\,
B \mbox{ is directed} \Rightarrow 
I_{\sqcup B} = \bigcup_{b \in B} I_b)\ .$
\end{Lemma}

\Proof
$\Rightarrow$. Consider $y \in I_{\sqcup B}$. 
$y \in I_{\sqcup B} \Leftrightarrow \sqcup B \in I_y$.
If $I_y$ is Scott open, $\exists b \in B.\ b \in I_y$,
and $b \in I_y \Leftrightarrow y \in I_b$.

$\Leftarrow$. Consider directed $B$, such that $\sqcup B \in I_y$.
$\sqcup B \in I_y \Leftrightarrow y \in I_{\sqcup B}$.
Condition $I_{\sqcup B} = \bigcup_{b \in B} I_b$ implies
that $\exists b \in B.\ y \in I_b$, and
$y \in I_b \Leftrightarrow b \in I_y$.
\eproof

One can informally restate this, by saying that all $I_y$ are
(Scott) observable~\cite{Smyth}, if and only if $I$ is a Scott
continuous function $A \rightarrow {\cal P}(A)$.

We will see that for some continuous dcpo's certain $I_y$'s are not
Scott open. This would be possible to tolerate, since, as we will see
later, co-continuity of valuations allows to extend those valuations to
Alexandrov open sets, and any $I_y$ is still Alexandrov open.
However, the corresponding breakdown of equality
$I_{\sqcup B} = \bigcup_{b \in B} I_b$ for directed sets $B$
cannot be tolerated, because it tends to lead to the loss of
Scott continuity for the resulting relaxed metrics.

We could have replaced $I_x$ with $\Int(I_x)$ (here and everywhere
in this chapter, interiors are being taken in Scott topology),
to rectify the problem of $I_x$ being not Scott open, but this
does nothing to fix the broken continuity property.
Somewhat surprisingly, the {\em negation duality} helps us to
resolve this problem. 

\subsection{Negation Duality and Stone Duality}

We consider equality $I_x = \{y \in A\,
|\,x \in I_y\}$, which is just another way to state the
negation duality. Then, instead of taking $\Int(I_x)$,
we consider $J_x = \{y \in A\,|\, x \in \Int(I_y)\}$.
Then the continuity property for the
resulting relaxed metric will be restored.
It is going to be technically convenient to replace $I_x$
not with $J_x$, but with $\Int(J_x)$, but we do not think that this
feature is principal.

We will see that the reasons for $J_x$ to work in this situation
can be best understood in terms of Stone duality.
In particular,
the map $y \mapsto \Int(I_y)$ gives rise to a map of the generators
$\uparrow\!\!\!\{y\}$
of the free frame of Scott open sets of the powerset of $A$ (ordered by
inclusion, the original ordering on $A$ is ignored) to the frame
of open sets of the domain $A$. 
Function $x \mapsto J_x$ turns out to be a Scott continuous function
$A \rightarrow {\cal P}(A)$, which is dual to the function
${\cal O}({\cal P}(A)) \rightarrow {\cal O}(A)$ obtained by extending
map $\uparrow\!\!\!\{y\} \mapsto \Int(I_y)$ onto the frame
${\cal O}({\cal P}(A))$.
We should emphasize here, that
what is going on in this chapter is a rather subtle interplay of
two {\em different} dualities --- negation duality and Stone duality ---
none of which seems to be reducible to another.

We will also see, that the function $x \mapsto J_x$ is the
largest ``negative'' Scott continuous function $A \rightarrow {\cal P}(A)$,
where the powerset of $A$, ${\cal P}(A)$, is ordered by the set-theoretic
inclusion.
This means that if $f: A \rightarrow {\cal P}(A)$ is
another Scott continuous function, 
such that for all $x \in A$, $f(x) \subseteq I_x$,
then for all $x \in A$, $f(x) \subseteq J_x$.

\subsection{Lawson Condition}

In general we only get deficient $\muInfo$-structures via the use of
$\Neginfo(x) = J_x$ or $\Neginfo(x) = \Int(J_x)$.
Namely, the totality property does not hold in general,
and thus we still do not get an induced metric on $\Total(A)$.
However, in this situation it helps to impose the {\em Lawson
condition}, that relative Scott and Lawson topologies
on $\Total(A)$ are equal. 

The Lawson condition was introduced in~\cite{Lawson}, and is
widely used lately. As Lawson writes in the
Introduction to~\cite{Lawson}, ``this turns out to be a very fruitful
notion that permit great generality, but at the same time permits
the derivation of many important structure results''. 
This condition is now a standard part of the notion of
a {\em computational model} for a topological space~\cite{Flagg}.

In this chapter, we present two equivalent formulations of the
Lawson condition: $\forall x \in A.\, I_x \cap \Total(A) =
\Int(I_x) \cap \Total(A)$ and $\forall x \in \Total(A).\, J_x = I_x$.

The second of these equivalent formulations implies
that if the Lawson condition holds
for the domain $A$ and, hence, for any 
maximal element $x$, $\Int(J_x) = J_x = I_x$,
then the totality property,
$C_x \cup \Neginfo(x) = A$, holds, and the induced metric on $\Total(A)$
results. Of course, the resulting metric topology is the same as
relative Scott and Lawson topologies on $\Total(A)$.

\subsection{Polish Spaces}

Lawson has shown in~\cite{Lawson}, that for every continuous
dcpo $A$ with countable basis, Lawson condition implies that
$\Total(A)$ is a Polish space, i.e. that it is homeomorphic to
a complete, separable metric space.

However, since completeness of metric spaces is not a topological
invariant, this does not mean that metrics obtained by our
present methods have to be complete. Indeed, our present methods,
based on assignment of converging systems of finite weights for
the case of algebraic domains, yield a non-complete metric on
$\Total(E')$ for the domain $E'$ of Section~\ref{sec:lawson}.

This raises a lot of open questions, ranging from the question
of when our construction yields a complete metric space to
the question of whether methods of Edalat and Heckmann, used to
approximate complete metric spaces (see~\cite{Heckmann} for
the variant of their approach using partial metrics and, thus,
closest to our methods), can be extended to describe certain
non-complete metric spaces, like an open interval of the real line
or set $\{1/2, 1/4, 1/8, \ldots \}$.

\subsection{Historical Remarks}

Bob Flagg noted, that our results from Section~\ref{sec:good}
can be generalized to continuous dcpo's with compact Lawson topology.
Mike Smyth observed that this is a corollary of the following two facts.
The first fact is
that the conditions that all $I_x$ are Scott open, that $I=J$,
and that the continuous dcpo is weakly Hausdorff are equivalent.
The second fact is that continuous dcpo's with compact
Lawson topology are weakly Hausdorff~\cite{Smyth2}.

Since $I=J$ is equivalent to the space in question being weakly
Hausdorff, and since the Lawson condition can be reformulated as
$I_x = J_x$ for maximal $x$, we can offer an alternative name for
the spaces, satisfying the Lawson condition --- {\em very weakly
Hausdorff spaces}.

Using this terminology, we can say that one of the discoveries
of this chapter is that continuous dcpo's do not have to be
weakly Hausdorff to be satisfactorily 
relaxed metrizable by our methods,
but it is enough to require that they be very weakly Hausdorff.

\subsection{Structure of the Chapter}

In Section~\ref{sec:good} we show that the old formulas, based on $I_x$,
work for the class of {\em coherently continuous} dcpo's with countable
basis,
because $I_x$'s are Scott open for this class of domains.

In Section~\ref{sec:bad} we analyze the pathologies of behavior
of $I_x$ on a specific example. We then study the properties of
$J_x$, which serves as a replacement for $I_x$, and explain those
properties from the viewpoint of Stone duality.

In Section~\ref{sec:lawson} we find equivalent formulations of
the Lawson condition and use these formulations to establish the
totality property of the resulting $\muInfo$-structures
with its ramifications for the induced metrics on $\Total(A)$.

In Section~\ref{sec:smyth} we talk about tolerances and prove
the Smyth Conjecture.

In Section~\ref{sec:lowerb} we establish that for the relaxed
metrics defined above, $x \not\sim y$ if and only if $l(x,y) \neq 0$,
and build a continuous family of tolerances.

\section{When $I_x$ Behaves Well}\label{sec:good}

In this section we study cases, when for all $x \in A$, $I_x$ is Scott open,
or, equivalently, when for all directed $B \subseteq A$,
$I_{\sqcup B} = \bigcup_{x \in B} I_x$. We already know, that this
situation takes place for continuous Scott domains.

Another class of domains, for which these properties can be established,
is the class of coherently continuous dcpo's with countable basis.
The term ``coherence'' here is understood in the weak sense
of~\cite{Abramsky} (weaker, than bounded completeness), and not in the
strong sense of~\cite{Gunter} (stronger, than bounded completeness).

\Def We say that a continuous dcpo $A$ with the basis $K$ is {\em
coherently
continuous}, if for any two basic elements $k, k' \in K$,
the set of their minimal upper bounds, $\MUB (k, k')$, is finite,
and for any $x \in A$, if $k \sqsubseteq x$ and $k' \sqsubseteq x$,
then there is $k'' \in \MUB (k, k')$, such that $k'' \sqsubseteq x$.

\begin{theorem} If $A$ is a coherently continuous dcpo with countable basis
K,
then for any $x \in A$, $I_x$ is open.
\end{theorem}
\Proof
Consider $y \in I_x$. Since the space has a countable basis,
we only need to show, that if there is a sequence of basic
elements, $k_1 \sqsubseteq k_2 \sqsubseteq \ldots$,
such that $y = \sqcup k_i$, then some $k_i$ belongs to $I_x$.

By contradiction, assume that this is not the case. Then
for all $i$, $k_i$ and $x$ have an upper bound. Using the
presence of a countable basis again, approximate $x$ with
a sequence of basic elements as well: $l_1 \sqsubseteq l_2
\sqsubseteq \ldots$, $x = \sqcup l_i$. Then for all $i$,
$k_i$ and $l_i$ have an upper bound.

Now we are going to build the sequence of (not necessarily basic)
elements, $u_1 \sqsubseteq u_2 \sqsubseteq \ldots$,
such that for all $i$, $k_i \sqsubseteq u_i$ and $l_i \sqsubseteq u_i$.
Then $\sqcup u_i$ would be an upper bound of $x$ and $y$, yielding
the desired contradiction.

Consider an element $v \in \MUB (k_i, l_i)$. Define the height
of $v$ as maximal $j$, such that there is $w \in \MUB (k_j, l_j)$,
such that $v \sqsubseteq w$. If there is no such maximal natural number,
we say that $v$ is of infinite height. Using coherence condition,
it is easy to see, that there is an element $u_1 \in \MUB (k_1, l_1)$
of infinite height. Now consider only elements $v \in \MUB (k_2, l_2)$,
such that $u_1 \sqsubseteq v$. Using coherence condition, it is easy to
see once again, that there is an element $u_2 \in \MUB (k_2, l_2)$ of
infinite height, such that $u_1 \sqsubseteq u_2$. Continuing this
process, we obtain the desired sequence.
\eproof

Therefore, one can use $N_x = I_x$ in order to obtain all results
of the previous section not only for continuous Scott domains,
but also for coherently continuous dcpo's with countable bases.

\section{When $I_x$ Behaves Badly}\label{sec:bad}

\subsection{Example}

Let start with the example. We define an algebraic countable dcpo $E$,
as a following subset of the powerset of $\Z$, ordered by subset inclusion.
$E = \{ \emptyset, \{1\}, \{1,2\}, \{1,2,3\}, \ldots,$
$\{1,2,3, \ldots\},
\{0\}$, $\{1, 0, -1\}, \{1,2, 0, -2\}$, $\{1,2,3,0, -3\}$, $\ldots \}$.
For convenience, we introduce a unique letter denotation for each of the
elements of $E$: $\bot_E = \emptyset$, $e_1 = \{1\}$, $e_2 = \{1,2\}$,
$e_3 = \{1,2,3\}, \ldots, e_{\infty} = \{1,2,3, \ldots\}$,
$0_E = \{0\}$, $f_1 = \{1,0, -1\}$, $f_2 = \{1,2, 0, -2\}$,
$f_3 = \{1,2,3,0, -3\}, \ldots$ We will use this notation throughout
the chapter. Observe, that all elements, except $e_{\infty}$, are
compact, that elements $e_{\infty}, f_1, f_2, \ldots$ are total,
and that $e_i \sqsubseteq f_j$ iff $i \leq j$.

\begin{figure}[h]
\begin{picture}(320,160)
\put(170,20){\circle*{3}}
\put(150,18){$\bot_E$}
\put(170,20){\line(0,1){20}}
\put(170,40){\circle*{3}}
\put(150,38){$e_1$}
\put(170,40){\line(0,1){20}}
\put(170,60){\circle*{3}}
\put(150,58){$e_2$}
\put(170,60){\line(0,1){20}}
\put(170,80){\circle*{3}}
\put(150,78){$e_3$}
\put(170,100){\circle{3}}
\put(170,110){\circle{3}}
\put(170,120){\circle{3}}
\put(170,140){\circle*{3}}
\put(150,138){$e_{\infty}$}
\put(230,40){\circle*{3}}
\qbezier(170,20)(200,30)(230,40)
\put(210,40){$0_E$}
\qbezier(230,40)(220,90)(210,140)
\put(210,140){\circle*{3}}
\put(208,150){$f_1$}
\qbezier(230,40)(230,90)(230,140)
\put(230,140){\circle*{3}}
\put(228,150){$f_2$}
\qbezier(230,40)(240,90)(250,140)
\put(250,140){\circle*{3}}
\put(248,150){$f_3$}
\put(270,140){\circle{3}}
\put(280,140){\circle{3}}
\put(290,140){\circle{3}}
\qbezier(170,40)(220,50)(210,140)
\qbezier(170,40)(220,50)(230,140)
\qbezier(170,40)(220,50)(250,140)
\qbezier(170,60)(220,80)(230,140)
\qbezier(170,60)(220,80)(250,140)
\qbezier(170,80)(210,110)(250,140)
\end{picture}
\caption{Domain $E$}
\end{figure}

Now we will see how {\em negation duality} works in this example.
In our previous terminology, $x = e_{\infty}$, $y=0_E$.
The role of a directed set $B$ is played by an increasing sequence,
$e_1 \sqsubseteq e_2 \sqsubseteq e_3 \sqsubseteq \ldots$.
Notice that $e_{\infty} = \sqcup B$.

Note also that $I_{0_E} = \{e_{\infty}\}$ and $0_E \in I_{e_{\infty}}$. 
You see, that $I_{0_E}$ is not Scott open (although 
$I_{e_{\infty}}$ is Scott open),
and dually, we obtain that $0_E \not\in \bigcup_{b \in B} I_b$
(due to observation, that $I_{e_1} = \emptyset,
I_{e_2} = \{f_1\}, I_{e_3} = \{f_1,f_2\},
\ldots$), thus breaking  $I_{\sqcup B} = \bigcup_{b \in B} I_b$.

The breaking of this equality prevents the resulting relaxed metrics
from being Scott continuous, as $0_E$ is compact and should naturally
carry a finite weight. Since all $I_b$ and $I_{\sqcup B}$ are Scott open,
taking $\Int(I_x)$ as $N_x$ instead of $I_x$ would not fix this problem. 

\subsection{Solution}

Let us rewrite the negation duality as $I_x = \{y\,|\,x \in I_y\}$.
What works, somewhat surprisingly, is taking a subset of $I_x$ via
taking the interior inside the right-hand side of this expression:
$J_x = \{y\,|\,x \in \Int(I_y)\}$.

\begin{Lemma}\label{sec:jgood}
If $B$ is a directed set, 
$J_{\sqcup B} = \bigcup_{b \in B} J_b$.
\end{Lemma}
\Proof
A potentially non-trivial part is to prove
$J_{\sqcup B} \subseteq \bigcup_{b \in B} J_b$.
Consider $y \in J_{\sqcup B}$. By definition of $J$,
$\sqcup B \in \Int(I_y)$. Since $\Int(I_y)$ is Scott open,
there is $b \in B$, such that $b \in \Int(I_y)$, that is
$y \in J_b$.
\eproof

In the next subsection, we explain this result in terms
of Stone duality. In our example domain $E$, $J_{e_{\infty}}$
does not include $0_E$, unlike $I_{e_{\infty}}$, yielding
$J_{e_{\infty}} = \bigcup_i J_{e_i}$.

In general, $J_x$ is Alexandrov open, but does not have to
be Scott open. E.g., in our example domain $E$,
we have that
$J_{0_E} = I_{0_E} = \{e_{\infty}\}$, and thus $J_{0_E}$ is not Scott open.
Due to co-continuity we can extend $\mu$ to Alexandrov open
sets $V$, by defining $\mu (V) = \mu (\Int(V))$.

However, in order to use a setup of Section~\ref{new_part_metric},
it is much more convenient to define $N_x = \Int(J_x)$ and to use
the following Lemma.

\begin{Lemma} If for arbitrary Alexandrov open sets $J, J_m, m \in M$,
the equality $J = \bigcup_{m \in M} J_m$ holds,
then $\Int(J) = \bigcup_{m \in M} \Int(J_m)$ holds as well.
\end{Lemma}
\Proof
The potentially non-trivial direction is to prove
$\Int(J) \subseteq \bigcup_{m \in M} \Int(J_m)$.
Consider $y \in \Int(J)$. By the Border Lemma (Lemma~\ref{border_lemma})
there is $x \in J$, such that $x \ll y$. Then, because of the
condition of the Lemma we are currently proving, there is $m \in M$,
such that $x \in J_m$. Then applying the Border Lemma again,
we obtain $y \in \Int(J_m)$.
\eproof

Hence, $N_x = \Int(J_x)$ enables us to satisfy all the requirements
of the setup of Section~\ref{new_part_metric}, except for the requirement
that for all $x \in \Total(A), C_x \cup N_x = A$, which does not hold
in general. E.g. consider our example domain $E$, and observe that
$e_{\infty} \in \Total(A)$, but 
$0_E \not\in C_x \cup N_x$. (Observe, also that changing
$N_x$ to $J_x$ does not fix this.)

Thus the Theorem~\ref{MainTheorem_a} holds, but the 
Theorem~\ref{OnTotals_a}
about the equality of $l(x,y)$ and $u(x,y)$
does not have to hold, and the resulting
induced metric on $\Total(A)$ cannot in general be obtained. E.g.,
in our example domain $E$,
we have that
$e_{\infty} \in \Total(E)$, but 
$u(e_{\infty},e_{\infty}) = \mu(I_{e_{\infty}}) -
\mu(J_{e_{\infty}}) = \mu(\{0_E\})$, which
is, in general, not zero, since $0_E$ is compact.

\subsection{Stone Duality}

The first Lemma in the previous subsection holds for the reasons,
which are not related to such specific
features of $J_x$ as the use of $\Int(I_y)$ (any open set
can be used instead) and the fact, that
$x$ and $y$ belong to the same set $A$.

We analyze this situation in the spirit of 
{\em Stone duality}~\cite{Johnstone,Vickers}, which is a contravariant
equivalence between categories of spatial frames (of open sets)
and sober topological spaces.

For the purpose of this subsection only, assume that there is
a continuous dcpo $A$ (Scott topologies of continuous dcpo's are 
sober~\cite{Johnstone})
and a set $D$, and that we are given a map $U: D \to {\cal O}(A)$,
where ${\cal O}(A)$ is the frame of Scott open sets of the
domain $A$.

Now generalize the construction of $J_x$ by considering the map
$J: A \to {\cal P}(D)$, where ${\cal P}(D)$ is a powerset of $D$
ordered by set-theoretic inclusion and equipped with the Scott topology.
Define $J$ by the formula: $J: x \mapsto \{y \in D\,|\,x \in U(y)\}$. 
Then observe that the proof of Lemma~\ref{sec:jgood}
still goes through, implying that $J$ is a Scott continuous
function from $A$ to ${\cal P}(D)$.

Now observe, if one applies $J^{-1}$ to a subbasic
open set $\uparrow\!\!\!\{y\}$, one obtains \linebreak
$J^{-1}(\uparrow\!\!\!\{y\}) = \{x\,|\,J(x) \in \uparrow\!\!\!\{y\}\} =
\{x\,|\,y \in J(x)\} = \{x\,|\,x \in U(y)\} = U(y)$.

Thus the map
$U$ can be thought of as defined on the generators
$\uparrow\!\!\!\{y\}$ of the
free frame of all Scott open sets on ${\cal P}(D)$
and giving raise to the frame homomorphism
$u: {\cal O}({\cal P}(D)) \to {\cal O}(A)$ (of course, $u = J^{-1}$,
e.g. for basic open sets, $u(\uparrow\!\!\!\{y_1, \ldots, y_n\}) =
U(y_1) \cap \ldots \cap U(y_n)$, and the similar thing goes
for unions of basic sets).

Now, since we are dealing with sober spaces, Stone duality means,
that not only $u=J^{-1}$ can be obtained from the continuous
function $J$, but also the continuous function $J$ can be restored
from the frame morphism $u$. And this is the essence of our
definition of $J$, when we think about $U$ as defined on the
generators of the frame ${\cal O}({\cal P}(D))$.

\subsection{$J_x$ Is the Largest Continuous Approximation of $I_x$}

Both $I_x$ and $J_x$ can be considered as functions from $A$ to the
powerset of $A$, ${\cal P}(A)$. However, in general, only $J_x$
is Scott continuous.
The following theorem shows that, in some sense, $J_x$ is the best
we can do.

\begin{theorem}
If $f: A \rightarrow {\cal P}(A)$ is a Scott continuous function
and $\forall x \in A.\,f(x) \subseteq I_x$, then
$\forall x \in A.\, f(x) \subseteq J_x$. 
\end{theorem}
\Proof
Assume that such Scott continuous function $f$ is given,
and for some $x$, there is $y \in f(x)$, such that $y \not\in J_x$,
i.e. $x \not\in \Int(I_y)$.
However, $y \in I_x$ means $x \in I_y$. Now consider a directed
set $B=K_x$. We have that $x = \sqcup B$ and that all $b \in B$
are way below $x$. Then, taking into account $x \not\in \Int(I_y)$
and applying the Border Lemma, we obtain that $\forall b \in B.\, b
\not\in I_y$.

However, the assumption of continuity of $f$ means, that
$f(\sqcup B) = \bigcup_{b \in B} f(b)$. Hence, since $y \in f(\sqcup B)$,
there is some $b \in B$, such that $y \in f(b)$, hence $y \in I_b$,
hence $b \in I_y$, contradicting the last formula of the previous
paragraph.
\eproof

A similar statement holds for $\Int(I_x)$ and $\Int(J_x)$,
understood as functions from $A$ to the dual domain of open sets
of $A$.

\section{The Use of the Lawson Condition}\label{sec:lawson}

We start with the equivalent formulation of the Lawson condition.

\begin{Lemma} Given a continuous dcpo $A$, the relative Scott and Lawson
topologies on $\Total(A)$ coincide if and only if for all $x \in A$,
$I_x \cap \Total(A) = \Int(I_x) \cap \Total(A)$.
\end{Lemma}

For example, in the domain $E$ from the previous section
Lawson condition does not hold. Indeed,
$I_{0_E} = \{e_{\infty}\}$ and $e_{\infty} \in \Total (E)$.  
However, $\Int(I_{0_E}) = \emptyset$. 
The set $\{e_{\infty}\}$ is open in the
relative Lawson topology, but not in the relative Scott topology.

Now we are going to modify the domain $E$, in order to obtain a
different example, which would satisfy the Lawson condition.
We add a new element, $\star$, to $\Z$, so that $E'$ will be
a subset of the powerset of $\Z \cup \{\star\}$, ordered by
the set-theoretic inclusion. Let $E' = E \cup \{e_{\star}\}$,
where $e_{\star} = \{\star, 1,2,3, \ldots\}$.

\begin{figure}[h]
\begin{picture}(320,180)
\put(170,20){\circle*{3}}
\put(150,18){$\bot_E$}
\put(170,20){\line(0,1){20}}
\put(170,40){\circle*{3}}
\put(150,38){$e_1$}
\put(170,40){\line(0,1){20}}
\put(170,60){\circle*{3}}
\put(150,58){$e_2$}
\put(170,60){\line(0,1){20}}
\put(170,80){\circle*{3}}
\put(150,78){$e_3$}
\put(170,100){\circle{3}}
\put(170,110){\circle{3}}
\put(170,120){\circle{3}}
\put(170,140){\circle*{3}}
\put(150,138){$e_{\infty}$}
\put(170,140){\line(0,1){20}}
\put(170,160){\circle*{3}}
\put(150,158){$e_{\star}$}
\put(230,40){\circle*{3}}
\qbezier(170,20)(200,30)(230,40)
\put(210,40){$0_E$}
\qbezier(230,40)(220,90)(210,140)
\put(210,140){\circle*{3}}
\put(208,150){$f_1$}
\qbezier(230,40)(230,90)(230,140)
\put(230,140){\circle*{3}}
\put(228,150){$f_2$}
\qbezier(230,40)(240,90)(250,140)
\put(250,140){\circle*{3}}
\put(248,150){$f_3$}
\put(270,140){\circle{3}}
\put(280,140){\circle{3}}
\put(290,140){\circle{3}}
\qbezier(170,40)(220,50)(210,140)
\qbezier(170,40)(220,50)(230,140)
\qbezier(170,40)(220,50)(250,140)
\qbezier(170,60)(220,80)(230,140)
\qbezier(170,60)(220,80)(250,140)
\qbezier(170,80)(210,110)(250,140)
\end{picture}
\caption{Domain $E'$}
\end{figure}

Now $I_{0_E} = \{e_{\infty},e_{\star}\}$, so 
this is still not a Scott open set,
however, $\Int(I_{0_E}) = \{e_{\star}\} = 
I_{0_E} \cap \Total(E') = \Int(I_{0_E}) \cap
\Total(E')$. Notice, that $J_{0_E} = I_{0_E}$ here, 
so $J_{0_E}$ is still only
Alexandrov open. $J_{e_{\infty}}$ 
is the same as in $E$, but now $e_{\infty}$ is not
a total element. However, $J_{e_{\star}} = J_{e_{\infty}} \cup 
\{0_E\}$, because
$e_{\star} \in \Int(I_{0_E})$, so $J_{e_{\star}} = 
I_{e_{\star}} = \Int(J_{e_{\star}})$, and
$C_{e_{\star}} \cup J_{e_{\star}} = 
C_{e_{\star}} \cup \Int(J_{e_{\star}}) = E'$. The set $\{e_{\star}\}$
is open in both Lawson and Scott relative topologies on
$\Total(E')$.

What is going on here is described by the following Theorem.

\begin{theorem} A continuous dcpo $A$ satisfies the Lawson condition
if and only if
for all $x \in \Total(A)$, $J_x = I_x$.
Hence, if the Lawson condition holds, then
$C_x \cup J_x = C_x \cup \Int(J_x) = A$.
\end{theorem}
\Proof
Assume that the Lawson condition holds and $x \in \Total(A)$.
Assume, that $y \not\sqsubseteq x$, i.e. $y \not\in C_x$ and
$y \in I_x$, using the totality of $x$.
Thus, by duality, $x \in I_y$. Because $x \in \Total(A)$ and
because due to the 
Lawson condition $I_y \cap \Total(A) = \Int(I_y) \cap \Total(A)$,
we obtain $x \in \Int(I_y)$, hence $y \in J_x$. 

Conversely, assume $\forall x \in \Total(A).\,J_x =I_x$.
Let us prove $I_y \cap \Total(A) = \Int(I_y) \cap \Total(A)$,
thus proving the Lawson condition.
Take $x \in I_y \cap \Total(A)$. By negation duality, 
$y \in I_x$, then, by totality 
of $x$ and our assumptions, $y \in J_x$, which, by definition of $J_x$,
means that $x \in \Int(I_y)$. 

The rest follows from the observation,
that if $x \in \Total(A)$, $I_x$ is Scott open.
\eproof

Hence if the Lawson condition holds, the resulting $\muInfo$-structure
is not deficient, and the Theorem~\ref{OnTotals_a} holds.

\section{Tolerances and Negative Information}\label{sec:smyth}

\subsection{Tolerances, Distinguishability, and Observability}

Smyth~\cite{Smyth2} requires that a tolerance relation is closed
in the product topology. Here are informal reasons for this.

The typical meaning of two points being in the relation of tolerance,
$x \sim y$, is that $x$ cannot be distinguished from $y$, i.e.
there is no way to establish, that $x$ and $y$ differ.

The natural way to interpret the statement, that $x$ and $y$ can
be distinguished, is to give some ``effective'' procedure for
making such a distinction. Thus, the property of being distinguishable
is observable~\cite{Smyth}. Correspondingly, the property $x \sim y$
is refutable, hence $\sim$ should be closed.

The fact that the least closed tolerance for a weakly Hausdorff
continuous dcpo is $\uparrow$ also is quite natural in this framework.
Indeed, domain elements are thought of as being only partially known
and dynamically increasing in the course of their lives. Hence the
fact that $x \uparrow y$, that is $\exists z.\, x \sqsubseteq z,
y \sqsubseteq z$, precisely means that $x$ and $y$ may approximate the
same ``genuine'' element $z$, hence we cannot distinguish between them.
Since $\uparrow$ is closed in the weakly Hausdorff case, its complement
is open, hence observable. That means that when $x \uparrow y$ does
not hold, there is some ``finite'' way to distinguish between $x$ and
$y$.

\subsection{$J'$ and the Least Closed Tolerance (a Smyth Conjecture)}

Consider a continuous dcpo $A$. In this subsection $x, y, v, w \in A$.
Recall that we defined
$x \not\sim y = \{\langle x,y \rangle\,|\, x \in J'_y\}$.

\begin{Lemma}
$x \not\sim y = \{\langle x, y \rangle\,|\, \exists \langle v, w
\rangle.\, v \ll x, w \ll y, \{v, w\}$ is unbounded$\}$.
\end{Lemma}

\Proof
Using the Border Lemma we get $x\in \Int(J_y)$ iff $\exists v \in
J_y.\, v \ll x$.  By the definition of $J_y$, $v \in J_y$ iff $y \in
\Int(I_v)$ i.e. $\exists w \in I_v.\, w \ll y$.  
Finally recall that $w \in
I_v $ iff $\{v,w\}$ is unbounded .
\eproof

It is an immediate corollary
that $\not\sim$ is symmetric. 

Let us show that $\not\sim$ is open in
the product topology. If we fix a pair $\langle v, w \rangle$ 
the set $\not\sim_{\langle v, w \rangle}=
\{x \,|\, v \ll x\}  \times \{ y \,|\, w \ll y \}$ is open, and our
$\not\sim$ is the union of all such sets for all unbounded pairs 
$\langle v, w \rangle$.

\begin{theorem}
The relation $\not\sim$ is the complement of the least closed tolerance.
\end{theorem}

\Proof 
Consider an open set $W \subseteq X \times X$, such that
$\not\sim \subset W$. Consider a pair 
$\langle x, y \rangle \in W \setminus
\not\sim$. Since $W$ is
open, we can choose two open sets $U \subseteq X$ and $V \subseteq X$, such that
$\langle x, y \rangle \in U \times V \subseteq W$. 
Consider a pair $\langle p, r \rangle, p \in U, r \in
V.  p \ll x, r \ll y$. 
The pair $\langle p, r \rangle$ is bounded, otherwise
$\langle x, y \rangle \in \not\sim$, 
so we can take $z$ such that $p \sqsubseteq z$, $r
\sqsubseteq z$, therefore $z \in U$, $z \in V$, so 
$\langle z, z \rangle \in U \times V$
and $\langle z, z \rangle \in W$. 
So the complement of $W$ is not a tolerance, because it
is not
reflexive.
\eproof

\subsection{Examples}

In our example domain $E$, the pair $\langle e_{\infty}, 0_E \rangle$ is
unbounded, but belongs to the least closed tolerance, since
these elements cannot be distinguished by looking at the approximation
pairs, $\langle e_1, 0_E \rangle,$ $\langle e_2, 0_E \rangle, \ldots$.

Moreover, by adding to domain $E$ elements
$g_2 = \{0, 2\}, g_3 = \{0, 2, 3\}, \ldots,
g_n = \{0, 2, 3, \ldots, n\},$ $\ldots$ and $g_{\infty} = 
\{0, 2, 3, 4, \ldots \}$,
we obtain a domain, where two different maximal elements, $e_{\infty}$ and
$g_{\infty}$, cannot be distinguished 
via finite observations, because all their
approximations are bounded.
Such a situation, when $\exists x, y \in \Total(A).\, x \sim y$,
where $\sim$ is the least closed tolerance, cannot
occur in the presence of the Lawson condition, because the Lawson condition
is equivalent to $\forall x \in \Total (A).\, I_x = J'_x$.

\section{Tolerances and Lower Bounds of Relaxed Metrics}\label{sec:lowerb}

We are going to prove the following statement.

\begin{theorem}
$x \not\sim y  \Leftrightarrow l(x, y) \neq 0$.
\end{theorem}

\Proof
$\Rightarrow$. Recall that
$l(x, y) = \mu (C_x \cap J'_y) + \mu (C_y \cap J'_x)$.
Notice that
$x \not\sim y$ implies $x \in C_x \cap J'_y$, which implies
that $C_x \cap J'_y \neq \emptyset$. Hence $J'_y \setminus C_x
\subset J'_y$, hence $\mu (C_x \cap J'_y) =
\mu (J'_y) - \mu (J'_y \setminus C_x) > 0$ due to the strong
non-degeneracy of $\mu$. Hence $l(x,y) > 0$.

$\Leftarrow$. $l(x,y) >0$ means $C_x \cap J'_y \neq \emptyset$ or
$C_y \cap J'_x \neq \emptyset$. It is enough to consider
$C_x \cap J'_y \neq \emptyset$. Since $x$ is the largest element
of $C_x$, we obtain $x \in J'_y$, hence $x \not\sim y$.
\eproof

Only upper bounds of relaxed metrics participate in the definition
of the relaxed metric topology. Hence lower bounds are usually considered
as only playing an auxiliary role in the computation of upper bounds.
Here we see an example of a quite different 
application of lower bounds.

\subsection{A Continuous Family of Tolerances}

A set $\{\langle x, y \rangle \,|\, l(x, y) \leq \epsilon\}$,
also forms a tolerance. Indeed, this is a symmetric, reflexive
relation. To see that it is closed, consider a Scott continuous
function $l: A \times A \rightarrow R^{+}$, where $R^{+}$ is
a segment $[0,1]$ with the usual ordering and the induced
Scott topology, and observe that the set in question is
the inverse image of a Scott closed set $[0, \epsilon] \subseteq R^{+}$
under $l^{-1}$.

The resulting family of tolerances parametrized by $\epsilon$ is
Scott continuous in the following sense. The dual domain for
$R^{+}$ is domain $R^{-}_{\top}$. Here $R^{-}$ is the same segment of numbers,
but with inverse ordering ($x \sqsubseteq y \Leftrightarrow x \geq y$),
an element $r \in R^{-}$ corresponds to the open set $(r, 1] \subseteq R^{+}$,
the element $\top \in R^{-}_{\top}$ corresponds to the open set $[0,1] = R^{+}$,
and domains $R^{-}$ and $R^{-}_{\top}$ are equipped with
the induced Scott topology.

The function $l^{-1} : R^{-}_{\top} \rightarrow {\cal O} (A \times A)$
is Scott continuous, and so is its restriction on $R^{-}$.
Then $l^{-1} (\epsilon)$ is the complement of the tolerance in
question, and we can think about this complement as representing
this tolerance in the dual domain ${\cal O} (A \times A)$.
 
\section{Open Problems}

It might be useful to extend the Stone duality analysis to
$\Int(J_x)$ and to be able to speak about the intuition behind
the Lawson condition in the spirit of~\cite{Smyth}.

Another open question is as follows.
If Lawson condition does not hold, can we obtain some negative
results about the existence of $\muInfo$-structures
with totality property, or, more generally,
about the existence of relaxed metrics
with the property $\forall x,y \in \Total(A).\,l(x,y) = u(x,y)$?
Obviously, this question allows a number of variations,
e.g. we know now, that when this
 question is restricted to the case of $\Info(x) = C_x$,
such negative results can indeed be obtained.

It seems that tolerances will play an increasingly important role
in domain theory. One particularly promising direction of development
is to use tolerances and especially their asymmetric generalizations
instead of transitivity of logical inference to formally
express and study the ideas of A.S.Esenin-Vol'pin and P.Vopenka,
that large numbers should be considered infinite,
and long proofs and computations should be considered 
meaningless~\cite{Vopenka}.

\part{Conclusion}\label{part:Conclusion}

In the preface we identified the key unsolved {\bf Problem A},
namely to learn how to find reasonable
approximations for a sufficiently wide class of definitions
of domain elements
and Scott continuous functions while spending realistic
amount of resources, as the main obstacle on the path
of wider practical applicability of domains with Scott
topology in software engineering.

In our work we achieved considerable progress in the
development of analysis on such domains. In conclusion,
we would like to stress the extreme importance of
finding analogs of various series decompositions
in domains with Scott topologies. The theory of
such decompositions should incorporate the Scott domain analogs
of such seemingly unrelated things as Fourier series
and decimal representations of real numbers and would
most likely represent a considerable step towards solving
the Problem A.

This direction of research is also closely related to
another aspect of Problem A, namely how to find
compact visual representations of domain elements. Such
representations, of course, should denote relatively
close approximations of the elements in question.

Failing the solution of Problem A, it is possible that
certain mathematical constructions in domain theory
might give us some hints on how to construct novel
algorithms.

\part{Bibliography}

\end{document}